\documentclass[a4paper]{article}
\usepackage{eurosym}
\usepackage{amsmath, amsfonts, amsthm,amssymb}
\usepackage{mathrsfs}
\usepackage[margin=2cm]{geometry}
\usepackage{xcolor}
\usepackage{colortbl}
\usepackage{graphicx}
\usepackage{caption}
\usepackage{multirow,makecell,multicol}
\usepackage{hhline,array,rotating}
\usepackage{stackengine}
\usepackage{lscape}
\usepackage[T1]{fontenc}
\usepackage{authblk}
\usepackage{rotating}
\usepackage{subcaption}

\newcolumntype{M}[1]{>{\centering\arraybackslash}m{#1}}
\newcommand\xrowht[2][0]{\addstackgap[.5\dimexpr#2\relax]{\vphantom{#1}}}
\allowdisplaybreaks
\makeatletter
\renewcommand*{\@fnsymbol}[1]{\ifcase#1 \or \@arabic{\numexpr#1 \relax}  \or \@arabic{\numexpr#1 \relax} \or \@arabic{\numexpr#1 \relax} \else \ * \fi}
\makeatother
\input{tcilatex}
\begin{document}

\title{Three-dimensional stress-gradient non-locality problem:\\
Displacement field for an arbitrary and a conservative force}
\author{Sla\dj an Jeli\'{c}\thanks{
Department of Physics, Faculty of Sciences, University of Novi Sad, Trg D.
Obradovi\'{c}a 4, 21000 Novi Sad, Serbia, sladjan.jelic@df.uns.ac.rs} \quad
Du\v{s}an Zorica\thanks{
Department of Physics, Faculty of Sciences, University of Novi Sad, Trg D.
Obradovi\'{c}a 4, 21000 Novi Sad, Serbia, dusan.zorica@df.uns.ac.rs} \ 
\thanks{%
Mathematical Institute, Serbian Academy of Arts and Sciences, Kneza Mihaila
36, 11000 Belgrade, Serbia, dusan\textunderscore zorica@mi.sanu.ac.rs} \ 
\thanks{%
Corresponding author}}
\maketitle

\begin{abstract}
\noindent In order to obtain the displacement field, both generally in the
case of an arbitrary body force and specifically in the case of a
conservative body force, the mechanical problem of statics of
three-dimensional stress-gradient non-local medium is analytically solved on
unbounded domain by considering a system of equations consisting of an
equilibrium equation, infinitesimal strain, and both scalar- and tensor-type
stress-gradient Eringen three-dimensional models. Since the system of
equations is linear, for each model the solution is obtained in the form of
a superposition of the local-type displacement field, common to each model,
and the non-local-type displacement field, specific to the non-locality
operator of the model. Among the three scalar-type and five tensor-type
considered models, six distinct analytical displacement field expressions
are obtained for an arbitrary body force: one expression specific to a
scalar-type model, three specific to individual tensor-type models, while
each of the remaining two expressions are common to one scalar- and one
tensor-type model. The displacement vector fields are graphically
represented in the case of gravitational body force, yielding five different
models, since two tensor-type models become the same when the force is
conservative.

\end{abstract}

\section{Introduction}

Non-local and gradient elasticity theories have been developed as
generalized framework in continuum mechanics in order to overcome some of
the limitations of classical elasticity, in particular its inability to
capture size effects, as well as singular behavior associated with
concentrated loads and material defects. Constitutive equations of non-local
elasticity, pioneered by Eringen, represented in integral form account for
long-range material interactions, in one-dimensional non-locality through
kernel $K$ as%
\begin{equation*}
\sigma \left( x\right) =\int_{%
\mathbb{R}
}K\left( x,x^{\prime }\right) \sigma ^{local} \left( x^{\prime }\right) 
\mathrm{d}x^{\prime },
\end{equation*}%
with non-local stress $\sigma $\ and local stress $\sigma ^{local}=E
\varepsilon$ if Hooke's law is imposed, where $E$ is Young's modulus and $%
\varepsilon$ is strain, whereas gradient elasticity introduces intrinsic
material length-scale parameters by incorporating higher-order spatial
gradients into the governing equations, in one-dimensional case represented
as stress-gradient Eringen model%
\begin{equation}
\left( 1-l^{2}\frac{\mathrm{d}^{2}}{\mathrm{d}x^{2}}\right) \sigma \left(
x\right) =\sigma ^{local}\left( x\right) ,  \label{1D-Eringen}
\end{equation}%
with the scalar quantity $l$ as a model/material/length-scale parameter,
see, e.g., \cite{AltanAifantis,ERINGEN19721,eringen1,eringen2,ERINGEN1977339}%
.

The present work builds upon the stress-gradient models formulated in \cite%
{SD-6}, where two systematically derived classes of three-dimensional
non-local models are established, comprising scalar- and tensor-type
formulations: scalar-type stress-gradient Eringen three-dimensional models%
\begin{equation}
\left( 1-\mathrm{S}_{i}\left( \nabla \right) \right) \boldsymbol{\hat{\sigma}%
}\left( \boldsymbol{r}\right) =\boldsymbol{\hat{\sigma}}^{local}\left( 
\boldsymbol{r}\right) ,\quad i\in \left\{ 1,2,3\right\} ,
\label{non-loc-scalar-type}
\end{equation}%
and tensor-type stress-gradient Eringen three-dimensional models%
\begin{equation}
\left( \boldsymbol{\hat{I}}-\mathbf{\hat{T}}_{i}\left( \nabla \right)
\right) \boldsymbol{\hat{\sigma}}\left( \boldsymbol{r}\right) =\boldsymbol{%
\hat{\sigma}}^{local}\left( \boldsymbol{r}\right) ,\quad i\in \left\{
1,2,3,4,5\right\} ,  \label{non-loc-tensor-type}
\end{equation}%
where $\boldsymbol{\hat{\sigma}}$ is a non-local Cauchy stress tensor, $%
\boldsymbol{\hat{\sigma}}^{local}$ is a local Cauchy stress tensor, and $%
\nabla $ denotes nabla operator, obtained within a unified theoretical
framework rather than through ad hoc constitutive assumptions. Namely,
three-dimensional generalization of one-dimensional stress-gradient Eringen
model (\ref{1D-Eringen}) is achieved by replacing the operator $l^{2}\frac{%
\mathrm{d}^{2}}{\mathrm{d}x^{2}}$ with scalar and tensor operators $\mathrm{S%
}_{i}=\mathrm{S}_{i}\left( \nabla \right) $ and $\mathbf{\hat{T}}_{i}=%
\mathbf{\hat{T}}_{i}\left( \nabla \right) $, listed in Table \ref{Modeli-new}%
. Additionally, the non-locality kernels of scalar- and tensor-type, namely $%
{\mathcal{K}}_{i}$ and $\boldsymbol{\hat{{\mathcal{K}}}}_{i}$, associated
with the proposed scalar- and tensor-type operators and also listed in Table %
\ref{Modeli-new}, are derived analytically and shown to characterize
non-local behavior of the material through the constitutive equations which
are equivalent with the gradient-type constitutive relations (\ref%
{non-loc-scalar-type}) and (\ref{non-loc-tensor-type}), namely 
\begin{equation*}
\boldsymbol{\hat{\sigma}}\left( \boldsymbol{r}\right) ={\mathcal{K}}%
_{i}\left( \boldsymbol{r}\right) \ast _{\boldsymbol{r}}\boldsymbol{\hat{%
\sigma}}^{local}\left( \boldsymbol{r}\right) \quad \text{and\quad }%
\boldsymbol{\hat{\sigma}}\left( \boldsymbol{r}\right) =\boldsymbol{\hat{{%
\mathcal{K}}}}_{i}\left( \boldsymbol{r}\right) \ast _{\boldsymbol{r}}%
\boldsymbol{\hat{\sigma}}^{local}\left( \boldsymbol{r}\right) ,
\end{equation*}%
where $\boldsymbol{r}\in 
\mathbb{R}
^{3}$ is a radius vector and $\ast _{\boldsymbol{r}}$ denotes the space
convolution, understood component-wise and defined by%
\begin{equation*}
f\left( \boldsymbol{r}\right) \ast _{\boldsymbol{r}}g\left( \boldsymbol{r}%
\right) =\int_{%
\mathbb{R}
^{3}}f\left( \boldsymbol{r}-\boldsymbol{r}^{\prime }\right) g\left( 
\boldsymbol{r}^{\prime }\right) \mathrm{d}_{\boldsymbol{r}^{\prime }}V,
\end{equation*}%
see \cite{SD-6} for details. Note, if one considers Hooke's law, then the
local Cauchy stress tensor $\boldsymbol{\hat{\sigma}}^{local}$ is connected
with the infinitesimal strain tensor $\boldsymbol{\hat{\varepsilon}}$
through $\boldsymbol{\hat{\sigma}}^{local}\left( \boldsymbol{r}\right)
=\lambda \,\mathrm{\limfunc{tr}}\boldsymbol{\hat{\varepsilon}}\left( 
\boldsymbol{r}\right) \boldsymbol{\hat{I}}+2\mu \,\boldsymbol{\hat{%
\varepsilon}}\left( \boldsymbol{r}\right) $, where $\lambda $ and $\mu $ are
Lam\'{e} coefficients.

The objective of the present study is to employ the three-dimensional
non-local stress-gradient models (\ref{non-loc-scalar-type}) and (\ref%
{non-loc-tensor-type}), proposed in \cite{SD-6}, in order to solve the
problem of statics of an unbounded three-dimensional non-local solid medium
under the influence of prescribed body force. Namely, complementing the
constitutive models (\ref{non-loc-scalar-type}) and (\ref%
{non-loc-tensor-type}) by the equilibrium equation and infinitesimal strain,
and by solving such obtained system of equations for the displacement vector
field $\boldsymbol{u}$, one obtains displacement vector field in integral
representation, in which the mechanical response of the three-dimensional
medium is expressed as the convolution of the resolvent (Green's) tensor $%
\boldsymbol{\hat{R}}$ and the body force distribution $\boldsymbol{f}_{b}$ as%
\begin{equation}
\boldsymbol{u}\left( \boldsymbol{r}\right) =\boldsymbol{\hat{R}}\left( 
\boldsymbol{r}\right) \ast _{\boldsymbol{r}}\boldsymbol{f}_{b}\left( 
\boldsymbol{r}\right) ,  \label{conv}
\end{equation}%
as it is shown in Section \ref{Displacement-field}. Such a formulation
naturally incorporates non-local interactions and internal length-scale
effects, through resolvent tensor, while simultaneously providing a
regularized description of the mechanical fields in comparison with
classical elasticity theories. Besides providing an analytical expression
for the problem solution, this representation establishes an unified
framework for analytical investigations and numerical methods based on
integral formulations, and it is applicable not only to elastic materials
but also to a broad class of non-local continuum models.

There are studies incorporating non-local effects within the framework of
three-dimensional non-local gradient elasticity theory, which are also
focused on analytical calculation of the displacement vector field as a
solution of various problems, including Helmholtz equation for the
displacement field 
\begin{equation*}
\left( 1-l^{2}\nabla ^{2}\right) \boldsymbol{u}\left( \boldsymbol{r}\right) =%
\boldsymbol{u}^{local}\left( \boldsymbol{r}\right)
\end{equation*}
in \cite{Aifantis,LazarMaugin}, equation containing displacement-driven
non-local constitutive model in which the strain equation is generalized as 
\begin{equation*}
\boldsymbol{\hat{\varepsilon}}\left( \boldsymbol{r}\right) =\frac{1}{2}%
\left( \bar{\nabla}\boldsymbol{u}\left( \boldsymbol{r}\right) +\left( \bar{%
\nabla}\boldsymbol{u}\left( \boldsymbol{r}\right) \right) ^{\mathrm{T}%
}\right) ,
\end{equation*}
by replacing nabla operator $\nabla $ with non-local operator $\bar{\nabla}$%
, see \cite{Patnaik}, non-local equilibrium equation 
\begin{equation*}
\left( \lambda +2\mu \right) \left( 1-l_{1}^{2}\nabla ^{2}\right) \left(
\nabla \left( \nabla \cdot \boldsymbol{u}\left( \boldsymbol{r}\right)
\right) \right) -\mu \left( 1-l_{2}^{2}\nabla ^{2}\right) \left( \nabla
\times \left( \nabla \times \boldsymbol{u}\left( \boldsymbol{r}\right)
\right) \right) =-\boldsymbol{f}_{b}\left( \boldsymbol{r}\right),
\end{equation*}
with Lam\'{e} coefficients $\lambda $ and $\mu ,$ see \cite%
{GourgiotisZisisGeorgiadis,Solyaev}, as well as twofold anisotropic
Helmholtz equation 
\begin{equation*}
LL_{ik}u_{k}\left( \boldsymbol{r}\right) +f_{i}=0,
\end{equation*}
with Navier operator $L_{ik}=C_{ijkl}\partial _{j}\partial _{l}$ regularized
by an additional Helmholtz-type operator $L=\Lambda _{mn}\partial
_{m}\partial _{n}$, yielding an integral solution given by convolution of
the form (\ref{conv}), see \cite{LazarPo1,LazarPo}. Additionally,
analytically derived solutions for the displacement field can be found for
certain specific cases, such as the non-singular Burgers formula considered
within the framework of gradient elasticity, see \cite{Lazar-2012,Lazar-2013}%
. Numerical methods are also developed for three-dimensional non-local
elasticity problems, see \cite{AnsariShahabodiniShojaei,SchwartzNianeNjiwa},
as well as an alternative iterative solution procedures, proposed in \cite%
{Shaat} for non-local boundary-value problems.

\begin{sidewaystable}
\begin{center}
\begin{tabular}{|c|c|c|c|c|c|} \hline \xrowht{25pt}
 & Model & \multicolumn{2}{c|}{Non-locality operator}  & \multicolumn{2}{c|}{Non-locality kernels}  \\ \cline{1-6} \xrowht{35pt}

\multirow{3}{*}{\multirowcell{1}{\rotatebox{90}{Scalar-type}}} & $S_1$ & \multirow{3}{*}{\multirowcell{4}{ ${\rm{S}}_{i}(\nabla)$ }} & $\left( \boldsymbol{L}\cdot \boldsymbol{L}\right) \left( \nabla \cdot \nabla \right)$  & \multirow{3}{*}{\multirowcell{4}{  ${\mathcal{K}}_{i}  \left( \boldsymbol{r}\right) $ }} & $\frac{1}{4\pi rL^{2}}\mathrm{e}^{-\frac{r}{L}}$ \\ \cline{2-2} \cline{4-4} \cline{6-6} \xrowht{35pt}

 & $S_2$  & & $\left( \boldsymbol{L}\cdot \nabla \right) ^{2} $ &  & $\frac{1}{2L}\delta \left( \boldsymbol{r}_{\perp }\right) \mathrm{e}^{-\frac{\left\vert r_{\parallel
}\right\vert }{L}}$ \\  \cline{2-2} \cline{4-4} \cline{6-6} \xrowht{35pt}

 & $S_3$ & & $\left(\boldsymbol{L}\times \nabla \right) \cdot \left( \boldsymbol{L}\times \nabla\right) $   &  & $\frac{1}{2\pi L^{2}}\delta \left( r_{\parallel }\right) K_{0}\left( \frac{r_{\perp }}{L}\right) $\\ \cline{1-6}  \xrowht{35pt}

\multirow{11.5}{*}{\multirowcell{3}{\rotatebox{90}{Tensor-type}}} & $T_{1}$ & \multirow{5}{*}{\multirowcell{10.75}{ $\boldsymbol{\hat{{\rm{T}}}}_{i}(\nabla)$ }} &$\left( \boldsymbol{L}\otimes \nabla \right) ^{\mathrm{T}}\left( \boldsymbol{%
L}\otimes \nabla \right) $ &  \multirow{5}{*}{\multirowcell{10.75}{  $\boldsymbol{\hat{{\mathcal{K}}}}_{i}  \left( \boldsymbol{r}\right) $ }} & $2\delta
\left( \boldsymbol{r}\right) \left( \boldsymbol{\hat{I}}-2\frac{\boldsymbol{r%
}\otimes \boldsymbol{r}}{r^{2}}\right) -\frac{1}{4\pi r^{3}}\mathrm{e}^{-%
\frac{r}{L}}\left( \left( 1+\frac{r}{L}\right) \boldsymbol{\hat{I}}+\left(
3\left( 1+\frac{r}{L}\right) +\frac{r^{2}}{L^{2}}\right) \frac{\boldsymbol{r}%
\otimes \boldsymbol{r}}{r^{2}}\right)$ \\  \cline{2-2} \cline{4-4} \cline{6-6} \xrowht{35pt}

 & $T_{2}$  & & $\left( \boldsymbol{L}\otimes \nabla \right) \left( \boldsymbol{L}\otimes
\nabla \right) ^{\mathrm{T}}$ &  &  $\delta
\left( \boldsymbol{r}\right) \left( \boldsymbol{\hat{I}}-\frac{\boldsymbol{L}%
\otimes \boldsymbol{L}}{L^{2}}\right) +\frac{1}{4\pi L^{2}r}\mathrm{e}^{-%
\frac{r}{L}}\frac{\boldsymbol{L}\otimes \boldsymbol{L}}{L^{2}}$ \\  \cline{2-2} \cline{4-4} \cline{6-6} \xrowht{35pt}

 & $T_3$ & & \makecell[l]{$\left( \boldsymbol{L}\otimes \nabla \right) \left( \boldsymbol{L}\otimes
\nabla \right) $ \smallskip \\ $ \quad  =\left( \boldsymbol{L}\otimes \boldsymbol{L}\right) \left(
\nabla \otimes \nabla \right) $}  & & $\delta
\left( \boldsymbol{r}\right) \left( \boldsymbol{\hat{I}}-\frac{\boldsymbol{L}%
\otimes \boldsymbol{L}}{L^{2}}\right) +\frac{1}{4L}\delta \left( \boldsymbol{%
r}_{\perp }\right) \mathrm{e}^{-\frac{\left\vert r_{\parallel }\right\vert }{%
L}}\frac{\boldsymbol{L}\otimes \boldsymbol{L}}{L^{2}}+\frac{\delta \left( 
\boldsymbol{r}_{\perp }\right) }{r_{\perp }}\mathrm{e}^{-\frac{\left\vert
r_{\parallel }\right\vert }{L}}\left( \frac{\boldsymbol{L}}{L}\otimes \frac{%
\boldsymbol{r}_{\perp }}{r_{\perp }}\right) $ \\ \cline{2-2} \cline{4-4} \cline{6-6} \xrowht{35pt}

 & $T_4$ & & \makecell[l]{$\left( \boldsymbol{L}\otimes \nabla \right) ^{\mathrm{T}}\left( \boldsymbol{%
L}\otimes \nabla \right) ^{\mathrm{T}} $ \smallskip \\   $ \quad =\left( \nabla \otimes \nabla \right)
\left( \boldsymbol{L}\otimes \boldsymbol{L}\right) $  } & & $\delta
\left( \boldsymbol{r}\right) \left( \boldsymbol{\hat{I}}-\frac{\boldsymbol{L}%
\otimes \boldsymbol{L}}{L^{2}}\right) +\frac{1}{4L}\delta \left( \boldsymbol{%
r}_{\perp }\right) \mathrm{e}^{-\frac{\left\vert r_{\parallel }\right\vert }{%
L}}\frac{\boldsymbol{L}\otimes \boldsymbol{L}}{L^{2}}+\frac{\delta \left( 
\boldsymbol{r}_{\perp }\right) }{r_{\perp }}\mathrm{e}^{-\frac{\left\vert
r_{\parallel }\right\vert }{L}}\left( \frac{%
\boldsymbol{r}_{\perp }}{r_{\perp }} \otimes  \frac{\boldsymbol{L}}{L}\right) $ \\ \cline{2-2} \cline{4-4} \cline{6-6} \xrowht{35pt}

 & $T_5$ & &  $\left( \boldsymbol{L}\times \nabla \right)
\otimes \left( \boldsymbol{L}\times \nabla \right)$ &  & 
\makecell{ $\delta \left( \boldsymbol{r}\right) \left( \boldsymbol{\hat{I}}-\left( 
\frac{\boldsymbol{L}}{L}\times \frac{\boldsymbol{r}_{\perp }}{r_{\perp }}%
\right) \otimes \left( \frac{\boldsymbol{L}}{L}\times \frac{\boldsymbol{r}%
_{\perp }}{r_{\perp }}\right) \right) -\frac{1}{2\pi L}\delta \left(
r_{\parallel }\right) \frac{1}{r_{\perp }}K_{1}\left( \frac{r_{\perp }}{L}%
\right) \frac{\boldsymbol{r}_{\perp }\otimes \boldsymbol{r}_{\perp }}{%
r_{\perp }^{2}} $\\
$+\frac{1}{2\pi L}\delta \left( r_{\parallel }\right) \left( \frac{1}{L}%
K_{0}\left( \frac{r_{\perp }}{L}\right) +\frac{1}{r_{\perp }}K_{1}\left( 
\frac{r_{\perp }}{L}\right) \right) \left( \left( \frac{\boldsymbol{L}}{L}%
\times \frac{\boldsymbol{r}_{\perp }}{r_{\perp }}\right) \otimes \left( 
\frac{\boldsymbol{L}}{L}\times \frac{\boldsymbol{r}_{\perp }}{r_{\perp }}%
\right) \right) $} \\ \cline{1-6}
\end{tabular}
\end{center}
\caption{Scalar and tensor operators, as well as non-locality kernels, corresponding to non-local models, where $\delta$ denotes Dirac delta distribution, while $K_0$ and $K_1$ are modified Bessel functions of the second kind (Macdonald functions) of the zeroth and first orders, as well as where $r_{\parallel }= \boldsymbol{r} \cdot \frac{\boldsymbol{L} }{L}$, $\boldsymbol{r}_{\perp }= \boldsymbol{r}-\boldsymbol{r}_{\parallel }$ with $\boldsymbol{r}_{\parallel }=r_{\parallel } \frac{\boldsymbol{L} }{L}$, and $r_{\perp }= \left\vert \boldsymbol{r}_{\perp } \right\vert $.}
\label{Modeli-new}
\end{sidewaystable}

\section{Displacement vector field\label{Displacement-field}}

In order to express the displacement vector field $\boldsymbol{u}$,
corresponding to the three-dimensional isotropic and elastic body, either of
local or of non-local type, one considers the problem of statics of such a
body for arbitrarily given body forces $\boldsymbol{f}_{b}$, by solving a
system of equations consisting of an equilibrium equation and infinitesimal
strain%
\begin{gather}
\func{div}\boldsymbol{\hat{\sigma}}\left( \boldsymbol{r}\right) +\boldsymbol{%
f}_{b}\left( \boldsymbol{r}\right) =\boldsymbol{0},  \label{equilibrium-eq}
\\
\boldsymbol{\hat{\varepsilon}}\left( \boldsymbol{r}\right) =\frac{1}{2}%
\left( \func{grad}\boldsymbol{u}\left( \boldsymbol{r}\right) +\left( \func{%
grad}\boldsymbol{u}\left( \boldsymbol{r}\right) \right) ^{\mathrm{T}}\right)
,  \label{str}
\end{gather}%
as well as of either Hooke's law as a constitutive equation, 
\begin{equation}
\boldsymbol{\hat{\sigma}}\left( \boldsymbol{r}\right) =\lambda \,\mathrm{%
\limfunc{tr}}\boldsymbol{\hat{\varepsilon}}\left( \boldsymbol{r}\right) 
\boldsymbol{\hat{I}}+2\mu \,\boldsymbol{\hat{\varepsilon}}\left( \boldsymbol{%
r}\right) ,  \label{Hook-loc}
\end{equation}%
describing local character of a body, or non-local scalar- and tensor-type
constitutive equations%
\begin{gather}
\left( 1-\mathrm{S}_{i}(\nabla )\right) \boldsymbol{\hat{\sigma}}\left( 
\boldsymbol{r}\right) =\lambda \,\mathrm{\limfunc{tr}}\boldsymbol{\hat{%
\varepsilon}}\left( \boldsymbol{r}\right) \boldsymbol{\hat{I}}+2\mu \,%
\boldsymbol{\hat{\varepsilon}}\left( \boldsymbol{r}\right) ,\quad i\in
\left\{ 1,2,3\right\} ,\quad \text{and}  \label{Hook-non-loc} \\
\left( \boldsymbol{\hat{I}}-\mathbf{\hat{T}}_{i}(\nabla )\right) \boldsymbol{%
\hat{\sigma}}\left( \boldsymbol{r}\right) =\lambda \,\mathrm{\limfunc{tr}}%
\boldsymbol{\hat{\varepsilon}}\left( \boldsymbol{r}\right) \boldsymbol{\hat{I%
}}+2\mu \,\boldsymbol{\hat{\varepsilon}}\left( \boldsymbol{r}\right) ,\quad
i\in \left\{ 1,2,3,4,5\right\} ,  \label{Hook-non-loc-t}
\end{gather}%
see (\ref{non-loc-scalar-type}) and (\ref{non-loc-tensor-type}), with scalar
operators $\mathrm{S}_{1}$ - $\mathrm{S}_{3}$ and tensor operators $\mathbf{%
\hat{T}}_{1}$ - $\mathbf{\hat{T}}_{5}$ given in Table \ref{Modeli-new},
describing non-local character of a body. The system of equations is solved
using a method of Fourier transform.

\subsection{Local displacement field}

In the case of local-type three-dimensional isotropic and elastic body,
described by Hooke's law (\ref{Hook-loc}), the local-type displacement
vector field $\boldsymbol{u}$ is given by%
\begin{align}
\boldsymbol{u}\left( \boldsymbol{r}\right) &=\frac{\lambda +3\mu }{8\pi \mu
\left( \lambda +2\mu \right) }\left( \frac{1}{r}\ast _{\boldsymbol{r}}%
\boldsymbol{f}_{b}\left( \boldsymbol{r}\right) \right) +\frac{\lambda +\mu }{%
8\pi \mu \left( \lambda +2\mu \right) }\frac{\boldsymbol{r}\otimes 
\boldsymbol{r}}{r^{3}}\ast _{\boldsymbol{r}}\boldsymbol{f}_{b}\left( 
\boldsymbol{r}\right) ,\quad \text{i.e.,}  \notag \\
&=\frac{\lambda +3\mu }{8\pi \mu \left( \lambda +2\mu \right) }\int_{%
\mathbb{R}
^{3}}\frac{1}{\left\vert \boldsymbol{r}-\boldsymbol{r}^{\prime }\right\vert }%
\boldsymbol{f}_{b}\left( \boldsymbol{r}^{\prime }\right) \mathrm{d}_{%
\boldsymbol{r}^{\prime }}V+\frac{\lambda +\mu }{8\pi \mu \left( \lambda
+2\mu \right) }\int_{%
\mathbb{R}
^{3}}\frac{\boldsymbol{r}-\boldsymbol{r}^{\prime }}{\left\vert \boldsymbol{r}%
-\boldsymbol{r}^{\prime }\right\vert ^{3}}\left( \boldsymbol{r}-\boldsymbol{r%
}^{\prime }\right) \boldsymbol{\cdot f}_{b}\left( \boldsymbol{r}^{\prime
}\right) \mathrm{d}_{\boldsymbol{r}^{\prime }}V,  \label{u-loc}
\end{align}%
which agrees with the result presented in \cite{WangXuGao}, here obtained by
the means of Fourier integral transforms.

Namely, the system of equations (\ref{equilibrium-eq}), (\ref{str}), and (%
\ref{Hook-loc}) reduces to%
\begin{gather}
\lambda \func{div}\left( \mathrm{\func{div}}\boldsymbol{u}\left( \boldsymbol{%
r}\right) \boldsymbol{\hat{I}}\right) +\mu \func{div}\func{grad}\boldsymbol{u%
}\left( \boldsymbol{r}\right) +\mu \func{div}\left( \func{grad}\boldsymbol{u}%
\left( \boldsymbol{r}\right) \right) ^{\mathrm{T}}=-\boldsymbol{f}_{b}\left( 
\boldsymbol{r}\right), \quad \text{i.e.,}  \notag \\
\mu \Delta \boldsymbol{u}\left( \boldsymbol{r}\right) +\left( \lambda +\mu
\right) \func{grad}\mathrm{\func{div}}\boldsymbol{u}\left( \boldsymbol{r}%
\right) =-\boldsymbol{f}_{b}\left( \boldsymbol{r}\right) ,  \label{equation}
\end{gather}%
since $\func{div}\left( \mathrm{\func{div}}\boldsymbol{u}\left( \boldsymbol{r%
}\right) \boldsymbol{\hat{I}}\right) =\func{div}\left( \func{grad}%
\boldsymbol{u}\left( \boldsymbol{r}\right) \right) ^{\mathrm{T}}=\func{grad}%
\mathrm{\func{div}}\boldsymbol{u}\left( \boldsymbol{r}\right) $ and $\func{%
div}\func{grad}\boldsymbol{u}\left( \boldsymbol{r}\right) =\Delta 
\boldsymbol{u}\left( \boldsymbol{r}\right) $. Further by applying the
Fourier transform defined by%
\begin{equation}
\bar{f}\left( \boldsymbol{k}\right) =\mathcal{F}\left[ f\left( \boldsymbol{r}%
\right) \right] \left( \boldsymbol{k}\right) =\int_{%
\mathbb{R}
^{3}}f\left( \boldsymbol{r}\right) \mathrm{e}^{-\mathrm{i}\boldsymbol{k\cdot
r}}\mathrm{d}_{\boldsymbol{r}}V,\quad \boldsymbol{k}\in 
\mathbb{R}
^{3},  \label{FT}
\end{equation}%
equation (\ref{equation}) takes the form%
\begin{align*}
& \mu \,k^{2}\boldsymbol{\bar{u}}\left( \boldsymbol{k}\right) -\left(
\lambda +\mu \right) \mathrm{i}\boldsymbol{k}\left( \mathrm{i}\boldsymbol{k}%
\cdot \boldsymbol{\bar{u}}\left( \boldsymbol{k}\right) \right) =\boldsymbol{%
\bar{f}}_{b}\left( \boldsymbol{k}\right), \quad \text{i.e.,} \\
& \mu \,k^{2}\boldsymbol{\bar{u}}\left( \boldsymbol{k}\right) +\left(
\lambda +\mu \right) \left( \boldsymbol{k}\otimes \boldsymbol{k}\right) 
\boldsymbol{\bar{u}}\left( \boldsymbol{k}\right) =\boldsymbol{\bar{f}}%
_{b}\left( \boldsymbol{k}\right) ,
\end{align*}%
since $\mathrm{i}\boldsymbol{k}\left( \mathrm{i}\boldsymbol{k}\cdot 
\boldsymbol{\bar{u}}\left( \boldsymbol{k}\right) \right) =\left( \left( 
\mathrm{i}\boldsymbol{k}\right) \otimes \left( \mathrm{i}\boldsymbol{k}%
\right) \right) \boldsymbol{\bar{u}}\left( \boldsymbol{k}\right) =-\left( 
\boldsymbol{k}\otimes \boldsymbol{k}\right) \boldsymbol{\bar{u}}\left( 
\boldsymbol{k}\right) $, which can be recast through the resolvent tensor 
\begin{equation}
\boldsymbol{\bar{\hat{R}}}_{\boldsymbol{\ell }}^{-1}\left( \boldsymbol{k}%
\right) \boldsymbol{\bar{u}}\left( \boldsymbol{k}\right) =\boldsymbol{\bar{f}%
}_{b}\left( \boldsymbol{k}\right) \quad \text{with\quad }\boldsymbol{\bar{%
\hat{R}}}_{\boldsymbol{\ell }}^{-1}\left( \boldsymbol{k}\right) =\mu
\,k^{2}\left( \boldsymbol{\hat{I}}+\frac{\lambda +\mu }{\mu }\frac{%
\boldsymbol{k}\otimes \boldsymbol{k}}{k^{2}}\right) ,
\label{R-l-na-minus-prvi}
\end{equation}%
so that, the displacement vector filed in the Fourier domain is expressed as%
\begin{equation}
\boldsymbol{\bar{u}}\left( \boldsymbol{k}\right) =\boldsymbol{\bar{\hat{R}}}%
_{\boldsymbol{\ell }}\left( \boldsymbol{k}\right) \boldsymbol{\bar{f}}%
_{b}\left( \boldsymbol{k}\right) \quad \text{with\quad }\boldsymbol{\bar{%
\hat{R}}}_{\boldsymbol{\ell }}\left( \boldsymbol{k}\right) =\frac{1}{\mu }%
\frac{1}{k^{2}}\boldsymbol{\hat{I}}-\frac{\lambda +\mu }{\mu \left( \lambda
+2\mu \right) }\frac{\boldsymbol{k}\otimes \boldsymbol{k}}{k^{4}},
\label{u-fd-loc}
\end{equation}%
while $\boldsymbol{\bar{\hat{R}}}_{\boldsymbol{\ell }}^{-1}$ and $%
\boldsymbol{\bar{\hat{R}}}_{\boldsymbol{\ell }}$ are mutually inverse
operators, i.e., $\boldsymbol{\bar{\hat{R}}}_{\boldsymbol{\ell }}\boldsymbol{%
\bar{\hat{R}}}_{\boldsymbol{\ell }}^{-1}=\boldsymbol{\hat{I}}$, which is due
to $\frac{\left( \boldsymbol{k}\otimes \boldsymbol{k}\right) \left( 
\boldsymbol{k}\otimes \boldsymbol{k}\right) }{k^{n}}=\frac{\boldsymbol{k}%
\otimes \boldsymbol{k}}{k^{n-2}}$.

Using the Fourier transform inversion, defined by%
\begin{equation}
f\left( \boldsymbol{r}\right) =\mathcal{F}^{-1}\left[ \bar{f}\left( 
\boldsymbol{k}\right) \right] \left( \boldsymbol{k}\right) =\frac{1}{\left(
2\pi \right) ^{3}}\int_{%
\mathbb{R}
^{3}}\bar{f}\left( \boldsymbol{k}\right) \mathrm{e}^{\mathrm{i}\boldsymbol{%
k\cdot r}}\mathrm{d}_{\boldsymbol{k}}V,\quad \boldsymbol{r}\in 
\mathbb{R}
^{3},  \label{IFT}
\end{equation}%
expression (\ref{u-fd-loc})$_{2}$ becomes%
\begin{align*}
\boldsymbol{\hat{R}}_{\ell }\left( \boldsymbol{r}\right) & =\frac{\lambda
+3\mu }{8\pi \mu \left( \lambda +2\mu \right) }\frac{1}{r}\boldsymbol{\hat{I}%
}+\frac{\lambda +\mu }{8\pi \mu \left( \lambda +2\mu \right) }\frac{1}{r}%
\frac{\boldsymbol{r}\otimes \boldsymbol{r}}{r^{2}},\quad \text{i.e.,} \\
& =\frac{1}{4\pi \mu }\frac{1}{r}\frac{\boldsymbol{r}\otimes \boldsymbol{r}}{%
r^{2}}+\frac{\lambda +3\mu }{8\pi \mu \left( \lambda +2\mu \right) }\frac{1}{%
r}\left( \boldsymbol{\hat{I}}-\frac{\boldsymbol{r}\otimes \boldsymbol{r}}{%
r^{2}}\right) ,
\end{align*}%
decomposing into compressive and shear component, since%
\begin{equation*}
\mathcal{F}^{-1}\left[ \frac{1}{k^{2}}\right] \left( \boldsymbol{r}\right) =%
\frac{1}{4\pi r}\quad \text{and\quad }\mathcal{F}^{-1}\left[ \frac{%
\boldsymbol{k}\otimes \boldsymbol{k}}{k^{4}}\right] \left( \boldsymbol{r}%
\right) =\frac{1}{8\pi r}\left( \boldsymbol{\hat{I}}-\frac{\boldsymbol{r}%
\otimes \boldsymbol{r}}{r^{2}}\right) ,
\end{equation*}%
see Appendices \ref{ik-kroz-k-na-kv} and \ref{dijadski-kroz-cetvrti} for
calculation, so that (\ref{u-fd-loc})$_{1}$ takes the form%
\begin{align*}
\boldsymbol{u}\left( \boldsymbol{r}\right) & =\boldsymbol{\hat{R}}_{%
\boldsymbol{\ell }}\left( \boldsymbol{r}\right) \ast _{\boldsymbol{r}}%
\boldsymbol{f}_{b}\left( \boldsymbol{r}\right) \\
& =\frac{\lambda +3\mu }{8\pi \mu \left( \lambda +2\mu \right) }\left( \frac{%
1}{r}\ast _{\boldsymbol{r}}\boldsymbol{f}_{b}\left( \boldsymbol{r}\right)
\right) +\frac{\lambda +\mu }{8\pi \mu \left( \lambda +2\mu \right) }\left( 
\frac{\boldsymbol{r}\otimes \boldsymbol{r}}{r^{3}}\ast _{\boldsymbol{r}}%
\boldsymbol{f}_{b}\left( \boldsymbol{r}\right) \right) \\
& =\frac{\lambda +3\mu }{8\pi \mu \left( \lambda +2\mu \right) }\int_{%
\mathbb{R}
^{3}}\frac{1}{\left\vert \boldsymbol{r}-\boldsymbol{r}^{\prime }\right\vert }%
\boldsymbol{f}_{b}\left( \boldsymbol{r}^{\prime }\right) \mathrm{d}_{%
\boldsymbol{r}^{\prime }}V+\frac{\lambda +\mu }{8\pi \mu \left( \lambda
+2\mu \right) }\int_{%
\mathbb{R}
^{3}}\frac{\boldsymbol{r}-\boldsymbol{r}^{\prime }}{\left\vert \boldsymbol{r}%
-\boldsymbol{r}^{\prime }\right\vert ^{3}}\left( \boldsymbol{r}-\boldsymbol{r%
}^{\prime }\right) \boldsymbol{\cdot f}_{b}\left( \boldsymbol{r}^{\prime
}\right) \mathrm{d}_{\boldsymbol{r}^{\prime }}V,
\end{align*}%
since the convolution integrand is $\frac{\left( \boldsymbol{r}-\boldsymbol{r%
}^{\prime }\right) \otimes \left( \boldsymbol{r}-\boldsymbol{r}^{\prime
}\right) }{\left\vert \boldsymbol{r}-\boldsymbol{r}^{\prime }\right\vert ^{3}%
}\boldsymbol{f}_{b}\left( \boldsymbol{r}^{\prime }\right) =\frac{\boldsymbol{%
r}-\boldsymbol{r}^{\prime }}{\left\vert \boldsymbol{r}-\boldsymbol{r}%
^{\prime }\right\vert ^{3}}\left( \boldsymbol{r}-\boldsymbol{r}^{\prime
}\right) \boldsymbol{\cdot f}_{b}\left( \boldsymbol{r}^{\prime }\right) ,$
becoming (\ref{u-loc}).

\subsection{Non-local displacement field}

In the case of non-local three-dimensional isotropic and elastic body,
described either by constitutive equation (\ref{Hook-non-loc}), containing
the scalar-type operator $\mathrm{S}_{1}$ - $\mathrm{S}_{3}$, or by
constitutive equation (\ref{Hook-non-loc-t}), containing the tensor-type
operator $\mathrm{T}_{1}$ - $\mathrm{T}_{5}$, the corresponding displacement
vector fields are given as a superposition of local and non-local
displacement vector fields by%
\begin{align}
\boldsymbol{u}& ^{\left( S_{1}/T_{1}\right) }\left( \boldsymbol{r}\right) =%
\boldsymbol{u}_{\ell }\left( \boldsymbol{r}\right) +\boldsymbol{u}_{n\ell
}^{\left( S_{1}/T_{1}\right) }\left( \boldsymbol{r}\right)  \notag \\
& =-\frac{L^{2}}{3}\frac{2\lambda -\mu }{\mu (\lambda +2\mu )}\boldsymbol{f}%
_{b}\left( \boldsymbol{r}\right) -\frac{4L^{2}}{3}\frac{\lambda +\mu }{\mu
(\lambda +2\mu )}\left( \boldsymbol{e}_{z}\cdot \boldsymbol{f}_{b}\left( 
\boldsymbol{r}\right) \right) \boldsymbol{e}_{z}  \notag \\
& \quad +\frac{\lambda +3\mu }{8\pi \mu \left( \lambda +2\mu \right) }\int_{%
\mathbb{R}
^{3}}\frac{1}{\left\vert \boldsymbol{r}-\boldsymbol{r}^{\prime }\right\vert }%
\boldsymbol{f}_{b}\left( \boldsymbol{r}^{\prime }\right) \mathrm{d}_{%
\boldsymbol{r}^{\prime }}V-\frac{L^{2}}{4\pi }\frac{\lambda +\mu }{\mu
(\lambda +2\mu )}\int_{%
\mathbb{R}
^{3}}\frac{1}{\left\vert \boldsymbol{r}-\boldsymbol{r}^{\prime }\right\vert
^{3}}\boldsymbol{f}_{b}\left( \boldsymbol{r}^{\prime }\right) \mathrm{d}_{%
\boldsymbol{r}^{\prime }}V  \notag \\
& \quad +\frac{\lambda +\mu }{8\pi \mu \left( \lambda +2\mu \right) }\int_{%
\mathbb{R}
^{3}}\frac{\boldsymbol{r}-\boldsymbol{r}^{\prime }}{\left\vert \boldsymbol{r}%
-\boldsymbol{r}^{\prime }\right\vert ^{3}}\left( \boldsymbol{r}-\boldsymbol{r%
}^{\prime }\right) \cdot \boldsymbol{f}_{b}\left( \boldsymbol{r}^{\prime
}\right) \mathrm{d}_{\boldsymbol{r}^{\prime }}V+\frac{3L^{2}}{4\pi }\frac{%
\lambda +\mu }{\mu (\lambda +2\mu )}\int_{%
\mathbb{R}
^{3}}\frac{\boldsymbol{r}-\boldsymbol{r}^{\prime }}{\left\vert \boldsymbol{r}%
-\boldsymbol{r}^{\prime }\right\vert ^{5}}\left( \boldsymbol{r}-\boldsymbol{r%
}^{\prime }\right) \cdot \boldsymbol{f}_{b}\left( \boldsymbol{r}^{\prime
}\right) \mathrm{d}_{\boldsymbol{r}^{\prime }}V  \label{u-non-loc-S1}
\end{align}%
in the case of models $S_{1}$ and $T_{1}$, further by%
\begin{align}
\boldsymbol{u}& ^{\left( S_{2}/T_{3}\right) }\left( \boldsymbol{r}\right) =%
\boldsymbol{u}_{\ell }\left( \boldsymbol{r}\right) +\boldsymbol{u}_{n\ell
}^{\left( S_{2}/T_{3}\right) }\left( \boldsymbol{r}\right)  \notag \\
& =L^{2}\frac{1}{\mu }\boldsymbol{f}_{b}\left( \boldsymbol{r}\right) -L^{2}%
\frac{\lambda +\mu }{\mu \left( \lambda +2\mu \right) }\left( \frac{%
\boldsymbol{L}}{L}\cdot \boldsymbol{f}_{b}\left( \boldsymbol{r}\right)
\right) \frac{\boldsymbol{L}}{L}  \notag \\
& \quad +\frac{\lambda +3\mu }{8\pi \mu \left( \lambda +2\mu \right) }\int_{%
\mathbb{R}
^{3}}\frac{1}{\left\vert \boldsymbol{r}-\boldsymbol{r}^{\prime }\right\vert }%
\boldsymbol{f}_{b}\left( \boldsymbol{r}^{\prime }\right) \mathrm{d}_{%
\boldsymbol{r}^{\prime }}V+\frac{L^{2}}{8\pi }\frac{\lambda +\mu }{\mu
\left( \lambda +2\mu \right) }\int_{%
\mathbb{R}
^{3}}\frac{1}{\left\vert \boldsymbol{r}-\boldsymbol{r}^{\prime }\right\vert
^{3}}\boldsymbol{f}_{b}\left( \boldsymbol{r}^{\prime }\right) \mathrm{d}_{%
\boldsymbol{r}^{\prime }}V  \notag \\
& \quad +\frac{\lambda +\mu }{8\pi \mu \left( \lambda +2\mu \right) }\int_{%
\mathbb{R}
^{3}}\frac{\boldsymbol{r}-\boldsymbol{r}^{\prime }}{\left\vert \boldsymbol{r}%
-\boldsymbol{r}^{\prime }\right\vert ^{3}}\left( \boldsymbol{r}-\boldsymbol{r%
}^{\prime }\right) \cdot \boldsymbol{f}_{b}\left( \boldsymbol{r}^{\prime
}\right) \mathrm{d}_{\boldsymbol{r}^{\prime }}V+\frac{3L^{2}}{8\pi }\frac{%
\lambda +\mu }{\mu \left( \lambda +2\mu \right) }\int_{%
\mathbb{R}
^{3}}\frac{\boldsymbol{r}-\boldsymbol{r}^{\prime }}{\left\vert \boldsymbol{r}%
-\boldsymbol{r}^{\prime }\right\vert ^{5}}\left( \boldsymbol{r}-\boldsymbol{r%
}^{\prime }\right) \cdot \boldsymbol{f}_{b}\left( \boldsymbol{r}^{\prime
}\right) \mathrm{d}_{\boldsymbol{r}^{\prime }}V  \notag \\
& \quad -\frac{L^{2}}{8\pi }\frac{\lambda +\mu }{\mu \left( \lambda +2\mu
\right) }  \notag \\
& \quad \quad \quad \times \left( 3\int_{%
\mathbb{R}
^{3}}\frac{1}{\left\vert \boldsymbol{r}-\boldsymbol{r}^{\prime }\right\vert
^{3}}\left( \frac{\boldsymbol{L}}{L}\cdot \frac{\boldsymbol{r}-\boldsymbol{r}%
^{\prime }}{\left\vert \boldsymbol{r}-\boldsymbol{r}^{\prime }\right\vert }%
\right) ^{2}\boldsymbol{f}_{b}\left( \boldsymbol{r}^{\prime }\right) \mathrm{%
d}_{\boldsymbol{r}^{\prime }}V\right.  \notag \\
& \quad \quad \quad \quad \quad \left. +2\frac{\boldsymbol{L}}{L}\int_{%
\mathbb{R}
^{3}}\frac{1}{\left\vert \boldsymbol{r}-\boldsymbol{r}^{\prime }\right\vert
^{3}}\left( \frac{\boldsymbol{L}}{L}\cdot \boldsymbol{f}_{b}\left( 
\boldsymbol{r}^{\prime }\right) \right) \mathrm{d}_{\boldsymbol{r}^{\prime
}}V-6\frac{\boldsymbol{L}}{L}\int_{%
\mathbb{R}
^{3}}\frac{1}{\left\vert \boldsymbol{r}-\boldsymbol{r}^{\prime }\right\vert
^{4}}\left( \frac{\boldsymbol{L}}{L}\cdot \frac{\boldsymbol{r}-\boldsymbol{r}%
^{\prime }}{\left\vert \boldsymbol{r}-\boldsymbol{r}^{\prime }\right\vert }%
\right) \left( \boldsymbol{r}-\boldsymbol{r}^{\prime }\right) \cdot 
\boldsymbol{f}_{b}\left( \boldsymbol{r}^{\prime }\right) \mathrm{d}_{%
\boldsymbol{r}^{\prime }}V\right.  \notag \\
& \quad \quad \quad \quad \quad \left. -6\int_{%
\mathbb{R}
^{3}}\frac{\boldsymbol{r}-\boldsymbol{r}^{\prime }}{\left\vert \boldsymbol{r}%
-\boldsymbol{r}^{\prime }\right\vert ^{4}}\left( \frac{\boldsymbol{L}}{L}%
\cdot \frac{\boldsymbol{r}-\boldsymbol{r}^{\prime }}{\left\vert \boldsymbol{r%
}-\boldsymbol{r}^{\prime }\right\vert }\right) \left( \frac{\boldsymbol{L}}{L%
}\cdot \boldsymbol{f}_{b}\left( \boldsymbol{r}^{\prime }\right) \right) 
\mathrm{d}_{\boldsymbol{r}^{\prime }}V\right.  \notag \\
& \quad \quad \quad \quad \quad \left. +15\int_{%
\mathbb{R}
^{3}}\frac{\boldsymbol{r}-\boldsymbol{r}^{\prime }}{\left\vert \boldsymbol{r}%
-\boldsymbol{r}^{\prime }\right\vert ^{5}}\left( \frac{\boldsymbol{L}}{L}%
\cdot \frac{\boldsymbol{r}-\boldsymbol{r}^{\prime }}{\left\vert \boldsymbol{r%
}-\boldsymbol{r}^{\prime }\right\vert }\right) ^{2}\left( \boldsymbol{r}-%
\boldsymbol{r}^{\prime }\right) \cdot \boldsymbol{f}_{b}\left( \boldsymbol{r}%
^{\prime }\right) \mathrm{d}_{\boldsymbol{r}^{\prime }}V\right)
\label{u-non-loc-S2}
\end{align}%
in the case of models $S_{2}$ and $T_{3}$, as well as by%
\begin{align}
\boldsymbol{u}& ^{\left( S_{3}\right) }\left( \boldsymbol{r}\right) =%
\boldsymbol{u}_{\ell }\left( \boldsymbol{r}\right) +\boldsymbol{u}_{n\ell
}^{\left( S_{3}\right) }\left( \boldsymbol{r}\right)  \notag \\
& =-\frac{5L^{2}}{3}\frac{\lambda +\mu }{\mu \left( \lambda +2\mu \right) }%
\boldsymbol{f}_{b}\left( \boldsymbol{r}\right) +L^{2}\frac{\lambda +\mu }{%
\mu \left( \lambda +2\mu \right) }\left( \frac{\boldsymbol{L}}{L}\cdot 
\boldsymbol{f}_{b}\left( \boldsymbol{r}\right) \right) \frac{\boldsymbol{L}}{%
L}-\frac{4L^{2}}{3}\frac{\lambda +\mu }{\mu \left( \lambda +2\mu \right) }%
\left( \boldsymbol{e}_{z}\cdot \boldsymbol{f}_{b}\left( \boldsymbol{r}%
\right) \right) \boldsymbol{e}_{z}  \notag \\
& \quad +\frac{\lambda +3\mu }{8\pi \mu \left( \lambda +2\mu \right) }\int_{%
\mathbb{R}
^{3}}\frac{1}{\left\vert \boldsymbol{r}-\boldsymbol{r}^{\prime }\right\vert }%
\boldsymbol{f}_{b}\left( \boldsymbol{r}^{\prime }\right) \mathrm{d}_{%
\boldsymbol{r}^{\prime }}V-\frac{3L^{2}}{8\pi }\frac{\lambda +\mu }{\mu
\left( \lambda +2\mu \right) }\int_{%
\mathbb{R}
^{3}}\frac{1}{\left\vert \boldsymbol{r}-\boldsymbol{r}^{\prime }\right\vert
^{3}}\boldsymbol{f}_{b}\left( \boldsymbol{r}^{\prime }\right) \mathrm{d}_{%
\boldsymbol{r}^{\prime }}V  \notag \\
& \quad +\frac{\lambda +\mu }{8\pi \mu \left( \lambda +2\mu \right) }\int_{%
\mathbb{R}
^{3}}\frac{\boldsymbol{r}-\boldsymbol{r}^{\prime }}{\left\vert \boldsymbol{r}%
-\boldsymbol{r}^{\prime }\right\vert ^{3}}\left( \boldsymbol{r}-\boldsymbol{r%
}^{\prime }\right) \boldsymbol{\cdot f}_{b}\left( \boldsymbol{r}^{\prime
}\right) \mathrm{d}_{\boldsymbol{r}^{\prime }}V+\frac{3L^{2}}{8\pi }\frac{%
\lambda +\mu }{\mu \left( \lambda +2\mu \right) }\int_{%
\mathbb{R}
^{3}}\frac{\boldsymbol{r}-\boldsymbol{r}^{\prime }}{\left\vert \boldsymbol{r}%
-\boldsymbol{r}^{\prime }\right\vert ^{5}}\left( \boldsymbol{r}-\boldsymbol{r%
}^{\prime }\right) \cdot \boldsymbol{f}_{b}\left( \boldsymbol{r}^{\prime
}\right) \mathrm{d}_{\boldsymbol{r}^{\prime }}V  \notag \\
& \quad +\frac{L^{2}}{8\pi }\frac{\lambda +\mu }{\mu \left( \lambda +2\mu
\right) }  \notag \\
& \quad \quad \quad \times \left( 3\int_{%
\mathbb{R}
^{3}}\frac{1}{\left\vert \boldsymbol{r}-\boldsymbol{r}^{\prime }\right\vert
^{3}}\left( \frac{\boldsymbol{L}}{L}\cdot \frac{\boldsymbol{r}-\boldsymbol{r}%
^{\prime }}{\left\vert \boldsymbol{r}-\boldsymbol{r}^{\prime }\right\vert }%
\right) ^{2}\boldsymbol{f}_{b}\left( \boldsymbol{r}^{\prime }\right) \mathrm{%
d}_{\boldsymbol{r}^{\prime }}V\right.  \notag \\
& \quad \quad \quad \quad \quad \left. 2\frac{\boldsymbol{L}}{L}\int_{%
\mathbb{R}
^{3}}\frac{1}{\left\vert \boldsymbol{r}-\boldsymbol{r}^{\prime }\right\vert
^{3}}\left( \frac{\boldsymbol{L}}{L}\cdot \boldsymbol{f}_{b}\left( 
\boldsymbol{r}^{\prime }\right) \right) \mathrm{d}_{\boldsymbol{r}^{\prime
}}V-6\frac{\boldsymbol{L}}{L}\int_{%
\mathbb{R}
^{3}}\frac{1}{\left\vert \boldsymbol{r}-\boldsymbol{r}^{\prime }\right\vert
^{4}}\left( \frac{\boldsymbol{L}}{L}\cdot \frac{\boldsymbol{r}-\boldsymbol{r}%
^{\prime }}{\left\vert \boldsymbol{r}-\boldsymbol{r}^{\prime }\right\vert }%
\right) \left( \boldsymbol{r}-\boldsymbol{r}^{\prime }\right) \cdot 
\boldsymbol{f}_{b}\left( \boldsymbol{r}^{\prime }\right) \mathrm{d}_{%
\boldsymbol{r}^{\prime }}V\right.  \notag \\
& \quad \quad \quad \quad \quad \left. -6\int_{%
\mathbb{R}
^{3}}\frac{\boldsymbol{r}-\boldsymbol{r}^{\prime }}{\left\vert \boldsymbol{r}%
-\boldsymbol{r}^{\prime }\right\vert ^{4}}\left( \frac{\boldsymbol{L}}{L}%
\cdot \frac{\boldsymbol{r}-\boldsymbol{r}^{\prime }}{\left\vert \boldsymbol{r%
}-\boldsymbol{r}^{\prime }\right\vert }\right) \left( \frac{\boldsymbol{L}}{L%
}\cdot \boldsymbol{f}_{b}\left( \boldsymbol{r}^{\prime }\right) \right) 
\mathrm{d}_{\boldsymbol{r}^{\prime }}V\right.  \notag \\
& \quad \quad \quad \quad \quad \left. +15\int_{%
\mathbb{R}
^{3}}\frac{\boldsymbol{r}-\boldsymbol{r}^{\prime }}{\left\vert \boldsymbol{r}%
-\boldsymbol{r}^{\prime }\right\vert ^{5}}\left( \frac{\boldsymbol{L}}{L}%
\cdot \frac{\boldsymbol{r}-\boldsymbol{r}^{\prime }}{\left\vert \boldsymbol{r%
}-\boldsymbol{r}^{\prime }\right\vert }\right) ^{2}\left( \boldsymbol{r}-%
\boldsymbol{r}^{\prime }\right) \cdot \boldsymbol{f}_{b}\left( \boldsymbol{r}%
^{\prime }\right) \mathrm{d}_{\boldsymbol{r}^{\prime }}V\right)
\label{u-non-loc-S3}
\end{align}%
in the case of model $S_{3}$, and next by%
\begin{align}
\boldsymbol{u}& ^{\left( T_{2}\right) }\left( \boldsymbol{r}\right) =%
\boldsymbol{u}_{\ell }\left( \boldsymbol{r}\right) +\boldsymbol{u}_{n\ell
}^{\left( T_{2}\right) }\left( \boldsymbol{r}\right)  \notag \\
& =-\frac{L^{2}}{3}\frac{2\lambda -\mu }{\mu \left( \lambda +2\mu \right) }%
\left( \frac{\boldsymbol{L}}{L}\cdot \boldsymbol{f}_{b}\left( \boldsymbol{r}%
\right) \right) \frac{\boldsymbol{L}}{L}-\frac{4L^{2}}{3}\frac{\lambda +\mu 
}{\mu \left( \lambda +2\mu \right) }\left( \boldsymbol{e}_{z}\cdot \frac{%
\boldsymbol{L}}{L}\right) \left( \frac{\boldsymbol{L}}{L}\cdot \boldsymbol{f}%
_{b}\left( \boldsymbol{r}\right) \right) \boldsymbol{e}_{z}  \notag \\
& \quad +\frac{\lambda +3\mu }{8\pi \mu \left( \lambda +2\mu \right) }\int_{%
\mathbb{R}
^{3}}\frac{1}{\left\vert \boldsymbol{r}-\boldsymbol{r}^{\prime }\right\vert }%
\boldsymbol{f}_{b}\left( \boldsymbol{r}^{\prime }\right) \mathrm{d}_{%
\boldsymbol{r}^{\prime }}V  \notag \\
& \quad -\frac{L^{2}}{4\pi }\frac{\lambda +\mu }{\mu (\lambda +2\mu )}\frac{%
\boldsymbol{L}}{L}\int_{%
\mathbb{R}
^{3}}\frac{1}{\left\vert \boldsymbol{r}-\boldsymbol{r}^{\prime }\right\vert
^{3}}\left( \frac{\boldsymbol{L}}{L}\cdot \boldsymbol{f}_{b}\left( 
\boldsymbol{r}^{\prime }\right) \right) \mathrm{d}_{\boldsymbol{r}^{\prime
}}V  \notag \\
& \quad +\frac{\lambda +\mu }{8\pi \mu \left( \lambda +2\mu \right) }\int_{%
\mathbb{R}
^{3}}\frac{\boldsymbol{r}-\boldsymbol{r}^{\prime }}{\left\vert \boldsymbol{r}%
-\boldsymbol{r}^{\prime }\right\vert ^{3}}\left( \boldsymbol{r}-\boldsymbol{r%
}^{\prime }\right) \boldsymbol{\cdot f}_{b}\left( \boldsymbol{r}^{\prime
}\right) \mathrm{d}_{\boldsymbol{r}^{\prime }}V  \notag \\
& \quad +\frac{3L^{2}}{4\pi }\frac{\lambda +\mu }{\mu (\lambda +2\mu )}\int_{%
\mathbb{R}
^{3}}\frac{\boldsymbol{r}-\boldsymbol{r}^{\prime }}{\left\vert \boldsymbol{r}%
-\boldsymbol{r}^{\prime }\right\vert ^{4}}\left( \frac{\boldsymbol{L}}{L}%
\cdot \frac{\boldsymbol{r}-\boldsymbol{r}^{\prime }}{\left\vert \boldsymbol{r%
}-\boldsymbol{r}^{\prime }\right\vert }\right) \left( \frac{\boldsymbol{L}}{L%
}\cdot \boldsymbol{f}_{b}\left( \boldsymbol{r}^{\prime }\right) \right) 
\mathrm{d}_{\boldsymbol{r}^{\prime }}V  \label{u-non-loc-T2}
\end{align}%
in the case of model $T_{2},$ further by%
\begin{align}
\boldsymbol{u}& ^{\left( T_{4}\right) }\left( \boldsymbol{r}\right) =%
\boldsymbol{u}_{\ell }\left( \boldsymbol{r}\right) +\boldsymbol{u}_{n\ell
}^{\left( T_{4}\right) }\left( \boldsymbol{r}\right)  \notag \\
& =-\frac{L^{2}}{3}\frac{2\lambda +\mu }{\mu \left( \lambda +2\mu \right) }%
\left( \frac{\boldsymbol{L}}{L}\cdot \boldsymbol{f}_{b}\left( \boldsymbol{r}%
\right) \right) \frac{\boldsymbol{L}}{L}+\frac{4L^{2}}{3}\frac{1}{\mu }%
\left( \boldsymbol{e}_{z}\cdot \frac{\boldsymbol{L}}{L}\right) \left( 
\boldsymbol{e}_{z}\cdot \boldsymbol{f}_{b}\left( \boldsymbol{r}\right)
\right) \frac{\boldsymbol{L}}{L}  \notag \\
& \quad +\frac{\lambda +3\mu }{8\pi \mu \left( \lambda +2\mu \right) }\int_{%
\mathbb{R}
^{3}}\frac{1}{\left\vert \boldsymbol{r}-\boldsymbol{r}^{\prime }\right\vert }%
\boldsymbol{f}_{b}\left( \boldsymbol{r}^{\prime }\right) \mathrm{d}_{%
\boldsymbol{r}^{\prime }}V-\frac{L^{2}}{8\pi }\frac{\lambda +\mu }{\mu
\left( \lambda +2\mu \right) }\int_{%
\mathbb{R}
^{3}}\frac{1}{\left\vert \boldsymbol{r}-\boldsymbol{r}^{\prime }\right\vert
^{3}}\boldsymbol{f}_{b}\left( \boldsymbol{r}^{\prime }\right) \mathrm{d}_{%
\boldsymbol{r}^{\prime }}V  \notag \\
& \quad +\frac{L^{2}}{4\pi }\frac{1}{\lambda +2\mu }\frac{\boldsymbol{L}}{L}%
\left( \int_{%
\mathbb{R}
^{3}}\frac{1}{\left\vert \boldsymbol{r}-\boldsymbol{r}^{\prime }\right\vert
^{3}}\left( \frac{\boldsymbol{L}}{L}\cdot \boldsymbol{f}_{b}\left( 
\boldsymbol{r}^{\prime }\right) \right) \mathrm{d}_{\boldsymbol{r}^{\prime
}}V-3\int_{%
\mathbb{R}
^{3}}\frac{1}{\left\vert \boldsymbol{r}-\boldsymbol{r}^{\prime }\right\vert
^{4}}\left( \frac{\boldsymbol{L}}{L}\cdot \frac{\boldsymbol{r}-\boldsymbol{r}%
^{\prime }}{\left\vert \boldsymbol{r}-\boldsymbol{r}^{\prime }\right\vert }%
\right) \left( \boldsymbol{r}-\boldsymbol{r}^{\prime }\right) \cdot 
\boldsymbol{f}_{b}\left( \boldsymbol{r}^{\prime }\right) \mathrm{d}_{%
\boldsymbol{r}^{\prime }}V\right)  \notag \\
& \quad +\frac{\lambda +\mu }{8\pi \mu \left( \lambda +2\mu \right) }\int_{%
\mathbb{R}
^{3}}\frac{\boldsymbol{r}-\boldsymbol{r}^{\prime }}{\left\vert \boldsymbol{r}%
-\boldsymbol{r}^{\prime }\right\vert ^{3}}\left( \boldsymbol{r}-\boldsymbol{r%
}^{\prime }\right) \cdot \boldsymbol{f}_{b}\left( \boldsymbol{r}^{\prime
}\right) \mathrm{d}_{\boldsymbol{r}^{\prime }}V+\frac{3L^{2}}{8\pi }\frac{%
\lambda +\mu }{\mu \left( \lambda +2\mu \right) }\int_{%
\mathbb{R}
^{3}}\frac{\boldsymbol{r}-\boldsymbol{r}^{\prime }}{\left\vert \boldsymbol{r}%
-\boldsymbol{r}^{\prime }\right\vert ^{5}}\left( \boldsymbol{r}-\boldsymbol{r%
}^{\prime }\right) \cdot \boldsymbol{f}_{b}\left( \boldsymbol{r}^{\prime
}\right) \mathrm{d}_{\boldsymbol{r}^{\prime }}V  \notag \\
& \quad +\frac{L^{2}}{8\pi }\frac{\lambda +\mu }{\mu \left( \lambda +2\mu
\right) }  \notag \\
& \quad \quad \quad \times \left( 3\int_{%
\mathbb{R}
^{3}}\frac{1}{\left\vert \boldsymbol{r}-\boldsymbol{r}^{\prime }\right\vert
^{3}}\left( \frac{\boldsymbol{L}}{L}\cdot \frac{\boldsymbol{r}-\boldsymbol{r}%
^{\prime }}{\left\vert \boldsymbol{r}-\boldsymbol{r}^{\prime }\right\vert }%
\right) ^{2}\boldsymbol{f}_{b}\left( \boldsymbol{r}^{\prime }\right) \mathrm{%
d}_{\boldsymbol{r}^{\prime }}V\right.  \notag \\
& \quad \quad \quad \quad \quad \left. +6\int_{%
\mathbb{R}
^{3}}\frac{\boldsymbol{r}-\boldsymbol{r}^{\prime }}{\left\vert \boldsymbol{r}%
-\boldsymbol{r}^{\prime }\right\vert ^{4}}\left( \frac{\boldsymbol{L}}{L}%
\cdot \frac{\boldsymbol{r}-\boldsymbol{r}^{\prime }}{\left\vert \boldsymbol{r%
}-\boldsymbol{r}^{\prime }\right\vert }\right) \left( \frac{\boldsymbol{L}}{L%
}\cdot \boldsymbol{f}_{b}\left( \boldsymbol{r}^{\prime }\right) \right) 
\mathrm{d}_{\boldsymbol{r}^{\prime }}V\right.  \notag \\
& \quad \quad \quad \quad \quad \left. -15\int_{%
\mathbb{R}
^{3}}\frac{\boldsymbol{r}-\boldsymbol{r}^{\prime }}{\left\vert \boldsymbol{r}%
-\boldsymbol{r}^{\prime }\right\vert ^{5}}\left( \frac{\boldsymbol{L}}{L}%
\cdot \frac{\boldsymbol{r}-\boldsymbol{r}^{\prime }}{\left\vert \boldsymbol{r%
}-\boldsymbol{r}^{\prime }\right\vert }\right) ^{2}\left( \boldsymbol{r}-%
\boldsymbol{r}^{\prime }\right) \cdot \boldsymbol{f}_{b}\left( \boldsymbol{r}%
^{\prime }\right) \mathrm{d}_{\boldsymbol{r}^{\prime }}V\right)
\label{u-non-loc-T4}
\end{align}%
in the case of model $T_{4}$, and finally by%
\begin{align}
\boldsymbol{u}& ^{\left( T_{5}\right) }\left( \boldsymbol{r}\right) =%
\boldsymbol{u}_{\ell }\left( \boldsymbol{r}\right) +\boldsymbol{u}_{n\ell
}^{\left( T_{5}\right) }\left( \boldsymbol{r}\right) =\boldsymbol{u}_{\ell
}\left( \boldsymbol{r}\right)  \notag \\
& =\frac{\lambda +3\mu }{8\pi \mu \left( \lambda +2\mu \right) }\int_{%
\mathbb{R}
^{3}}\frac{1}{\left\vert \boldsymbol{r}-\boldsymbol{r}^{\prime }\right\vert }%
\boldsymbol{f}_{b}\left( \boldsymbol{r}^{\prime }\right) \mathrm{d}_{%
\boldsymbol{r}^{\prime }}V+\frac{\lambda +\mu }{8\pi \mu \left( \lambda
+2\mu \right) }\int_{%
\mathbb{R}
^{3}}\frac{\boldsymbol{r}-\boldsymbol{r}^{\prime }}{\left\vert \boldsymbol{r}%
-\boldsymbol{r}^{\prime }\right\vert ^{3}}\left( \boldsymbol{r}-\boldsymbol{r%
}^{\prime }\right) \boldsymbol{\cdot f}_{b}\left( \boldsymbol{r}^{\prime
}\right) \mathrm{d}_{\boldsymbol{r}^{\prime }}V  \label{u-non-loc-T5}
\end{align}%
in the case of model $T_{5},$ where $\boldsymbol{u}_{\ell }$ is given by (%
\ref{u-loc}).

Namely, after the application of Fourier transform (\ref{FT}), the
equilibrium equation (\ref{equilibrium-eq}) and infinitesimal strain (\ref%
{str}) become%
\begin{gather*}
\mathrm{i}\boldsymbol{k\,\bar{\hat{\sigma}}}\left( \boldsymbol{k}\right) +%
\boldsymbol{\bar{f}}_{b}\left( \boldsymbol{k}\right) =\boldsymbol{0}\quad 
\text{and} \\
\boldsymbol{\bar{\hat{\varepsilon}}}\left( \boldsymbol{k}\right) =\frac{1}{2}%
\left( \left( \mathrm{i}\boldsymbol{k}\right) \otimes \boldsymbol{\bar{u}}%
\left( \boldsymbol{k}\right) +\boldsymbol{\bar{u}}\left( \boldsymbol{k}%
\right) \otimes \left( \mathrm{i}\boldsymbol{k}\right) \right) ,
\end{gather*}%
while non-local scalar- and tensor-type constitutive equations (\ref%
{Hook-non-loc}) and (\ref{Hook-non-loc-t}) become%
\begin{gather*}
\left( 1+\bar{S}_{i}(\boldsymbol{k})\right) \boldsymbol{\bar{\hat{\sigma}}}%
\left( \boldsymbol{k}\right) =\lambda \left( \mathrm{i}\boldsymbol{k}\cdot 
\boldsymbol{\bar{u}}\left( \boldsymbol{k}\right) \right) \boldsymbol{\hat{I}}%
+2\mu \,\boldsymbol{\bar{\hat{\varepsilon}}}\left( \boldsymbol{k}\right)
\quad \text{and} \\
\left( \boldsymbol{\hat{I}}+\boldsymbol{\bar{\hat{T}}}_{i}(\boldsymbol{k}%
)\right) \boldsymbol{\bar{\hat{\sigma}}}\left( \boldsymbol{k}\right)
=\lambda \left( \mathrm{i}\boldsymbol{k}\cdot \boldsymbol{\bar{u}}\left( 
\boldsymbol{k}\right) \right) \boldsymbol{\hat{I}}+2\mu \,\boldsymbol{\bar{%
\hat{\varepsilon}}}\left( \boldsymbol{k}\right) ,
\end{gather*}%
with functions $\bar{S}_{1}$ - $\bar{S}_{3}$ and tensors $\boldsymbol{\bar{%
\hat{T}}}_{1}$ - $\boldsymbol{\bar{\hat{T}}}_{5}$ given in Table \ref{Modeli}%
, since $\mathcal{F}\left[ \func{div}\boldsymbol{\hat{\sigma}}\left( 
\boldsymbol{r}\right) \right] \left( \boldsymbol{k}\right) =\left( \mathrm{i}%
\boldsymbol{k}\right) \boldsymbol{\,\bar{\hat{\sigma}}}\left( \boldsymbol{k}%
\right) ,$ $\mathcal{F}\left[ \func{grad}\boldsymbol{u}\left( \boldsymbol{r}%
\right) \right] \left( \boldsymbol{k}\right) =\left( \mathrm{i}\boldsymbol{k}%
\right) \otimes \boldsymbol{\bar{u}}\left( \boldsymbol{k}\right) ,$ $%
\mathcal{F}\left[ \left( \func{grad}\boldsymbol{u}\left( \boldsymbol{r}%
\right) \right) ^{\mathrm{T}}\right] \left( \boldsymbol{k}\right) =\left(
\left( \mathrm{i}\boldsymbol{k}\right) \otimes \boldsymbol{\bar{u}}\left( 
\boldsymbol{k}\right) \right) ^{\mathrm{T}}=\boldsymbol{\bar{u}}\left( 
\boldsymbol{k}\right) \otimes \left( \mathrm{i}\boldsymbol{k}\right) ,$ and $%
\mathcal{F}\left[ \mathrm{\limfunc{tr}}\boldsymbol{\hat{\varepsilon}}\left( 
\boldsymbol{r}\right) \right] \left( \boldsymbol{k}\right) =\mathcal{F}\left[
\mathrm{\limfunc{div}}\boldsymbol{u}\left( \boldsymbol{r}\right) \right]
\left( \boldsymbol{k}\right) =\left( \mathrm{i}\boldsymbol{k}\right) \cdot 
\boldsymbol{\bar{u}}\left( \boldsymbol{k}\right) $, so that previously
mentioned system of equations reduces to either%
\begin{gather*}
\left( 1+\bar{S}_{i}(\boldsymbol{k})\right) \mathrm{i}\boldsymbol{k}\,%
\boldsymbol{\bar{\hat{\sigma}}}\left( \boldsymbol{k}\right) =\lambda \,%
\mathrm{i}\boldsymbol{k}\left( \mathrm{i}\boldsymbol{k}\cdot \boldsymbol{%
\bar{u}}\left( \boldsymbol{k}\right) \right) \boldsymbol{\hat{I}}+\mu \,%
\mathrm{i}\boldsymbol{k}\left( \left( \mathrm{i}\boldsymbol{k}\right)
\otimes \boldsymbol{\bar{u}}\left( \boldsymbol{k}\right) \right) +\mu \,%
\mathrm{i}\boldsymbol{k}\left( \boldsymbol{\bar{u}}\left( \boldsymbol{k}%
\right) \otimes \left( \mathrm{i}\boldsymbol{k}\right) \right) ,\quad \text{%
i.e.,} \\
\left( 1+\bar{S}_{i}(\boldsymbol{k})\right) \boldsymbol{\bar{f}}_{b}\left( 
\boldsymbol{k}\right) =\mu \,k^{2}\boldsymbol{\bar{u}}\left( \boldsymbol{k}%
\right) +\left( \lambda +\mu \right) \left( \boldsymbol{k}\otimes 
\boldsymbol{k}\right) \boldsymbol{\bar{u}}\left( \boldsymbol{k}\right) \\
\left( 1+\bar{S}_{i}(\boldsymbol{k})\right) \boldsymbol{\bar{f}}_{b}\left( 
\boldsymbol{k}\right) =\mu \,k^{2}\left( \boldsymbol{\hat{I}}+\frac{\lambda
+\mu }{\mu }\frac{\boldsymbol{k}\otimes \boldsymbol{k}}{k^{2}}\right) 
\boldsymbol{\bar{u}}\left( \boldsymbol{k}\right) =\boldsymbol{\bar{\hat{R}}}%
_{\boldsymbol{\ell }}^{-1}\left( \boldsymbol{k}\right) \boldsymbol{\bar{u}}%
\left( \boldsymbol{k}\right) ,
\end{gather*}%
in the case of scalar-type models, with $\boldsymbol{\bar{\hat{R}}}_{%
\boldsymbol{\ell }}^{-1}$ given by (\ref{R-l-na-minus-prvi}), or to%
\begin{gather*}
\mathrm{i}\boldsymbol{k}\left( \boldsymbol{\hat{I}}+\boldsymbol{\bar{\hat{T}}%
}_{i}(\boldsymbol{k})\right) \boldsymbol{\bar{\hat{\sigma}}}\left( 
\boldsymbol{k}\right) =\lambda \,\mathrm{i}\boldsymbol{k}\left( \mathrm{i}%
\boldsymbol{k}\cdot \boldsymbol{\bar{u}}\left( \boldsymbol{k}\right) \right) 
\boldsymbol{\hat{I}}+\mu \,\mathrm{i}\boldsymbol{k}\left( \left( \mathrm{i}%
\boldsymbol{k}\right) \otimes \boldsymbol{\bar{u}}\left( \boldsymbol{k}%
\right) \right) +\mu \,\mathrm{i}\boldsymbol{k}\left( \boldsymbol{\bar{u}}%
\left( \boldsymbol{k}\right) \otimes \left( \mathrm{i}\boldsymbol{k}\right)
\right) ,\quad \text{i.e.,} \\
-\left( \boldsymbol{\hat{I}}+\boldsymbol{\bar{\hat{\mathcal{T}}}}_{i}(%
\boldsymbol{k})\right) \mathrm{i}\boldsymbol{k\,\bar{\hat{\sigma}}}\left( 
\boldsymbol{k}\right) =\left( \boldsymbol{\hat{I}}+\boldsymbol{\bar{\hat{%
\mathcal{T}}}}_{i}(\boldsymbol{k})\right) \boldsymbol{\bar{f}}_{b}\left( 
\boldsymbol{k}\right) =\mu \,k^{2}\boldsymbol{\bar{u}}\left( \boldsymbol{k}%
\right) +\left( \lambda +\mu \right) \left( \boldsymbol{k}\otimes 
\boldsymbol{k}\right) \boldsymbol{\bar{u}}\left( \boldsymbol{k}\right) \\
\left( \boldsymbol{\hat{I}}+\boldsymbol{\bar{\hat{\mathcal{T}}}}_{i}(%
\boldsymbol{k})\right) \boldsymbol{\bar{f}}_{b}\left( \boldsymbol{k}\right)
=\mu \,k^{2}\left( \boldsymbol{\hat{I}}+\frac{\lambda +\mu }{\mu }\frac{%
\boldsymbol{k}\otimes \boldsymbol{k}}{k^{2}}\right) \boldsymbol{\bar{u}}%
\left( \boldsymbol{k}\right) =\boldsymbol{\bar{\hat{R}}}_{\boldsymbol{\ell }%
}^{-1}\left( \boldsymbol{k}\right) \boldsymbol{\bar{u}}\left( \boldsymbol{k}%
\right) ,
\end{gather*}%
in the case of tensor-type models, with $\mathrm{i}\boldsymbol{k\,\bar{\hat{T%
}}}_{i}(\boldsymbol{k})=\boldsymbol{\bar{\hat{\mathcal{T}}}}_{i}(\boldsymbol{%
k})\,\mathrm{i}\boldsymbol{k}$, see Table \ref{Modeli}, since $\mathrm{i}%
\boldsymbol{k}\left( \boldsymbol{\bar{u}}\left( \boldsymbol{k}\right)
\otimes \left( \mathrm{i}\boldsymbol{k}\right) \right) =\mathrm{i}%
\boldsymbol{k}\left( \mathrm{i}\boldsymbol{k}\cdot \boldsymbol{\bar{u}}%
\left( \boldsymbol{k}\right) \right) =\left( \left( \mathrm{i}\boldsymbol{k}%
\right) \otimes \left( \mathrm{i}\boldsymbol{k}\right) \right) \boldsymbol{%
\bar{u}}\left( \boldsymbol{k}\right) =-\left( \boldsymbol{k}\otimes 
\boldsymbol{k}\right) \boldsymbol{\bar{u}}\left( \boldsymbol{k}\right) $ and 
$\mathrm{i}\boldsymbol{k}\left( \left( \mathrm{i}\boldsymbol{k}\right)
\otimes \boldsymbol{\bar{u}}\left( \boldsymbol{k}\right) \right) =-k^{2}%
\boldsymbol{\bar{u}}\left( \boldsymbol{k}\right) ,$ which can be written in
terms of local resolvent tensor either as 
\begin{gather*}
\boldsymbol{\bar{\hat{R}}}_{\boldsymbol{\ell }}^{-1}\left( \boldsymbol{k}%
\right) \boldsymbol{\bar{u}}\left( \boldsymbol{k}\right) =\left( 1+\bar{S}%
_{i}(\boldsymbol{k})\right) \boldsymbol{\bar{f}}_{b}\left( \boldsymbol{k}%
\right) ,\quad \text{i.e.,}\quad \boldsymbol{\bar{u}}\left( \boldsymbol{k}%
\right) =\boldsymbol{\bar{\hat{R}}}_{\boldsymbol{\ell }}\left( \boldsymbol{k}%
\right) \left( 1+\bar{S}_{i}(\boldsymbol{k})\right) \boldsymbol{\bar{f}}%
_{b}\left( \boldsymbol{k}\right) ,\quad \text{or as}  \label{u-bar-sk} \\
\boldsymbol{\bar{\hat{R}}}_{\boldsymbol{\ell }}^{-1}\left( \boldsymbol{k}%
\right) \boldsymbol{\bar{u}}\left( \boldsymbol{k}\right) =\left( \boldsymbol{%
\hat{I}}+\boldsymbol{\bar{\hat{\mathcal{T}}}}_{i}(\boldsymbol{k})\right) 
\boldsymbol{\bar{f}}_{b}\left( \boldsymbol{k}\right) ,\quad \text{i.e.,}%
\quad \boldsymbol{\bar{u}}\left( \boldsymbol{k}\right) =\boldsymbol{\bar{%
\hat{R}}}_{\boldsymbol{\ell }}\left( \boldsymbol{k}\right) \left( 
\boldsymbol{\hat{I}}+\boldsymbol{\bar{\hat{\mathcal{T}}}}_{i}(\boldsymbol{k}%
)\right) \boldsymbol{\bar{f}}_{b}\left( \boldsymbol{k}\right) ,
\label{u-bar-ten}
\end{gather*}%
with $\boldsymbol{\bar{\hat{R}}}_{\boldsymbol{\ell }}$ given by (\ref%
{u-fd-loc})$_{2}$, implying%
\begin{align}
& \boldsymbol{\bar{u}}\left( \boldsymbol{k}\right) =\boldsymbol{\bar{u}}%
_{\ell }\left( \boldsymbol{k}\right) +\boldsymbol{\bar{u}}_{n\ell }^{\left(
S_{i}/T_{i}\right) }\left( \boldsymbol{k}\right) ,\quad \text{with}
\label{u-l+u-ul} \\
& \boldsymbol{\bar{u}}_{\ell }\left( \boldsymbol{k}\right) =\boldsymbol{\bar{%
\hat{R}}}_{\boldsymbol{\ell }}\left( \boldsymbol{k}\right) \boldsymbol{\bar{f%
}}_{b}\left( \boldsymbol{k}\right) \quad \text{and\quad }\boldsymbol{\bar{u}}%
_{n\ell }^{\left( S_{i}/T_{i}\right) }\left( \boldsymbol{k}\right) =%
\boldsymbol{\bar{\hat{R}}}_{n\ell }^{\left( S_{i}/T_{i}\right) }\left( 
\boldsymbol{k}\right) \boldsymbol{\bar{f}}_{b}\left( \boldsymbol{k}\right)
,\quad \text{ where}  \label{u-l-i-u-ul} \\
& \boldsymbol{\bar{\hat{R}}}_{n\ell }^{\left( S_{i}\right) }\left( 
\boldsymbol{k}\right) =\boldsymbol{\bar{\hat{R}}}_{\boldsymbol{\ell }}\left( 
\boldsymbol{k}\right) \bar{S}_{i}(\boldsymbol{k})\text{\quad and\quad }%
\boldsymbol{\bar{\hat{R}}}_{n\ell }^{\left( T_{i}\right) }\left( \boldsymbol{%
k}\right) =\boldsymbol{\bar{\hat{R}}}_{\boldsymbol{\ell }}\left( \boldsymbol{%
k}\right) \boldsymbol{\bar{\hat{\mathcal{T}}}}_{i}(\boldsymbol{k}).
\label{R-nl-bar}
\end{align}

Therefore, for displacement vector field $\boldsymbol{u}$ one obtains%
\begin{align}
& \boldsymbol{u}\left( \boldsymbol{r}\right) =\boldsymbol{u}_{\ell }\left( 
\boldsymbol{r}\right) +\boldsymbol{u}_{n\ell }^{\left( S_{i}/T_{i}\right)
}\left( \boldsymbol{r}\right), \quad \text{with}  \label{u-superpoz} \\
& \boldsymbol{u}_{\ell }\left( \boldsymbol{r}\right) =\boldsymbol{\hat{R}}_{%
\boldsymbol{\ell }}\left( \boldsymbol{r}\right) \ast \boldsymbol{f}%
_{b}\left( \boldsymbol{r}\right) \quad \text{and\quad }\boldsymbol{u}_{n\ell
}^{\left( S_{i}/T_{i}\right) }\left( \boldsymbol{r}\right) =\boldsymbol{\hat{%
R}}_{n\ell }^{\left( S_{i}/T_{i}\right) }\left( \boldsymbol{r}\right) \ast 
\boldsymbol{f}_{b}\left( \boldsymbol{r}\right) ,  \label{u-konv}
\end{align}%
by applying the Fourier transform inversion (\ref{IFT}) to (\ref{u-l+u-ul})
and (\ref{u-l-i-u-ul}).

\begin{table}[h]
\begin{center}
\begin{tabular}{|c|c|c|c|c|c|c|c|} \cline{1-6} \xrowht{20pt}
 & Model & \multicolumn{2}{c|}{Non-locality operator}  & \multicolumn{2}{c|}{\makecell{Non-locality operator \\ in Fourier domain}}    &    \multicolumn{2}{c}{\multirow{4}{*}{}} \\ \cline{1-6} \xrowht{20pt}

\multirow{2.5}{*}{\multirowcell{3}{\rotatebox{90}{Scalar-type}}} & $S_1$ & \multirow{3}{*}{\multirowcell{3.25}{ ${\rm{S}}_{i}(\nabla)$ }} & $\left( \boldsymbol{L}\cdot \boldsymbol{L}\right) \left( \nabla \cdot \nabla \right)$  & \multirow{3}{*}{\multirowcell{3.25}{ $\bar{S}_{i}(\boldsymbol{k})$ }} & $L^{2}k^{2}$ &  \multicolumn{2}{c}{} \\ \cline{2-2} \cline{4-4} \cline{6-6}  \xrowht{20pt}

 & $S_2$  & & $\left( \boldsymbol{L}\cdot \nabla \right) ^{2} $ &  & $\left( \boldsymbol{L}\cdot 
\boldsymbol{k}\right) ^{2}$ &  \multicolumn{2}{c}{}   \\  \cline{2-2} \cline{4-4} \cline{6-6}  \xrowht{20pt}

 & $S_3$ & & $\left(\boldsymbol{L}\times \nabla \right) \cdot \left( \boldsymbol{L}\times \nabla\right) $   &  & $L^{2}k^{2}-\left( \boldsymbol{L}\cdot 
\boldsymbol{k}\right) ^{2}$  &  \multicolumn{2}{c}{}   \\ \cline{1-8}  \xrowht{20pt}

\multirow{6.7}{*}{\multirowcell{4}{\rotatebox{90}{Tensor-type}}} & $T_{1}$ & \multirow{4}{*}{\multirowcell{6.8}{ $\boldsymbol{\hat{{\rm{T}}}}_{i}(\nabla)$ }} & $\left( \boldsymbol{L}\cdot\boldsymbol{L}\right) \left( \nabla \otimes \nabla \right) $ &  \multirow{4}{*}{\multirowcell{6.8}{ $\boldsymbol{\bar{\hat{{T}}}}_{i}(\boldsymbol{k})$ }} & $ L^{2} (\boldsymbol{k} \otimes \boldsymbol{k}) $ & \multirow{4}{*}{\multirowcell{6.8}{  $\boldsymbol{\bar{\hat{\mathcal{T}}}}_{i}(\boldsymbol{k})$}} & $ L^{2} k^{2}  \boldsymbol{\hat{I}} $ \\  \cline{2-2} \cline{4-4} \cline{6-6} \cline{8-8} \xrowht{20pt}

 & $T_{2}$  & & $\left( \boldsymbol{L}\otimes \boldsymbol{L}\right) \left( \nabla \cdot
\nabla \right) $ &  &  $ k^2 (\boldsymbol{L} \otimes \boldsymbol{L}) $ & & $ k^{2} (\boldsymbol{L} \otimes \boldsymbol{L}) $ \\  \cline{2-2} \cline{4-4} \cline{6-6} \cline{8-8} \xrowht{20pt}

 & $T_3$ & &$\left( \boldsymbol{L}\cdot \nabla \right) \left( \boldsymbol{L}\otimes
\nabla \right) $  & & $ (\boldsymbol{L} \cdot \boldsymbol{k}) (\boldsymbol{L} \otimes \boldsymbol{k}) $ & & $ (\boldsymbol{L} \cdot \boldsymbol{k})^{2}  \boldsymbol{\hat{I}} $ \\ \cline{2-2} \cline{4-4} \cline{6-6} \cline{8-8} \xrowht{20pt}

 & $T_4$ & &$\left( \boldsymbol{L}\cdot \nabla \right)\left( \boldsymbol{L}\otimes \nabla \right) ^{\mathrm{T}} $ & & $ (\boldsymbol{L} \cdot \boldsymbol{k}) ( \boldsymbol{k}  \otimes \boldsymbol{L}) $ & & $ (\boldsymbol{L} \cdot \boldsymbol{k}) (\boldsymbol{L} \otimes \boldsymbol{k}) $ \\ \cline{2-2} \cline{4-4} \cline{6-6} \cline{8-8} \xrowht{20pt}

 & $T_5$ & &  $\left( \boldsymbol{L}\times \nabla \right)
\otimes \left( \boldsymbol{L}\times \nabla \right)$ &  & $ (\boldsymbol{L} \times \boldsymbol{k}) \otimes (\boldsymbol{L} \times \boldsymbol{k}) $ & & $ \boldsymbol{\hat{0}} $ \\ \cline{1-8}
\end{tabular}
\end{center}
\caption{Scalar and tensor operators in both spatial and Fourier domains, as well as tensor function $\boldsymbol{\bar{\hat{\mathcal{T}}}}_{i}$, where $\bar{S}_{i}(\boldsymbol{k})=-\mathcal{F}\left[ {\rm{S}}_{i}(\nabla)\right] \left(\boldsymbol{ k}\right) $, $\boldsymbol{\bar{\hat{{T}}}}_{i}(\boldsymbol{k})=-\mathcal{F}\left[ \boldsymbol{\hat{{\rm{T}}}}_{i}(\nabla)\right] \left(\boldsymbol{ k}\right) $, and   
 $\mathrm{i}\boldsymbol{k\,\bar{\hat{T}}}_{i}(\boldsymbol{k})=%
\boldsymbol{\bar{\hat{\mathcal{T}}}}_{i}(\boldsymbol{k})\,\mathrm{i}%
\boldsymbol{k}$.}
\label{Modeli}
\end{table}

\subsubsection{Models $S_{1}$ and $T_{1}$}

According to (\ref{u-konv})$_{2},$ the non-local displacement vector field
in the case of models $S_{1}$ and $T_{1}$ takes the form%
\begin{align}
\boldsymbol{u}_{n\ell }^{\left( S_{1}/T_{1}\right) }\left( \boldsymbol{r}%
\right) & =\boldsymbol{\hat{R}}_{n\ell }^{\left( S_{1}/T_{1}\right) }\left( 
\boldsymbol{r}\right) \ast _{\boldsymbol{r}}\boldsymbol{f}_{b}\left( 
\boldsymbol{r}\right)  \notag \\
& =L^{2}\frac{2\lambda +3\mu }{\mu \left( \lambda +2\mu \right) }\boldsymbol{%
f}_{b}\left( \boldsymbol{r}\right) -4L^{2}\frac{\lambda +\mu }{\mu \left(
\lambda +2\mu \right) }\left( \delta \left( \boldsymbol{r}\right) \frac{%
\boldsymbol{r}\otimes \boldsymbol{r}}{r^{2}}\right) \ast _{\boldsymbol{r}}%
\boldsymbol{f}_{b}\left( \boldsymbol{r}\right)  \notag \\
& \quad -\frac{L^{2}}{4\pi }\frac{\lambda +\mu }{\mu \left( \lambda +2\mu
\right) }\left( \frac{1}{r^{3}}\left( \boldsymbol{\hat{I}}-3\frac{%
\boldsymbol{r}\otimes \boldsymbol{r}}{r^{2}}\right) \right) \ast _{%
\boldsymbol{r}}\boldsymbol{f}_{b}\left( \boldsymbol{r}\right)
\label{nltdf-S1} \\
& =-\frac{L^{2}}{3}\frac{2\lambda -\mu }{\mu \left( \lambda +2\mu \right) }%
\boldsymbol{f}_{b}\left( \boldsymbol{r}\right) -\frac{4L^{2}}{3}\frac{%
\lambda +\mu }{\mu \left( \lambda +2\mu \right) }\left( \boldsymbol{e}%
_{z}\cdot \boldsymbol{f}_{b}\left( \boldsymbol{r}\right) \right) \boldsymbol{%
e}_{z}  \notag \\
& \quad -\frac{L^{2}}{4\pi }\frac{\lambda +\mu }{\mu \left( \lambda +2\mu
\right) }\left( \frac{1}{r^{3}}\left( \boldsymbol{\hat{I}}-3\frac{%
\boldsymbol{r}\otimes \boldsymbol{r}}{r^{2}}\right) \right) \ast _{%
\boldsymbol{r}}\boldsymbol{f}_{b}\left( \boldsymbol{r}\right) ,  \notag
\end{align}%
where the non-local resolvent tensor%
\begin{equation}
\boldsymbol{\hat{R}}_{n\ell }^{\left( S_{1}/T_{1}\right) }\left( \boldsymbol{%
r}\right) =L^{2}\frac{2\lambda +3\mu }{\mu \left( \lambda +2\mu \right) }%
\,\delta \left( \boldsymbol{r}\right) \boldsymbol{\hat{I}}-4L^{2}\frac{%
\lambda +\mu }{\mu \left( \lambda +2\mu \right) }\delta \left( \boldsymbol{r}%
\right) \frac{\boldsymbol{r}\otimes \boldsymbol{r}}{r^{2}}-\frac{L^{2}}{4\pi 
}\frac{\lambda +\mu }{\mu \left( \lambda +2\mu \right) }\frac{1}{r^{3}}%
\left( \boldsymbol{\hat{I}}-3\frac{\boldsymbol{r}\otimes \boldsymbol{r}}{%
r^{2}}\right) ,  \label{R-nl-S1}
\end{equation}%
is obtained by applying the Fourier transform inversion (\ref{IFT}) to the
expression for the non-local resolvent tensor in Fourier domain, given by (%
\ref{R-nl-bar}), so that, according to (\ref{u-superpoz}), for the
displacement vector field one has (\ref{u-non-loc-S1}).

Note, in the first term of (\ref{nltdf-S1}) the property $\boldsymbol{f}%
_{b}\left( \boldsymbol{r}\right) =\delta \left( \boldsymbol{r}\right) \ast _{%
\boldsymbol{r}}\boldsymbol{f}_{b}\left( \boldsymbol{r}\right) $ of Dirac
delta distribution is used, while in the second term of (\ref{nltdf-S1}) the
expression $\left( \delta \left( \boldsymbol{r}\right) \frac{\boldsymbol{r}%
\otimes \boldsymbol{r}}{r^{2}}\right) \ast _{\boldsymbol{r}}\boldsymbol{f}%
_{b}\left( \boldsymbol{r}\right) $ can be rewritten in the integral form
using Cartesian coordinate system for representation and expressing
components in spherical coordinate system, yielding%
\begin{align}
\left( \delta \left( \boldsymbol{r}\right) \frac{\boldsymbol{r}\otimes 
\boldsymbol{r}}{r^{2}}\right) \ast _{\boldsymbol{r}}\boldsymbol{f}_{b}\left( 
\boldsymbol{r}\right) & =\int_{%
\mathbb{R}
^{3}}\delta \left( \boldsymbol{r}^{\prime }\right) \frac{\boldsymbol{r}%
^{\prime }\otimes \boldsymbol{r}^{\prime }}{r^{\prime 2}}\boldsymbol{f}%
_{b}\left( \boldsymbol{r}-\boldsymbol{r}^{\prime }\right) \mathrm{d}_{%
\boldsymbol{r}^{\prime }}V  \notag \\
& =\int_{0}^{2\pi }\!\!\!\int_{0}^{\infty }\!\!\!\int_{0}^{\pi }\frac{\delta
\left( r^{\prime }\right) }{2\pi r^{\prime 2}}\Bigl(\sin ^{2}\theta \cos
^{2}\varphi \left( \boldsymbol{e}_{x}\otimes \boldsymbol{e}_{x}\right) +\sin
^{2}\theta \sin ^{2}\varphi \left( \boldsymbol{e}_{y}\otimes \boldsymbol{e}%
_{y}\right) +\cos ^{2}\theta \left( \boldsymbol{e}_{z}\otimes \boldsymbol{e}%
_{z}\right) \Bigr.  \notag \\
& \quad \quad \quad \quad \quad +\sin ^{2}\theta \sin \varphi \cos \varphi
\left( \boldsymbol{e}_{x}\otimes \boldsymbol{e}_{y}+\boldsymbol{e}%
_{y}\otimes \boldsymbol{e}_{x}\right) +\sin \theta \cos \theta \cos \varphi
\left( \boldsymbol{e}_{x}\otimes \boldsymbol{e}_{z}+\boldsymbol{e}%
_{z}\otimes \boldsymbol{e}_{x}\right)  \notag \\
& \quad \quad \quad \quad \quad \quad \Bigl.+\sin \theta \cos \theta \sin
\varphi \left( \boldsymbol{e}_{y}\otimes \boldsymbol{e}_{z}+\boldsymbol{e}%
_{z}\otimes \boldsymbol{e}_{y}\right) \Bigr)  \notag \\
& \quad \quad \quad \quad \quad \quad \quad \times \boldsymbol{f}_{b}\left(
x-r^{\prime }\sin \theta \cos \varphi ,y-r^{\prime }\sin \theta \sin \varphi
,z-r^{\prime }\cos \theta \right) r^{\prime 2}\sin \theta \mathrm{d}\theta 
\mathrm{d}r^{\prime }\mathrm{d}\varphi  \notag \\
& =\frac{1}{2\pi }\int_{0}^{2\pi }\!\!\!\int_{0}^{\pi }\Bigl(\sin ^{3}\theta
\cos ^{2}\varphi \left( \boldsymbol{e}_{x}\otimes \boldsymbol{e}_{x}\right)
+\sin ^{3}\theta \sin ^{2}\varphi \left( \boldsymbol{e}_{y}\otimes 
\boldsymbol{e}_{y}\right) +\sin \theta \cos ^{2}\theta \left( \boldsymbol{e}%
_{z}\otimes \boldsymbol{e}_{z}\right) \Bigr.  \notag \\
& \quad \quad \quad \quad \quad +\sin ^{3}\theta \sin \varphi \cos \varphi
\left( \boldsymbol{e}_{x}\otimes \boldsymbol{e}_{y}+\boldsymbol{e}%
_{y}\otimes \boldsymbol{e}_{x}\right) +\sin ^{2}\theta \cos \theta \cos
\varphi \left( \boldsymbol{e}_{x}\otimes \boldsymbol{e}_{z}+\boldsymbol{e}%
_{z}\otimes \boldsymbol{e}_{x}\right)  \notag \\
& \quad \quad \quad \quad \quad \quad \Bigl.+\sin ^{2}\theta \cos \theta
\sin \varphi \left( \boldsymbol{e}_{y}\otimes \boldsymbol{e}_{z}+\boldsymbol{%
e}_{z}\otimes \boldsymbol{e}_{y}\right) \Bigr.\boldsymbol{f}_{b}\left( 
\boldsymbol{r}\right) \mathrm{d}\theta \mathrm{d}\varphi  \notag \\
& =\frac{1}{2\pi }\left( \frac{4\pi }{3}\left( \boldsymbol{e}_{x}\otimes 
\boldsymbol{e}_{x}\right) +\frac{4\pi }{3}\left( \boldsymbol{e}_{y}\otimes 
\boldsymbol{e}_{y}\right) +2\pi \left( \boldsymbol{e}_{z}\otimes \boldsymbol{%
e}_{z}\right) \right) \boldsymbol{f}_{b}\left( \boldsymbol{r}\right)  \notag
\\
& =\left( \frac{2}{3}\boldsymbol{\hat{I}}+\frac{1}{3}\left( \boldsymbol{e}%
_{z}\otimes \boldsymbol{e}_{z}\right) \right) \boldsymbol{f}_{b}\left( 
\boldsymbol{r}\right)  \notag \\
& =\frac{2}{3}\boldsymbol{f}_{b}\left( \boldsymbol{r}\right) +\frac{1}{3}%
\left( \boldsymbol{e}_{z}\cdot \boldsymbol{f}_{b}\left( \boldsymbol{r}%
\right) \right) \boldsymbol{e}_{z},  \label{delta-projec-conv}
\end{align}%
where $\boldsymbol{f}_{b}\left( \boldsymbol{r}-\boldsymbol{r}^{\prime
}\right) =\boldsymbol{f}_{b}\left( x-x^{\prime },y-y^{\prime },z-z^{\prime
}\right) =\boldsymbol{f}_{b}\left( x-r^{\prime }\sin \theta \cos \varphi
,y-r^{\prime }\sin \theta \sin \varphi ,z-r^{\prime }\cos \theta \right) $.

Namely, in the case of models $S_{1}$ and $T_{1},$ the scalar and tensor
functions $\bar{S}_{1}$ and $\boldsymbol{\bar{\hat{\mathcal{T}}}}_{1}$,
taking the form%
\begin{equation*}
\bar{S}_{1}\left( \boldsymbol{k}\right) =L^{2}k^{2}\quad \text{and\quad }%
\boldsymbol{\bar{\hat{\mathcal{T}}}}_{1}=L^{2}k^{2}\boldsymbol{\hat{I}},
\end{equation*}%
since%
\begin{equation*}
\mathrm{i}\boldsymbol{k\,\bar{\hat{T}}}_{1}(\boldsymbol{k})=L^{2}\,\mathrm{i}%
\boldsymbol{k}\left( \boldsymbol{k}\otimes \boldsymbol{k}\right)
=L^{2}k^{2}\,\mathrm{i}\boldsymbol{k}=\boldsymbol{\bar{\hat{\mathcal{T}}}}%
_{1}\,\mathrm{i}\boldsymbol{k},
\end{equation*}%
see Table \ref{Modeli}, along with tensor $\boldsymbol{\bar{\hat{R}}}_{%
\boldsymbol{\ell }}$, given by (\ref{u-fd-loc})$_{2}$, yield 
\begin{equation}
\boldsymbol{\bar{\hat{R}}}_{n\ell }^{\left( S_{1}/T_{1}\right) }\left( 
\boldsymbol{k}\right) =L^{2}\frac{1}{\mu }\boldsymbol{\hat{I}}-L^{2}\frac{%
\lambda +\mu }{\mu \left( \lambda +2\mu \right) }\frac{\boldsymbol{k}\otimes 
\boldsymbol{k}}{k^{2}}  \label{R-nl-bar-S1}
\end{equation}%
for the non-local resolvent tensor in the Fourier domain (\ref{R-nl-bar}).
After the Fourier transform inversion, defined by (\ref{IFT}), the previous
expression transforms into (\ref{R-nl-S1}), since $\mathcal{F}^{-1}\left[ 1%
\right] \left( \boldsymbol{r}\right) =\delta \left( \boldsymbol{r}\right) $
because of $\mathcal{F}\left[ \delta \left( \boldsymbol{r}\right) \right]
\left( \boldsymbol{k}\right) =\int_{%
\mathbb{R}
^{3}}\delta \left( \boldsymbol{r}\right) \mathrm{e}^{-\mathrm{i}\boldsymbol{%
k\cdot r}}\mathrm{d}_{\boldsymbol{r}}V=1$ and since the Fourier transform
inversion of tensor $\frac{\boldsymbol{k}\otimes \boldsymbol{k}}{k^{2}}$
yields%
\begin{equation}
\mathcal{F}^{-1}\left[ \frac{\boldsymbol{k}\otimes \boldsymbol{k}}{k^{2}}%
\right] \left( \boldsymbol{r}\right) =-\delta \left( \boldsymbol{r}\right)
\left( \boldsymbol{\hat{I}}-4\frac{\boldsymbol{r}\otimes \boldsymbol{r}}{%
r^{2}}\right) +\frac{1}{4\pi r^{3}}\left( \boldsymbol{\hat{I}}-3\frac{%
\boldsymbol{r}\otimes \boldsymbol{r}}{r^{2}}\right) ,
\label{dijadski-kroz-kv-rezultat}
\end{equation}%
see (\ref{k-dij-k-kroz-drugi}), as well as calculation in Appendix \ref%
{dijadski-kroz-kvadrat}.

\subsubsection{Model $S_{2}$ and $T_{3}$}

According to (\ref{u-konv})$_{2},$ the non-local displacement vector field
in the case of models $S_{2}$ and $T_{3}$ takes the form%
\begin{align}
\boldsymbol{u}_{n\ell }^{\left( S_{2}/T_{3}\right) }\left( \boldsymbol{r}%
\right) & =\boldsymbol{\hat{R}}_{n\ell }^{\left( S_{2}/T_{3}\right) }\left( 
\boldsymbol{r}\right) \ast _{\boldsymbol{r}}\boldsymbol{f}_{b}\left( 
\boldsymbol{r}\right)  \notag \\
& =L^{2}\frac{1}{\mu }\left( \boldsymbol{\hat{I}}-\frac{\lambda +\mu }{%
\lambda +2\mu }\frac{\boldsymbol{L}\otimes \boldsymbol{L}}{L^{2}}\right) 
\boldsymbol{f}_{b}\left( \boldsymbol{r}\right)  \notag \\
& \quad +\frac{L^{2}}{8\pi }\frac{\lambda +\mu }{\mu \left( \lambda +2\mu
\right) }\left( \left( \frac{1}{r^{3}}\left( 1-3\left( \frac{\boldsymbol{L}}{%
L}\cdot \frac{\boldsymbol{r}}{r}\right) ^{2}\right) \right) \ast _{%
\boldsymbol{r}}\boldsymbol{f}_{b}\left( \boldsymbol{r}\right) \right.  \notag
\\
& \quad \quad \quad \quad \quad \quad \quad \quad \quad \left. -2\left( 
\frac{1}{r^{3}}\frac{\boldsymbol{L}\otimes \boldsymbol{L}}{L^{2}}\right)
\ast _{\boldsymbol{r}}\boldsymbol{f}_{b}\left( \boldsymbol{r}\right)
+6\left( \frac{1}{r^{3}}\left( \frac{\boldsymbol{L}}{L}\cdot \frac{%
\boldsymbol{r}}{r}\right) \left( \frac{\boldsymbol{L}}{L}\otimes \frac{%
\boldsymbol{r}}{r}+\frac{\boldsymbol{r}}{r}\otimes \frac{\boldsymbol{L}}{L}%
\right) \right) \ast _{\boldsymbol{r}}\boldsymbol{f}_{b}\left( \boldsymbol{r}%
\right) \right.  \notag \\
& \quad \quad \quad \quad \quad \quad \quad \quad \quad \left. +3\left( 
\frac{1}{r^{3}}\left( 1-5\left( \frac{\boldsymbol{L}}{L}\cdot \frac{%
\boldsymbol{r}}{r}\right) ^{2}\right) \frac{\boldsymbol{r}\otimes 
\boldsymbol{r}}{r^{2}}\right) \ast _{\boldsymbol{r}}\boldsymbol{f}_{b}\left( 
\boldsymbol{r}\right) \right) ,  \label{nltdf-S2}
\end{align}%
where the non-local resolvent tensor%
\begin{align}
\boldsymbol{\hat{R}}_{n\ell }^{\left( S_{2}/T_{3}\right) }\left( \boldsymbol{%
r}\right) & =L^{2}\frac{1}{\mu }\delta \left( \boldsymbol{r}\right) \left( 
\boldsymbol{\hat{I}}-\frac{\lambda +\mu }{\lambda +2\mu }\frac{\boldsymbol{L}%
\otimes \boldsymbol{L}}{L^{2}}\right)  \notag \\
& \quad +\frac{L^{2}}{8\pi }\frac{\lambda +\mu }{\mu \left( \lambda +2\mu
\right) }\frac{1}{r^{3}}\left( \left( 1-3\left( \frac{\boldsymbol{L}}{L}%
\cdot \frac{\boldsymbol{r}}{r}\right) ^{2}\right) \boldsymbol{\hat{I}}-2%
\frac{\boldsymbol{L}\otimes \boldsymbol{L}}{L^{2}}\right.  \notag \\
& \quad \quad \quad \quad \quad \quad \quad \quad \quad \quad \left.
+6\left( \frac{\boldsymbol{L}}{L}\cdot \frac{\boldsymbol{r}}{r}\right)
\left( \frac{\boldsymbol{L}}{L}\otimes \frac{\boldsymbol{r}}{r}+\frac{%
\boldsymbol{r}}{r}\otimes \frac{\boldsymbol{L}}{L}\right) +3\left( 1-5\left( 
\frac{\boldsymbol{L}}{L}\cdot \frac{\boldsymbol{r}}{r}\right) ^{2}\right) 
\frac{\boldsymbol{r}\otimes \boldsymbol{r}}{r^{2}}\right) ,  \label{R-nl-S2}
\end{align}%
is obtained by applying the Fourier transform inversion (\ref{IFT}) to the
expression for the non-local resolvent tensor in Fourier domain, given by (%
\ref{R-nl-bar}), so that, according to (\ref{u-superpoz}), for the
displacement vector field one has (\ref{u-non-loc-S2}).

Namely, in the case of models $S_{2}$ and $T_{3},$ the scalar and tensor
functions $\bar{S}_{2}$ and $\boldsymbol{\bar{\hat{\mathcal{T}}}}_{3}$,
taking the form%
\begin{equation*}
\bar{S}_{2}\left( \boldsymbol{k}\right) =\left( \boldsymbol{L}\cdot 
\boldsymbol{k}\right) ^{2}\quad \text{and\quad }\boldsymbol{\bar{\hat{%
\mathcal{T}}}}_{3}=\left( \boldsymbol{L}\cdot \boldsymbol{k}\right) ^{2}%
\boldsymbol{\hat{I}},
\end{equation*}%
since%
\begin{equation*}
\mathrm{i}\boldsymbol{k\,\bar{\hat{T}}}_{3}(\boldsymbol{k})=\left( 
\boldsymbol{L}\cdot \boldsymbol{k}\right) \mathrm{i}\boldsymbol{k\,}\left( 
\boldsymbol{L}\otimes \boldsymbol{k}\right) =\left( \boldsymbol{L}\cdot 
\boldsymbol{k}\right) ^{2}\,\mathrm{i}\boldsymbol{k}=\boldsymbol{\bar{\hat{%
\mathcal{T}}}}_{3}\,\mathrm{i}\boldsymbol{k},
\end{equation*}%
see Table \ref{Modeli}, along with tensor $\boldsymbol{\bar{\hat{R}}}_{%
\boldsymbol{\ell }}$, given by (\ref{u-fd-loc})$_{2}$, yield 
\begin{equation}
\boldsymbol{\bar{\hat{R}}}_{n\ell }^{\left( S_{2}/T_{3}\right) }\left( 
\boldsymbol{k}\right) =\frac{1}{\mu }\frac{\left( \boldsymbol{L}\cdot 
\boldsymbol{k}\right) ^{2}}{k^{2}}\boldsymbol{\hat{I}}-\frac{\lambda +\mu }{%
\mu \left( \lambda +2\mu \right) }\left( \boldsymbol{L}\cdot \boldsymbol{k}%
\right) ^{2}\frac{\boldsymbol{k}\otimes \boldsymbol{k}}{k^{4}},
\label{R-nl-bar-S2}
\end{equation}%
for the non-local resolvent tensor in the Fourier domain (\ref{R-nl-bar}).
After the Fourier transform inversion, defined by (\ref{IFT}), the previous
expression transforms into (\ref{R-nl-S2}), since the Fourier transform
inversion of function $\frac{\left( \boldsymbol{L}\cdot \boldsymbol{k}%
\right) ^{2}}{k^{2}},$ respectively tensor $\left( \boldsymbol{L}\cdot 
\boldsymbol{k}\right) ^{2}\frac{\boldsymbol{k}\otimes \boldsymbol{k}}{k^{4}}$
yields%
\begin{equation*}
\mathcal{F}^{-1}\left[ \frac{\left( \boldsymbol{L}\cdot \boldsymbol{k}%
\right) ^{2}}{k^{2}}\right] \left( \boldsymbol{r}\right) =L^{2}\delta \left( 
\boldsymbol{r}\right) +\frac{L^{2}}{4\pi r^{3}}\left( 1-3\left( \frac{%
\boldsymbol{L}}{L}\cdot \frac{\boldsymbol{r}}{r}\right) ^{2}\right) ,
\end{equation*}%
respectively%
\begin{align}
\mathcal{F}^{-1}\left[ \left( \boldsymbol{L}\cdot \boldsymbol{k}\right) ^{2}%
\frac{\boldsymbol{k}\otimes \boldsymbol{k}}{k^{4}}\right] \left( \boldsymbol{%
r}\right) & =\delta \left( \boldsymbol{r}\right) \left( \boldsymbol{L}%
\otimes \boldsymbol{L}\right) +\frac{L^{2}}{8\pi r^{3}}\left( \left(
1-3\left( \frac{\boldsymbol{L}}{L}\cdot \frac{\boldsymbol{r}}{r}\right)
^{2}\right) \boldsymbol{\hat{I}}+2\frac{\boldsymbol{L}\otimes \boldsymbol{L}%
}{L^{2}}\right.  \notag \\
& \quad \left. -6\left( \frac{\boldsymbol{L}}{L}\cdot \frac{\boldsymbol{r}}{r%
}\right) \left( \frac{\boldsymbol{L}}{L}\otimes \frac{\boldsymbol{r}}{r}+%
\frac{\boldsymbol{r}}{r}\otimes \frac{\boldsymbol{L}}{L}\right) -3\left(
1-5\left( \frac{\boldsymbol{L}}{L}\cdot \frac{\boldsymbol{r}}{r}\right)
^{2}\right) \frac{\boldsymbol{r}\otimes \boldsymbol{r}}{r^{2}}\right) ,
\label{dodat}
\end{align}%
see (\ref{skalarni-kroz-kv}) and (\ref{operator-D}), as well as calculations
in Appendices \ref{skalarni-kroz-kvadrat} and \ref%
{dijadski-skalarni-kroz-cetvrti}.

\subsubsection{Model $S_{3}$}

According to (\ref{u-konv})$_{2},$ the non-local displacement vector field
in the case of model $S_{3}$ takes the form%
\begin{align}
\boldsymbol{u}_{n\ell }^{\left( S_{3}\right) }\left( \boldsymbol{r}\right) &
=\boldsymbol{\hat{R}}_{n\ell }^{\left( S_{3}\right) }\left( \boldsymbol{r}%
\right) \ast _{\boldsymbol{r}}\boldsymbol{f}_{b}\left( \boldsymbol{r}\right)
=\left( \boldsymbol{\hat{R}}_{n\ell }^{\left( S_{1}\right) }\left( 
\boldsymbol{r}\right) -\boldsymbol{\hat{R}}_{n\ell }^{\left( S_{2}\right)
}\left( \boldsymbol{r}\right) \right) \ast _{\boldsymbol{r}}\boldsymbol{f}%
_{b}\left( \boldsymbol{r}\right) =\boldsymbol{u}_{n\ell }^{\left(
S_{1}\right) }\left( \boldsymbol{r}\right) -\boldsymbol{u}_{n\ell }^{\left(
S_{2}\right) }\left( \boldsymbol{r}\right)  \notag \\
& =L^{2}\frac{\lambda +\mu }{\mu \left( \lambda +2\mu \right) }\left( 
\boldsymbol{\hat{I}}+\frac{\boldsymbol{L}\otimes \boldsymbol{L}}{L^{2}}%
\right) \boldsymbol{f}_{b}\left( \boldsymbol{r}\right) -4L^{2}\frac{\lambda
+\mu }{\mu \left( \lambda +2\mu \right) }\delta \left( \boldsymbol{r}\right) 
\frac{\boldsymbol{r}\otimes \boldsymbol{r}}{r^{2}}\ast _{\boldsymbol{r}}%
\boldsymbol{f}_{b}\left( \boldsymbol{r}\right)  \notag \\
& \quad -\frac{L^{2}}{8\pi }\frac{\lambda +\mu }{\mu \left( \lambda +2\mu
\right) }\left( 3\left( \frac{1}{r^{3}}\left( 1-\left( \frac{\boldsymbol{L}}{%
L}\cdot \frac{\boldsymbol{r}}{r}\right) ^{2}\right) \right) \ast _{%
\boldsymbol{r}}\boldsymbol{f}_{b}\left( \boldsymbol{r}\right) -2\left( \frac{%
1}{r^{3}}\frac{\boldsymbol{L}\otimes \boldsymbol{L}}{L^{2}}\right) \ast _{%
\boldsymbol{r}}\boldsymbol{f}_{b}\left( \boldsymbol{r}\right) \right.  \notag
\\
& \quad \quad \quad \quad \quad \quad \quad \quad \quad \left. +6\left( 
\frac{1}{r^{3}}\left( \frac{\boldsymbol{L}}{L}\cdot \frac{\boldsymbol{r}}{r}%
\right) \left( \frac{\boldsymbol{L}}{L}\otimes \frac{\boldsymbol{r}}{r}+%
\frac{\boldsymbol{r}}{r}\otimes \frac{\boldsymbol{L}}{L}\right) \right) \ast
_{\boldsymbol{r}}\boldsymbol{f}_{b}\left( \boldsymbol{r}\right) \right. 
\notag \\
& \quad \quad \quad \quad \quad \quad \quad \quad \quad \left. -3\left( 
\frac{1}{r^{3}}\left( 1+5\left( \frac{\boldsymbol{L}}{L}\cdot \frac{%
\boldsymbol{r}}{r}\right) ^{2}\right) \frac{\boldsymbol{r}\otimes 
\boldsymbol{r}}{r^{2}}\right) \ast _{\boldsymbol{r}}\boldsymbol{f}_{b}\left( 
\boldsymbol{r}\right) \right) ,  \label{nltdf-S3}
\end{align}%
where the non-local resolvent tensor%
\begin{align}
\boldsymbol{\hat{R}}_{n\ell }^{\left( S_{3}\right) }\left( \boldsymbol{r}%
\right) & =\boldsymbol{\hat{R}}_{n\ell }^{\left( S_{1}\right) }\left( 
\boldsymbol{r}\right) -\boldsymbol{\hat{R}}_{n\ell }^{\left( S_{2}\right)
}\left( \boldsymbol{r}\right)  \notag \\
& =L^{2}\frac{\lambda +\mu }{\mu \left( \lambda +2\mu \right) }\delta \left( 
\boldsymbol{r}\right) \left( \boldsymbol{\hat{I}}+\frac{\boldsymbol{L}%
\otimes \boldsymbol{L}}{L^{2}}\right) -4L^{2}\frac{\lambda +\mu }{\mu \left(
\lambda +2\mu \right) }\delta \left( \boldsymbol{r}\right) \frac{\boldsymbol{%
r}\otimes \boldsymbol{r}}{r^{2}}  \notag \\
& \quad -\frac{L^{2}}{8\pi }\frac{\lambda +\mu }{\mu \left( \lambda +2\mu
\right) }\frac{1}{r^{3}}\left( 3\left( 1-\left( \frac{\boldsymbol{L}}{L}%
\cdot \frac{\boldsymbol{r}}{r}\right) ^{2}\right) \boldsymbol{\hat{I}}-2%
\frac{\boldsymbol{L}\otimes \boldsymbol{L}}{L^{2}}\right.  \notag \\
& \quad \quad \quad \quad \quad \quad \quad \quad \quad \quad \left.
+6\left( \frac{\boldsymbol{L}}{L}\cdot \frac{\boldsymbol{r}}{r}\right)
\left( \frac{\boldsymbol{L}}{L}\otimes \frac{\boldsymbol{r}}{r}+\frac{%
\boldsymbol{r}}{r}\otimes \frac{\boldsymbol{L}}{L}\right) -3\left( 1+5\left( 
\frac{\boldsymbol{L}}{L}\cdot \frac{\boldsymbol{r}}{r}\right) ^{2}\right) 
\frac{\boldsymbol{r}\otimes \boldsymbol{r}}{r^{2}}\right)  \label{R-nl-S3}
\end{align}%
is obtained by applying the Fourier transform inversion (\ref{IFT}) to the
expression for the non-local resolvent tensor in Fourier domain, given by (%
\ref{R-nl-bar})$_{1}$, so that, according to (\ref{u-superpoz}), for the
displacement vector field one has (\ref{u-non-loc-S3}).

Namely, in the case of model $S_{3}$, the scalar function $\bar{S}_{3}$,
taking the form%
\begin{equation*}
\bar{S}_{3}\left( \boldsymbol{k}\right) =\left( \boldsymbol{L}\times 
\boldsymbol{k}\right) ^{2}=L^{2}k^{2}-\left( \boldsymbol{L}\cdot \boldsymbol{%
k}\right) ^{2}=\bar{S}_{1}\left( \boldsymbol{k}\right) -\bar{S}_{2}\left( 
\boldsymbol{k}\right) ,
\end{equation*}%
according to%
\begin{align*}
\left( \boldsymbol{L}\times \boldsymbol{k}\right) ^{2}& =\left( \boldsymbol{L%
}\times \boldsymbol{k}\right) \cdot \left( \boldsymbol{L}\times \boldsymbol{k%
}\right) =\boldsymbol{L}\cdot \left( \boldsymbol{k}\times \left( \boldsymbol{%
L}\times \boldsymbol{k}\right) \right) \\
& =\boldsymbol{L}\cdot \left( \boldsymbol{L}\left( \boldsymbol{k}\cdot 
\boldsymbol{k}\right) -\boldsymbol{k}\left( \boldsymbol{L}\cdot \boldsymbol{k%
}\right) \right) =L^{2}k^{2}-\left( \boldsymbol{L}\cdot \boldsymbol{k}%
\right) ^{2},
\end{align*}%
see Table \ref{Modeli}, and along with $\boldsymbol{\bar{\hat{R}}}_{%
\boldsymbol{\ell }}$, given by (\ref{u-fd-loc})$_{2}$, yields%
\begin{equation}
\boldsymbol{\bar{\hat{R}}}_{n\ell }^{\left( S_{3}\right) }\left( \boldsymbol{%
k}\right) =\boldsymbol{\bar{\hat{R}}}_{\boldsymbol{\ell }}\left( \boldsymbol{%
k}\right) \left( \bar{S}_{1}\left( \boldsymbol{k}\right) -\bar{S}_{2}\left( 
\boldsymbol{k}\right) \right) =\boldsymbol{\bar{\hat{R}}}_{n\boldsymbol{\ell 
}}^{\left( S_{1}\right) }\left( \boldsymbol{k}\right) -\boldsymbol{\bar{\hat{%
R}}}_{n\boldsymbol{\ell }}^{\left( S_{2}\right) }\left( \boldsymbol{k}\right)
\label{R-nl-bar-1-minus-2}
\end{equation}%
with tensors $\boldsymbol{\bar{\hat{R}}}_{n\boldsymbol{\ell }}^{\left(
S_{1}\right) }$ and $\boldsymbol{\bar{\hat{R}}}_{n\boldsymbol{\ell }%
}^{\left( S_{2}\right) },$ see (\ref{R-nl-bar-S1}) and (\ref{R-nl-bar-S2}),
for the non-local resolvent tensor in the Fourier domain (\ref{R-nl-bar})$%
_{1}$. After Fourier transform inversion, defined by (\ref{IFT}), previous
expression transforms into (\ref{R-nl-S3}), since $\boldsymbol{\hat{R}}%
_{n\ell }^{\left( S_{1}\right) }$ and $\boldsymbol{\hat{R}}_{n\ell }^{\left(
S_{2}\right) }$ are given by (\ref{R-nl-S1}) and (\ref{R-nl-S2}).

\subsubsection{Model $T_{2}$}

According to (\ref{u-konv})$_{2},$ the non-local displacement vector field
in the case of model $T_{2}$ takes the form%
\begin{align}
\boldsymbol{u}_{n\ell }^{\left( T_{2}\right) }\left( \boldsymbol{r}\right) &
=\boldsymbol{\hat{R}}_{n\ell }^{\left( T_{2}\right) }\left( \boldsymbol{r}%
\right) \ast _{\boldsymbol{r}}\boldsymbol{f}_{b}\left( \boldsymbol{r}\right)
\notag \\
& =\frac{2\lambda +3\mu }{\mu \left( \lambda +2\mu \right) }\left( 
\boldsymbol{L}\otimes \boldsymbol{L}\right) \boldsymbol{f}_{b}\left( 
\boldsymbol{r}\right)  \notag \\
& \quad -\frac{L^{2}}{4\pi }\frac{\lambda +\mu }{\mu \left( \lambda +2\mu
\right) }\left( \left( \frac{1}{r^{3}}\frac{\boldsymbol{L}\otimes 
\boldsymbol{L}}{L^{2}}\right) \ast _{\boldsymbol{r}}\boldsymbol{f}_{b}\left( 
\boldsymbol{r}\right) -3\left( \frac{1}{r^{3}}\left( \frac{\boldsymbol{r}}{r}%
\cdot \frac{\boldsymbol{L}}{L}\right) \left( \frac{\boldsymbol{r}}{r}\otimes 
\frac{\boldsymbol{L}}{L}\right) \right) \ast _{\boldsymbol{r}}\boldsymbol{f}%
_{b}\left( \boldsymbol{r}\right) \right)  \notag \\
& \quad -4L^{2}\frac{\lambda +\mu }{\mu \left( \lambda +2\mu \right) }\left(
\delta \left( \boldsymbol{r}\right) \frac{\boldsymbol{r}\otimes \boldsymbol{r%
}}{r^{2}}\frac{\boldsymbol{L}\otimes \boldsymbol{L}}{L^{2}}\right) \ast _{%
\boldsymbol{r}}\boldsymbol{f}_{b}\left( \boldsymbol{r}\right) ,  \notag \\
& =\frac{2\lambda +3\mu }{\mu \left( \lambda +2\mu \right) }\left( 
\boldsymbol{L}\otimes \boldsymbol{L}\right) \boldsymbol{f}_{b}\left( 
\boldsymbol{r}\right)  \notag \\
& \quad -\frac{L^{2}}{4\pi }\frac{\lambda +\mu }{\mu \left( \lambda +2\mu
\right) }\left( \left( \frac{1}{r^{3}}\frac{\boldsymbol{L}\otimes 
\boldsymbol{L}}{L^{2}}\right) \ast _{\boldsymbol{r}}\boldsymbol{f}_{b}\left( 
\boldsymbol{r}\right) -3\left( \frac{1}{r^{3}}\left( \frac{\boldsymbol{r}}{r}%
\cdot \frac{\boldsymbol{L}}{L}\right) \left( \frac{\boldsymbol{r}}{r}\otimes 
\frac{\boldsymbol{L}}{L}\right) \right) \ast _{\boldsymbol{r}}\boldsymbol{f}%
_{b}\left( \boldsymbol{r}\right) \right)  \notag \\
& \quad -\frac{8L^{2}}{3}\frac{\lambda +\mu }{\mu \left( \lambda +2\mu
\right) }\left( \frac{\boldsymbol{L}}{L}\cdot \boldsymbol{f}_{b}\left( 
\boldsymbol{r}\right) \right) \frac{\boldsymbol{L}}{L}-\frac{4L^{2}}{3}\frac{%
\lambda +\mu }{\mu \left( \lambda +2\mu \right) }\left( \boldsymbol{e}%
_{z}\cdot \frac{\boldsymbol{L}}{L}\right) \left( \frac{\boldsymbol{L}}{L}%
\cdot \boldsymbol{f}_{b}\left( \boldsymbol{r}\right) \right) \boldsymbol{e}%
_{z},  \label{nltdf-T2}
\end{align}%
where the non-local resolvent tensor%
\begin{align}
\boldsymbol{\hat{R}}_{n\ell }^{\left( T_{2}\right) }\left( \boldsymbol{r}%
\right) & =\frac{2\lambda +3\mu }{\mu \left( \lambda +2\mu \right) }\delta
\left( \boldsymbol{r}\right) \left( \boldsymbol{L}\otimes \boldsymbol{L}%
\right) -\frac{1}{4\pi }\frac{\lambda +\mu }{\mu \left( \lambda +2\mu
\right) }\frac{1}{r^{3}}\left( \boldsymbol{\hat{I}}-3\frac{\boldsymbol{r}%
\otimes \boldsymbol{r}}{r^{2}}\right) \left( \boldsymbol{L}\otimes 
\boldsymbol{L}\right) -4\frac{\lambda +\mu }{\mu \left( \lambda +2\mu
\right) }\frac{\boldsymbol{r}\otimes \boldsymbol{r}}{r^{2}}\left( 
\boldsymbol{L}\otimes \boldsymbol{L}\right)  \notag \\
& =\frac{2\lambda +3\mu }{\mu \left( \lambda +2\mu \right) }\delta \left( 
\boldsymbol{r}\right) \left( \boldsymbol{L}\otimes \boldsymbol{L}\right) -%
\frac{L^{2}}{4\pi }\frac{\lambda +\mu }{\mu \left( \lambda +2\mu \right) }%
\frac{1}{r^{3}}\left( \frac{\boldsymbol{L}\otimes \boldsymbol{L}}{L^{2}}%
-3\left( \frac{\boldsymbol{r}}{r}\cdot \frac{\boldsymbol{L}}{L}\right)
\left( \frac{\boldsymbol{r}}{r}\otimes \frac{\boldsymbol{L}}{L}\right)
\right)  \notag \\
& \quad -4L^{2}\frac{\lambda +\mu }{\mu \left( \lambda +2\mu \right) }\frac{%
\boldsymbol{r}\otimes \boldsymbol{r}}{r^{2}}\frac{\boldsymbol{L}\otimes 
\boldsymbol{L}}{L^{2}},  \label{R-nl-T2}
\end{align}%
is obtained by applying the Fourier transform inversion (\ref{IFT}) to the
expression for the non-local resolvent tensor in Fourier domain, given by (%
\ref{R-nl-bar}), so that, according to (\ref{u-superpoz}), for the
displacement vector field one has (\ref{u-non-loc-T2}). Note, the expression%
\begin{align*}
\left( \delta \left( \boldsymbol{r}\right) \frac{\boldsymbol{r}\otimes 
\boldsymbol{r}}{r^{2}}\frac{\boldsymbol{L}\otimes \boldsymbol{L}}{L^{2}}%
\right) \ast _{\boldsymbol{r}}\boldsymbol{f}_{b}\left( \boldsymbol{r}\right)
&=\left( \delta \left( \boldsymbol{r}\right) \frac{\boldsymbol{r}\otimes 
\boldsymbol{r}}{r^{2}}\right) \ast _{\boldsymbol{r}}\left( \frac{\boldsymbol{%
L}\otimes \boldsymbol{L}}{L^{2}}\boldsymbol{f}_{b}\left( \boldsymbol{r}%
\right) \right) \\
&=\left( \delta \left( \boldsymbol{r}\right) \frac{\boldsymbol{r}\otimes 
\boldsymbol{r}}{r^{2}}\right) \ast _{\boldsymbol{r}}\left( \left( \frac{%
\boldsymbol{L}}{L}\cdot \boldsymbol{f}_{b}\left( \boldsymbol{r}\right)
\right) \frac{\boldsymbol{L}}{L}\right) \\
&=\frac{2}{3}\left( \frac{\boldsymbol{L}}{L}\cdot \boldsymbol{f}_{b}\left( 
\boldsymbol{r}\right) \right) \frac{\boldsymbol{L}}{L}+\frac{1}{3}\left( 
\boldsymbol{e}_{z}\cdot \frac{\boldsymbol{L}}{L}\right) \left( \frac{%
\boldsymbol{L}}{L}\cdot \boldsymbol{f}_{b}\left( \boldsymbol{r}\right)
\right) \boldsymbol{e}_{z},
\end{align*}%
transformed according to (\ref{delta-projec-conv}), in which $\boldsymbol{f}%
_{b}\left( \boldsymbol{r}\right) $ is replaced with $\left( \frac{%
\boldsymbol{L}}{L}\cdot \boldsymbol{f}_{b}\left( \boldsymbol{r}\right)
\right) \frac{\boldsymbol{L}}{L},$ is used in obtaining the last term in (%
\ref{nltdf-T2}).

Namely, in the case of model $T_{2},$ the tensor function $\boldsymbol{\bar{%
\hat{\mathcal{T}}}}_{2}$, taking the form%
\begin{equation*}
\boldsymbol{\bar{\hat{\mathcal{T}}}}_{2}=k^{2}\left( \boldsymbol{L}\otimes 
\boldsymbol{L}\right) ,
\end{equation*}%
since%
\begin{equation*}
\mathrm{i}\boldsymbol{k\,\bar{\hat{T}}}_{2}(\boldsymbol{k})=k^{2}\mathrm{i}%
\boldsymbol{k\,}\left( \boldsymbol{L}\otimes \boldsymbol{L}\right) =k^{2}%
\boldsymbol{L}\left( \boldsymbol{L}\cdot \mathrm{i}\boldsymbol{k}\right)
=k^{2}\left( \boldsymbol{L}\otimes \boldsymbol{L}\right) \,\mathrm{i}%
\boldsymbol{k}=\boldsymbol{\bar{\hat{\mathcal{T}}}}_{2}\,\mathrm{i}%
\boldsymbol{k},
\end{equation*}%
see Table \ref{Modeli}, along with tensor $\boldsymbol{\bar{\hat{R}}}_{%
\boldsymbol{\ell }}$, given by (\ref{u-fd-loc})$_{2}$, yields%
\begin{equation}
\boldsymbol{\bar{\hat{R}}}_{n\ell }^{\left( T_{2}\right) }\left( \boldsymbol{%
k}\right) =\frac{1}{\mu }\left( \boldsymbol{L}\otimes \boldsymbol{L}\right) -%
\frac{\lambda +\mu }{\mu \left( \lambda +2\mu \right) }\frac{\boldsymbol{k}%
\otimes \boldsymbol{k}}{k^{2}}\left( \boldsymbol{L}\otimes \boldsymbol{L}%
\right) ,  \label{R-nl-bar-T2}
\end{equation}%
for the non-local resolvent tensor in the Fourier domain (\ref{R-nl-bar}).
After Fourier transform inversion, defined by (\ref{IFT}), the previous
expression transforms into (\ref{R-nl-T2}), since $\mathcal{F}^{-1}\left[ 1%
\right] \left( \boldsymbol{r}\right) =\delta \left( \boldsymbol{r}\right) $
and since the Fourier transform inversion of tensor $\frac{\boldsymbol{k}%
\otimes \boldsymbol{k}}{k^{2}}$ is given by (\ref{dijadski-kroz-kv-rezultat}%
), as calculated in Appendix \ref{dijadski-kroz-kvadrat}.

\subsubsection{Model $T_{4}$}

According to (\ref{u-konv})$_{2},$ the non-local displacement vector field
in the case of model $T_{4}$ takes the form%
\begin{align}
\boldsymbol{u}_{n\ell }^{\left( T_{4}\right) }\left( \boldsymbol{r}\right) &
=\boldsymbol{\hat{R}}_{n\ell }^{\left( T_{4}\right) }\left( \boldsymbol{r}%
\right) \ast _{\boldsymbol{r}}\boldsymbol{f}_{b}\left( \boldsymbol{r}\right)
\notag \\
& =-\frac{2\lambda +\mu }{3\mu \left( \lambda +2\mu \right) }\left( 
\boldsymbol{L}\otimes \boldsymbol{L}\right) \boldsymbol{f}_{b}\left( 
\boldsymbol{r}\right) +\frac{4}{3}\frac{1}{\mu }\left( \boldsymbol{L}\otimes 
\boldsymbol{L}\right) \left( \left( \boldsymbol{e}_{z}\cdot \boldsymbol{f}%
_{b}\left( \boldsymbol{r}\right) \right) \boldsymbol{e}_{z}\right)  \notag \\
& \quad +\frac{1}{4\pi }\frac{1}{\lambda +2\mu }\left( \frac{1}{r^{3}}\left( 
\boldsymbol{L}\otimes \boldsymbol{L}\right) \ast _{\boldsymbol{r}}%
\boldsymbol{f}_{b}\left( \boldsymbol{r}\right) \right) -\frac{L^{2}}{8\pi }%
\frac{\lambda +\mu }{\mu \left( \lambda +2\mu \right) }\left( \frac{1}{r^{3}}%
\left( 1-3\left( \frac{\boldsymbol{L}}{L}\cdot \frac{\boldsymbol{r}}{r}%
\right) ^{2}\right) \ast _{\boldsymbol{r}}\boldsymbol{f}_{b}\left( 
\boldsymbol{r}\right) \right)  \notag \\
& \quad -\frac{3L^{2}}{4\pi }\frac{1}{\lambda +2\mu }\left( \frac{1}{r^{3}}%
\left( \frac{\boldsymbol{L}}{L}\cdot \frac{\boldsymbol{r}}{r}\right) \left( 
\frac{\boldsymbol{L}}{L}\otimes \frac{\boldsymbol{r}}{r}\right) \ast _{%
\boldsymbol{r}}\boldsymbol{f}_{b}\left( \boldsymbol{r}\right) \right)  \notag
\\
& \quad +\frac{3L^{2}}{4\pi }\frac{\lambda +\mu }{\mu \left( \lambda +2\mu
\right) }\left( \frac{1}{r^{3}}\left( \frac{\boldsymbol{L}}{L}\cdot \frac{%
\boldsymbol{r}}{r}\right) \left( \frac{\boldsymbol{r}}{r}\otimes \frac{%
\boldsymbol{L}}{L}\right) \ast _{\boldsymbol{r}}\boldsymbol{f}_{b}\left( 
\boldsymbol{r}\right) \right)  \notag \\
& \quad +\frac{3L^{2}}{8\pi }\frac{\lambda +\mu }{\mu \left( \lambda +2\mu
\right) }\left( \frac{1}{r^{3}}\left( 1-5\left( \frac{\boldsymbol{L}}{L}%
\cdot \frac{\boldsymbol{r}}{r}\right) ^{2}\right) \frac{\boldsymbol{r}%
\otimes \boldsymbol{r}}{r^{2}}\right) \ast _{\boldsymbol{r}}\boldsymbol{f}%
_{b}\left( \boldsymbol{r}\right)  \label{nltdf-T4}
\end{align}%
where the non-local resolvent tensor%
\begin{align}
\boldsymbol{\hat{R}}_{n\ell }^{\left( T_{4}\right) }\left( \boldsymbol{r}%
\right) & =-\frac{2\lambda +3\mu }{\mu \left( \lambda +2\mu \right) }\delta
\left( \boldsymbol{r}\right) \left( \boldsymbol{L}\otimes \boldsymbol{L}%
\right) +4\frac{1}{\mu }\left( \boldsymbol{L}\otimes \boldsymbol{L}\right)
\delta \left( \boldsymbol{r}\right) \frac{\boldsymbol{r}\otimes \boldsymbol{r%
}}{r^{2}}  \notag \\
& \quad +\frac{1}{4\pi }\frac{1}{\lambda +2\mu }\frac{1}{r^{3}}\left( 
\boldsymbol{L}\otimes \boldsymbol{L}\right) -\frac{L^{2}}{8\pi }\frac{%
\lambda +\mu }{\mu \left( \lambda +2\mu \right) }\frac{1}{r^{3}}\left(
1-3\left( \frac{\boldsymbol{L}}{L}\cdot \frac{\boldsymbol{r}}{r}\right)
^{2}\right) \boldsymbol{\hat{I}}  \notag \\
& \quad -\frac{3L^{2}}{4\pi }\frac{1}{\lambda +2\mu }\frac{1}{r^{3}}\left( 
\frac{\boldsymbol{L}}{L}\cdot \frac{\boldsymbol{r}}{r}\right) \left( \frac{%
\boldsymbol{L}}{L}\otimes \frac{\boldsymbol{r}}{r}\right) +\frac{3L^{2}}{%
4\pi }\frac{\lambda +\mu }{\mu \left( \lambda +2\mu \right) }\frac{1}{r^{3}}%
\left( \frac{\boldsymbol{L}}{L}\cdot \frac{\boldsymbol{r}}{r}\right) \left( 
\frac{\boldsymbol{r}}{r}\otimes \frac{\boldsymbol{L}}{L}\right)  \notag \\
& \quad +\frac{3L^{2}}{8\pi }\frac{\lambda +\mu }{\mu \left( \lambda +2\mu
\right) }\frac{1}{r^{3}}\left( 1-5\left( \frac{\boldsymbol{L}}{L}\cdot \frac{%
\boldsymbol{r}}{r}\right) ^{2}\right) \frac{\boldsymbol{r}\otimes 
\boldsymbol{r}}{r^{2}}  \label{R-nl-T4}
\end{align}%
is obtained by applying the Fourier transform inversion (\ref{IFT}) to the
expression for the non-local resolvent tensor in Fourier domain, given by (%
\ref{R-nl-bar}), so that, according (\ref{u-superpoz}), for the displacement
vector field one has (\ref{u-non-loc-T4}), where in obtaining the second
term of (\ref{nltdf-T4}) the expression $\left( \delta \left( \boldsymbol{r}%
\right) \frac{\boldsymbol{r}\otimes \boldsymbol{r}}{r^{2}}\right) \ast _{%
\boldsymbol{r}}\boldsymbol{f}_{b}\left( \boldsymbol{r}\right) $ is evaluated
using (\ref{delta-projec-conv}).

Namely, in the case of model $T_{4},$ the tensor function $\boldsymbol{\bar{%
\hat{\mathcal{T}}}}_{4}$, taking the form%
\begin{equation*}
\boldsymbol{\bar{\hat{\mathcal{T}}}}_{4}=\left( \boldsymbol{L}\cdot 
\boldsymbol{k}\right) \left( \boldsymbol{L}\otimes \boldsymbol{k}\right) ,
\end{equation*}%
since%
\begin{equation*}
\mathrm{i}\boldsymbol{k\,\bar{\hat{T}}}_{4}(\boldsymbol{k})=\left( 
\boldsymbol{L}\cdot \boldsymbol{k}\right) \mathrm{i}\boldsymbol{k}\left( 
\boldsymbol{k}\otimes \boldsymbol{L}\right) =\left( \boldsymbol{L}\cdot 
\boldsymbol{k}\right) \boldsymbol{L}\left( \boldsymbol{k}\cdot \mathrm{i}%
\boldsymbol{k}\right) =\left( \boldsymbol{L}\cdot \boldsymbol{k}\right)
\left( \boldsymbol{L}\otimes \boldsymbol{k}\right) \,\mathrm{i}\boldsymbol{k}%
=\boldsymbol{\bar{\hat{\mathcal{T}}}}_{4}\,\mathrm{i}\boldsymbol{k},
\end{equation*}%
see Table \ref{Modeli}, along with tensor $\boldsymbol{\bar{\hat{R}}}_{%
\boldsymbol{\ell }}$, given by (\ref{u-fd-loc})$_{2}$, yields%
\begin{align}
\boldsymbol{\bar{\hat{R}}}_{n\ell }^{\left( T_{4}\right) }\left( \boldsymbol{%
k}\right) & =\frac{1}{\mu }\frac{1}{k^{2}}\left( \boldsymbol{L}\cdot 
\boldsymbol{k}\right) \left( \boldsymbol{L}\otimes \boldsymbol{k}\right) -%
\frac{\lambda +\mu }{\mu \left( \lambda +2\mu \right) }\left( \boldsymbol{L}%
\cdot \boldsymbol{k}\right) \frac{\boldsymbol{k}\otimes \boldsymbol{k}}{k^{4}%
}\left( \boldsymbol{L}\otimes \boldsymbol{k}\right)  \notag \\
& =\frac{1}{\mu }\left( \boldsymbol{L}\otimes \boldsymbol{L}\right) \frac{%
\boldsymbol{k}\otimes \boldsymbol{k}}{k^{2}}-\frac{\lambda +\mu }{\mu \left(
\lambda +2\mu \right) }\left( \boldsymbol{L}\cdot \boldsymbol{k}\right) ^{2}%
\frac{\boldsymbol{k}\otimes \boldsymbol{k}}{k^{4}},  \label{R-nl-bar-T4}
\end{align}%
for the non-local resolvent tensor in the Fourier domain (\ref{R-nl-bar}).
After Fourier transform inversion, defined by (\ref{IFT}), the previous
expression transforms into (\ref{R-nl-T4}), since the Fourier transform
inversion of tensor $\frac{\boldsymbol{k}\otimes \boldsymbol{k}}{k^{2}}$ is
given by (\ref{dijadski-kroz-kv-rezultat}), as well as the Fourier transform
inversion of tensor $\left( \boldsymbol{L}\cdot \boldsymbol{k}\right) ^{2}%
\frac{\boldsymbol{k}\otimes \boldsymbol{k}}{k^{4}}$ is given by (\ref{dodat}%
), as calculated in Appendices \ref{dijadski-kroz-kvadrat} and \ref%
{dijadski-skalarni-kroz-cetvrti}.

\subsubsection{Model $T_{5}$}

In the case of model $T_{5},$ the tensor function $\boldsymbol{\bar{\hat{%
\mathcal{T}}}}_{5}$ is a zero tensor, i.e., $\boldsymbol{\bar{\hat{\mathcal{T%
}}}}_{5}=\boldsymbol{\hat{0}}$, see Table \ref{Modeli}, since%
\begin{equation*}
\mathrm{i}\boldsymbol{k\,\bar{\hat{T}}}_{5}(\boldsymbol{k})=\mathrm{i}%
\boldsymbol{k}\left( \left( \boldsymbol{L}\times \boldsymbol{k}\right)
\otimes \left( \boldsymbol{L}\times \boldsymbol{k}\right) \right) =\left( 
\mathrm{i}\boldsymbol{k}\cdot \left( \boldsymbol{L}\times \boldsymbol{k}%
\right) \right) \left( \boldsymbol{L}\times \boldsymbol{k}\right) =%
\boldsymbol{0},
\end{equation*}%
yielding that the non-local resolvent tensor in the Fourier domain (\ref%
{R-nl-bar})$_2$ is a zero tensor as well, i.e., $\boldsymbol{\bar{\hat{R}}}%
_{n\ell }^{\left( T_{5}\right) }\left( \boldsymbol{k}\right) =\boldsymbol{%
\hat{0}},$ implying that the non-local resolvent tensor is $\boldsymbol{\hat{%
R}}_{n\ell }^{\left( T_{5}\right) }\left( \boldsymbol{r}\right) =\boldsymbol{%
\hat{0}},$ so as the non-local displacement vector field%
\begin{equation}
\boldsymbol{u}_{n\ell }^{\left( T_{5}\right) }\left( \boldsymbol{r}\right) =%
\boldsymbol{0}.  \label{nltdf-T5}
\end{equation}

\section{Displacement vector field for a conservative force}

In the case when body force $\boldsymbol{f}_{b}$ corresponds to a
conservative force, taking the well known form%
\begin{equation}
\boldsymbol{f}_{b}\left( \boldsymbol{r}\right) =-\func{grad}\phi \left( 
\boldsymbol{r}\right) ,  \label{f-konz}
\end{equation}%
where $\phi $ is the scalar field corresponding to potential energy, the
displacement field in Fourier domain, according to (\ref{u-l+u-ul}) and (\ref%
{u-l-i-u-ul}),\ can be expressed as 
\begin{align}
& \boldsymbol{\bar{u}}^{\left( S_{i}/T_{i}\right) }\left( \boldsymbol{k}%
\right) =\boldsymbol{\bar{u}}_{\ell }\left( \boldsymbol{k}\right) +%
\boldsymbol{\bar{u}}_{n\ell }^{\left( S_{i}/T_{i}\right) }\left( \boldsymbol{%
k}\right) \quad \text{with}  \label{u-bar-conserv} \\
& \boldsymbol{\bar{u}}_{\ell }\left( \boldsymbol{k}\right) =\boldsymbol{\bar{%
R}}_{\boldsymbol{\ell }}\left( \boldsymbol{k}\right) \bar{\phi}\left( 
\boldsymbol{k}\right) \text{\quad and\quad }\boldsymbol{\bar{u}}_{n\ell
}^{\left( S_{i}/T_{i}\right) }\left( \boldsymbol{k}\right) =\boldsymbol{\bar{%
R}}_{n\boldsymbol{\ell }}^{\left( S_{i}/T_{i}\right) }\left( \boldsymbol{k}%
\right) \bar{\phi}\left( \boldsymbol{k}\right) \quad \text{where}
\label{u-bar-conserv-2} \\
&\boldsymbol{\bar{R}}_{\boldsymbol{\ell }}\left( \boldsymbol{k}\right) =-%
\boldsymbol{\bar{\hat{R}}}_{\boldsymbol{\ell }}\left( \boldsymbol{k}\right)
\,\mathrm{i}\boldsymbol{k}\text{\quad and\quad }\boldsymbol{\bar{R}}_{n%
\boldsymbol{\ell }}^{\left( S_{i}/T_{i}\right) }\left( \boldsymbol{k}\right)
=-\boldsymbol{\bar{\hat{R}}}_{n\ell }^{\left( S_{i}/T_{i}\right) }\left( 
\boldsymbol{k}\right) \,\mathrm{i}\boldsymbol{k},  \label{R-ovi-bar}
\end{align}%
with the local resolvent vector $\boldsymbol{\bar{R}}_{\boldsymbol{\ell }}$
in Fourier domain and the non-local resolvent vector $\boldsymbol{\bar{R}}_{n%
\boldsymbol{\ell }}^{\left( S_{i}/T_{i}\right) }$ in Fourier domain, since
conservative body force in Fourier domain takes the form $\boldsymbol{\bar{f}%
}_{b}\left( \boldsymbol{k}\right) =\mathcal{F}\left[ \boldsymbol{f}%
_{b}\left( \boldsymbol{r}\right) \right] \left( \boldsymbol{k}\right) =-%
\mathrm{i}\boldsymbol{k\,}\bar{\phi}\left( k\right) ,$ where the local
resolvent tensor in Fourier domain $\boldsymbol{\bar{\hat{R}}}_{\boldsymbol{%
\ell }}$ is given by (\ref{u-fd-loc})$_{2}$, as well as the non-local
resolvent tensor in Fourier domain $\boldsymbol{\bar{\hat{R}}}_{n\boldsymbol{%
\ell }}$ is given by (\ref{R-nl-bar})$,$ and where non-locallity model is
accounted through scalar and tensor functions $\bar{S}_{i}$ and $\boldsymbol{%
\bar{\hat{\mathcal{T}}}}_{i}$, listed in Table \ref{Modeli}. After the
inverse Fourier transform, the expression (\ref{u-bar-conserv}) yields%
\begin{equation}
\boldsymbol{u}\left( \boldsymbol{r}\right) =\boldsymbol{u}_{\ell }\left( 
\boldsymbol{r}\right) +\boldsymbol{u}_{n\ell }^{\left( S_{i}/T_{i}\right)
}\left( \boldsymbol{r}\right)  \label{u-loc-i-u-non-loc}
\end{equation}%
for the displacement field in the presence of a conservative force as a
particular case of (\ref{u-superpoz}) with (\ref{u-konv}). It turns out that
the non-local resolvent vector $\boldsymbol{R}_{n\ell }^{\left(
S_{i}/T_{i}\right) }$ is an operator rather than a function and therefore
non-local displacement vector field $\boldsymbol{u}_{n\ell }^{\left(
S_{i}/T_{i}\right)}$ is expressed in terms of body force rather than in
terms of potential energy.

\subsection{Local displacement field}

The local resolvent vector $\boldsymbol{\bar{R}}_{\boldsymbol{\ell }}$ in
Fourier domain, see (\ref{R-ovi-bar})$_{1}$, corresponds to local
three-dimensional elastic body described by Hooke's low (\ref{Hook-loc}),
but it is also common for any three-dimensional scalar- and tensor-type
model, listed in Table \ref{Modeli-new}, and takes the form%
\begin{equation}
\boldsymbol{\bar{R}}_{\boldsymbol{\ell }}\left( \boldsymbol{k}\right) =-%
\boldsymbol{\bar{\hat{R}}}_{\boldsymbol{\ell }}\left( \boldsymbol{k}\right)
\,\mathrm{i}\boldsymbol{k}=-\frac{1}{\lambda +2\mu }\frac{\mathrm{i}%
\boldsymbol{k}}{k^{2}},  \label{R-nl-bar-vector}
\end{equation}%
since the local resolvent tensor $\boldsymbol{\bar{\hat{R}}}_{\boldsymbol{%
\ell }} $ is given by (\ref{u-fd-loc})$_2$, yielding%
\begin{equation*}
\boldsymbol{R}_{\boldsymbol{\ell }}\left( \boldsymbol{r}\right) =\frac{1}{%
4\pi \left( \lambda +2\mu \right) }\frac{\boldsymbol{r}}{r^{3}},
\end{equation*}%
according to the Fourier transform inversion given by (\ref{ik-kroz-kv})$_1$
in Appendix \ref{ik-kroz-k-na-kv}, so that the local term of displacement
field in (\ref{u-loc-i-u-non-loc}) takes the form%
\begin{equation*}
\boldsymbol{u}_{\ell }\left( \boldsymbol{r}\right) =\boldsymbol{R}_{%
\boldsymbol{\ell }}\left( \boldsymbol{r}\right) \ast _{\boldsymbol{r}}\phi
\left( \boldsymbol{r}\right) =\frac{1}{4\pi \left( \lambda +2\mu \right) }%
\frac{\boldsymbol{r}}{r^{3}}\ast _{\boldsymbol{r}}\phi \left( \boldsymbol{r}%
\right),
\end{equation*}%
by the inverse Fourier transform of (\ref{u-bar-conserv-2})$_1$. On the
other hand, the local displacement field in Fourier domain, according to (%
\ref{u-bar-conserv-2})$_{1}$ and (\ref{R-nl-bar-vector}), is%
\begin{equation}
\boldsymbol{\bar{u}}_{\ell }\left( \boldsymbol{k}\right) =-\frac{1}{\lambda
+2\mu }\frac{\mathrm{i}\boldsymbol{k}}{k^{2}}\bar{\phi}\left( \boldsymbol{k}%
\right) =\frac{1}{\lambda +2\mu }\frac{1}{k^{2}}\boldsymbol{\bar{f}}%
_{b}\left( \boldsymbol{k}\right) ,  \label{u-l-bar}
\end{equation}%
yielding%
\begin{equation*}
\boldsymbol{u}_{\ell }\left( \boldsymbol{r}\right) =\frac{1}{4\pi \left(
\lambda +2\mu \right) }\frac{1}{r}\ast _{\boldsymbol{r}}\boldsymbol{f}%
_{b}\left( \boldsymbol{r}\right) ,
\end{equation*}%
after Fourier transform inversion.

In the following, the non-local displacement vector fields corresponding to
scalar- and tensor-type stress-gradient Eringen three-dimensional models (%
\ref{Hook-non-loc}) and (\ref{Hook-non-loc-t}) are calculated in order to
obtain the displacement vector fields (\ref{u-loc-i-u-non-loc})
corresponding to these models, by superposition with the local displacement
vector field.

\subsection{Non-local displacement field}

In the case of non-local-type three-dimensional isotropic and elastic body,
described either by constitutive equation (\ref{Hook-non-loc}), containing
the scalar-type operator $\mathrm{S}_{i}$, or by constitutive equation (\ref%
{Hook-non-loc-t}), containing the tensor-type operator $\mathrm{\hat{T}}_{i}$%
, the corresponding displacement vector fields in the presence of a
conservative body force are given as a superposition of local and non-local
displacement vector fields by%
\begin{align}
\boldsymbol{u}^{\left( S_{1}/T_{1}\right) }\left( \boldsymbol{r}\right) & =%
\boldsymbol{u}_{\ell }\left( \boldsymbol{r}\right) +\boldsymbol{u}_{n\ell
}^{\left( S_{1}/T_{1}\right) }\left( \boldsymbol{r}\right)  \notag \\
& =\frac{L^{2}}{\lambda +2\mu }\boldsymbol{f}_{b}\left( \boldsymbol{r}%
\right) +\frac{1}{4\pi \left( \lambda +2\mu \right) }\int_{%
\mathbb{R}
^{3}}\frac{1}{\left\vert \boldsymbol{r}-\boldsymbol{r}^{\prime }\right\vert }%
\boldsymbol{f}_{b}\left( \boldsymbol{r}^{\prime }\right) \mathrm{d}_{%
\boldsymbol{r}^{\prime }}V  \label{u-non-loc-S1-konz}
\end{align}%
in the case of models $S_{1}$ and $T_{1}$, further by%
\begin{align}
\boldsymbol{u}^{\left( S_{2}/T_{3}\right) }\left( \boldsymbol{r}\right) & =%
\boldsymbol{u}_{\ell }\left( \boldsymbol{r}\right) +\boldsymbol{u}_{n\ell
}^{\left( S_{2}/T_{3}\right) }\left( \boldsymbol{r}\right)  \notag \\
& =\frac{L^{2}}{\lambda +2\mu }\boldsymbol{f}_{b}\left( \boldsymbol{r}%
\right) +\frac{1}{4\pi \left( \lambda +2\mu \right) }\int_{%
\mathbb{R}
^{3}}\frac{1}{\left\vert \boldsymbol{r}-\boldsymbol{r}^{\prime }\right\vert }%
\boldsymbol{f}_{b}\left( \boldsymbol{r}^{\prime }\right) \mathrm{d}_{%
\boldsymbol{r}^{\prime }}V  \notag \\
& \quad +\frac{L^{2}}{4\pi }\frac{1}{\lambda +2\mu }\int_{%
\mathbb{R}
^{3}}\frac{1}{\left\vert \boldsymbol{r}-\boldsymbol{r}^{\prime }\right\vert
^{3}}\left( 1-3\left( \frac{\boldsymbol{L}}{L}\cdot \frac{\boldsymbol{r}-%
\boldsymbol{r}^{\prime }}{\left\vert \boldsymbol{r}-\boldsymbol{r}^{\prime
}\right\vert }\right) ^{2}\right) \boldsymbol{f}_{b}\left( \boldsymbol{r}%
^{\prime }\right) \mathrm{d}_{\boldsymbol{r}^{\prime }}V
\label{u-non-loc-S2-konz}
\end{align}%
in the case of models $S_{2}$ and $T_{3}$, as well as by%
\begin{align}
\boldsymbol{u}^{\left( S_{3}\right) }\left( \boldsymbol{r}\right) & =%
\boldsymbol{u}_{\ell }\left( \boldsymbol{r}\right) +\boldsymbol{u}_{n\ell
}^{\left( S_{3}\right) }\left( \boldsymbol{r}\right)  \notag \\
& =\frac{1}{4\pi \left( \lambda +2\mu \right) }\int_{%
\mathbb{R}
^{3}}\frac{1}{\left\vert \boldsymbol{r}-\boldsymbol{r}^{\prime }\right\vert }%
\boldsymbol{f}_{b}\left( \boldsymbol{r}^{\prime }\right) \mathrm{d}_{%
\boldsymbol{r}^{\prime }}V  \notag \\
& \quad - \frac{L^{2}}{4\pi }\frac{1}{\lambda +2\mu }\int_{%
\mathbb{R}
^{3}}\frac{1}{\left\vert \boldsymbol{r}-\boldsymbol{r}^{\prime }\right\vert
^{3}} \left( 1-3\left( \frac{\boldsymbol{L}}{L}\cdot \frac{\boldsymbol{r} - 
\boldsymbol{r}^{\prime }}{\left\vert \boldsymbol{r} -\boldsymbol{r}^{\prime
}\right\vert }\right) ^{2}\right) \boldsymbol{f}_{b}\left( \boldsymbol{r}%
^{\prime }\right) \mathrm{d}_{\boldsymbol{r}^{\prime }}V
\label{u-non-loc-S3-konz}
\end{align}%
in the case of model $S_{3}$, and next by%
\begin{align}
\boldsymbol{u}^{\left( T_{2}/T_{4}\right) }\left( \boldsymbol{r}\right) & =%
\boldsymbol{u}_{\ell }\left( \boldsymbol{r}\right) +\boldsymbol{u}_{n\ell
}^{\left( T_{2}/T_{4}\right) }\left( \boldsymbol{r}\right)  \notag \\
& =\frac{L^2}{\mu }\left( \frac{\boldsymbol{L}}{L}\cdot \boldsymbol{f}%
_{b}\left( \boldsymbol{r}\right) \right) \frac{\boldsymbol{L}}{L} +\frac{1}{%
4\pi \left( \lambda +2\mu \right) }\int_{%
\mathbb{R}
^{3}}\frac{1}{\left\vert \boldsymbol{r}-\boldsymbol{r}^{\prime }\right\vert }%
\boldsymbol{f}_{b}\left( \boldsymbol{r}^{\prime }\right) \mathrm{d}_{%
\boldsymbol{r}^{\prime }}V  \notag \\
& \quad -\frac{L^{2}}{4\pi }\frac{\lambda +\mu }{\mu \left( \lambda +2\mu
\right) }\int_{%
\mathbb{R}
^{3}}\frac{1}{\left\vert \boldsymbol{r}-\boldsymbol{r}^{\prime }\right\vert
^{3}}\left( 1-3\left( \frac{\boldsymbol{L}}{L}\cdot \frac{\boldsymbol{r}-%
\boldsymbol{r}^{\prime }}{\left\vert \boldsymbol{r}-\boldsymbol{r}^{\prime
}\right\vert }\right) ^{2}\right) \boldsymbol{f}_{b}\left( \boldsymbol{r}%
^{\prime }\right) \mathrm{d}_{\boldsymbol{r}^{\prime }}V
\label{u-non-loc-T2-konz}
\end{align}%
in the case of models $T_{2}$ and $T_{4},$ and finally by%
\begin{align}
\boldsymbol{u}^{\left( T_{5}\right) }\left( \boldsymbol{r}\right) & =%
\boldsymbol{u}_{\ell }\left( \boldsymbol{r}\right) +\boldsymbol{u}_{n\ell
}^{\left( T_{5}\right) }\left( \boldsymbol{r}\right) =\boldsymbol{u}_{\ell
}\left( \boldsymbol{r}\right)  \notag \\
& =\frac{1}{4\pi \left( \lambda +2\mu \right) }\int_{%
\mathbb{R}
^{3}}\frac{1}{\left\vert \boldsymbol{r}-\boldsymbol{r}^{\prime }\right\vert }%
\boldsymbol{f}_{b}\left( \boldsymbol{r}^{\prime }\right) \mathrm{d}_{%
\boldsymbol{r}^{\prime }}V  \label{u-non-loc-T5-konz}
\end{align}%
in the case of model $T_{5},$ where $\boldsymbol{u}_{\ell }$ is given by (%
\ref{u-loc}).

In the sequel, the non-local displacement fields, corresponding to scalar-
and tensor-type stress-gradient Eringen three-dimensional models (\ref%
{Hook-non-loc}) and (\ref{Hook-non-loc-t}), are calculated.

\subsubsection{Models $S_{1}$ and $T_{1}$}

In the case of models $S_{1}$ and $T_{1},$ the non-local resolvent vector in
Fourier domain becomes%
\begin{equation}
\boldsymbol{\bar{R}}_{n\boldsymbol{\ell }}^{\left( S_{1}/T_{1}\right)
}\left( \boldsymbol{k}\right) =-\frac{L^{2}}{\lambda +2\mu }\mathrm{i}%
\boldsymbol{k}  \label{R-nl-bar-S1-T1}
\end{equation}%
according to (\ref{R-ovi-bar})$_{2}$ and (\ref{R-nl-bar-S1}), since%
\begin{equation*}
\boldsymbol{\bar{R}}_{n\boldsymbol{\ell }}^{\left( S_{1}/T_{1}\right)
}\left( \boldsymbol{k}\right) =\boldsymbol{\bar{\hat{R}}}_{n\ell }^{\left(
S_{1}/T_{1}\right) }\left( \boldsymbol{k}\right) \,\mathrm{i}\boldsymbol{k}%
=L^{2}\frac{1}{\mu }\,\mathrm{i}\boldsymbol{k}-L^{2}\frac{\lambda +\mu }{\mu
\left( \lambda +2\mu \right) }\frac{\boldsymbol{k}\otimes \boldsymbol{k}}{%
k^{2}}\,\mathrm{i}\boldsymbol{k},
\end{equation*}%
where $\left( \boldsymbol{k}\otimes \boldsymbol{k}\right) \mathrm{i}%
\boldsymbol{k}=k^{2}\,\mathrm{i}\boldsymbol{k}$ is used. The Fourier
transform inversion of the expression (\ref{R-nl-bar-S1-T1}) is an operator%
\begin{equation*}
\boldsymbol{R}_{n\ell }^{\left( S_{1}/T_{1}\right) }\left( \boldsymbol{r}%
\right) =-\frac{L^{2}}{\lambda +2\mu }\nabla _{\boldsymbol{r}},
\end{equation*}%
and therefore the non-local displacement vector field, by the use of inverse
Fourier transform of (\ref{u-bar-conserv-2})$_{2},$ is obtained as the
action of resolvent vector, being operator, to the scalar field $\phi $,
rather than a convolution, so that one has%
\begin{equation}
\boldsymbol{u}_{n\ell }^{\left( S_{1}/T_{1}\right) }\left( \boldsymbol{r}%
\right) =\boldsymbol{R}_{n\boldsymbol{\ell }}^{\left( S_{1}/T_{1}\right)
}\left( \boldsymbol{r}\right) \phi \left( \boldsymbol{r}\right) =-\frac{L^{2}%
}{\lambda +2\mu }\nabla \phi \left( \boldsymbol{r}\right) =\frac{L^{2}}{%
\lambda +2\mu }\boldsymbol{f}_{b}\left( \boldsymbol{r}\right) .
\label{u-nl-S1-T1-fi}
\end{equation}%
The same result is obtained by inverting the non-local displacement vector
field in Fourier domain (\ref{u-bar-conserv-2})$_{2}$ directly, thus taking
the form 
\begin{equation}
\boldsymbol{\bar{u}}_{n\ell }^{\left( S_{1}/T_{1}\right) }\left( \boldsymbol{%
k}\right) =-\frac{L^{2}}{\lambda +2\mu }\mathrm{i}\boldsymbol{k\,}\bar{\phi}%
\left( \boldsymbol{k}\right) =\frac{L^{2}}{\lambda +2\mu }\boldsymbol{\bar{f}%
}_{b}\left( \boldsymbol{k}\right) ,  \label{u-nl-bar-S1}
\end{equation}%
so that, after inverse Fourier transform, one obtains%
\begin{equation*}
\boldsymbol{u}_{n\ell }^{\left( S_{1}/T_{1}\right) }\left( \boldsymbol{r}%
\right) =\frac{L^{2}}{\lambda +2\mu }\boldsymbol{f}_{b}\left( \boldsymbol{r}%
\right) .
\end{equation*}

\subsubsection{Model $S_{2}$ and $T_{3}$}

In the case of models $S_{2}$ and $T_{3},$ the non-local resolvent vector in
Fourier domain becomes%
\begin{equation}
\boldsymbol{\bar{R}}_{n\ell }^{\left( S_{2}/T_{3}\right) }\left( \boldsymbol{%
k}\right) =-\frac{1}{\lambda +2\mu }\frac{\left( \boldsymbol{L}\cdot 
\boldsymbol{k}\right) ^{2}}{k^{2}}\mathrm{i}\boldsymbol{k},
\label{R-nl-bar-S2-T3}
\end{equation}%
according to (\ref{R-ovi-bar})$_{2}$ and (\ref{R-nl-bar-S2}), since%
\begin{equation*}
\boldsymbol{\bar{R}}_{n\ell }^{\left( S_{2}/T_{3}\right) }\left( \boldsymbol{%
k}\right) =\boldsymbol{\bar{\hat{R}}}_{n\ell }^{\left( S_{2}/T_{3}\right)
}\left( \boldsymbol{k}\right) \,\mathrm{i}\boldsymbol{k}=\frac{1}{\mu }\frac{%
\left( \boldsymbol{L}\cdot \boldsymbol{k}\right) ^{2}}{k^{2}}\,\mathrm{i}%
\boldsymbol{k}-\frac{\lambda +\mu }{\mu \left( \lambda +2\mu \right) }\left( 
\boldsymbol{L}\cdot \boldsymbol{k}\right) ^{2}\frac{\boldsymbol{k}\otimes 
\boldsymbol{k}}{k^{4}}\,\mathrm{i}\boldsymbol{k},
\end{equation*}%
where $\left( \boldsymbol{k}\otimes \boldsymbol{k}\right) \mathrm{i}%
\boldsymbol{k}=k^{2}\,\mathrm{i}\boldsymbol{k}$ is used. The non-local
displacement vector field in Fourier domain (\ref{u-bar-conserv-2})$_{2}$,
using the non-local resolvent vector in Fourier domain (\ref{R-nl-bar-S2-T3}%
),\ takes the form%
\begin{equation}
\boldsymbol{\bar{u}}_{n\ell }^{\left( S_{2}/T_{3}\right) }\left( \boldsymbol{%
k}\right) =-\frac{1}{\lambda +2\mu }\frac{\left( \boldsymbol{L}\cdot 
\boldsymbol{k}\right) ^{2}}{k^{2}}\mathrm{i}\boldsymbol{k\,}\bar{\phi}\left( 
\boldsymbol{k}\right) =\frac{1}{\lambda +2\mu }\frac{\left( \boldsymbol{L}%
\cdot \boldsymbol{k}\right) ^{2}}{k^{2}}\boldsymbol{\bar{f}}_{b}\left( 
\boldsymbol{k}\right) ,  \label{u-nl-bar-S2}
\end{equation}%
and becomes%
\begin{equation*}
\boldsymbol{u}_{n\ell }^{\left( S_{2}/T_{3}\right) }\left( \boldsymbol{r}%
\right) =\frac{L^{2}}{\lambda +2\mu }\boldsymbol{f}_{b}\left( \boldsymbol{r}%
\right) +\frac{L^{2}}{4\pi }\frac{1}{\lambda +2\mu }\left( \frac{1}{r^{3}}%
\left( 1-3\left( \frac{\boldsymbol{L}}{L}\cdot \frac{\boldsymbol{r}}{r}%
\right) ^{2}\right) \right) \ast _{\boldsymbol{r}}\boldsymbol{f}_{b}\left( 
\boldsymbol{r}\right) ,
\end{equation*}%
after Fourier transform inversion, where inverse Fourier transform of $\frac{%
\left( \boldsymbol{L}\cdot \boldsymbol{k}\right) ^{2}}{k^{2}},$ calculated
in Appendix \ref{skalarni-kroz-kvadrat}, see (\ref{skalarni-kroz-kv}), is
used.

Equivalently, one may express the non-local displacement vector field in
terms of scalar function $\phi $ by inverting the expression (\ref%
{u-bar-conserv-2})$_{2}$, yielding%
\begin{align}
\boldsymbol{u}_{n\ell }^{\left( S_{2}/T_{3}\right) }\left( \boldsymbol{r}%
\right)  &=\boldsymbol{R}_{n\boldsymbol{\ell }}^{\left( S_{2}/T_{3}\right)
}\left( \boldsymbol{r}\right) \ast \phi \left( \boldsymbol{r}\right)   \notag
\\
&=\frac{1}{4\pi }\frac{1}{\lambda +2\mu }\left( \left( \left( \frac{%
\boldsymbol{L}}{L}\cdot \frac{\boldsymbol{r}}{r}\right) \frac{L^{2}}{r^{2}}%
\right) \ast \phi \left( \boldsymbol{r}\right) \right) \frac{\boldsymbol{L}}{%
L}+\frac{1}{8\pi }\frac{1}{\lambda +2\mu }\left( \left( 1-3\left( \frac{%
\boldsymbol{L}}{L}\cdot \frac{\boldsymbol{r}}{r}\right) ^{2}\right) \frac{%
L^{2}}{r^{2}}\frac{\boldsymbol{r}}{r}\right) \ast \phi \left( \boldsymbol{r}%
\right) ,  \label{u-nl-S2-T3-fi}
\end{align}%
where%
\begin{equation*}
\boldsymbol{R}_{n\boldsymbol{\ell }}^{\left( S_{2}/T_{3}\right) }\left( 
\boldsymbol{r}\right) =\mathcal{F}^{-1}\left[ \boldsymbol{\bar{R}}_{n\ell
}^{\left( S_{2}/T_{3}\right) }\left( \boldsymbol{k}\right) \right] \left( 
\boldsymbol{r}\right) =-\frac{1}{\lambda +2\mu }\mathcal{F}^{-1}\left[ \frac{%
\left( \boldsymbol{L}\cdot \boldsymbol{k}\right) ^{2}}{k^{2}}\mathrm{i}%
\boldsymbol{k}\right] \left( \boldsymbol{r}\right) ,
\end{equation*}%
according to calculations in Appendix \ref{skalarni-kroz-kvadrat-ik}, see (%
\ref{izraz}). Note, the non-local displacement field, given by (\ref%
{u-nl-S2-T3-fi}), is the Fourier transform inversion of expression (\ref%
{u-nl-bar-S2})$_{1}$.

\subsubsection{Model $S_{3}$}

In the case of model $S_{3}$, the non-local resolvent vector in Fourier
domain becomes%
\begin{equation*}
\boldsymbol{\bar{R}}_{n\ell }^{\left( S_{3}\right) }\left( \boldsymbol{k}%
\right) =-\boldsymbol{\bar{\hat{R}}}_{n\ell }^{\left( S_{3}\right) }\left( 
\boldsymbol{k}\right) \,\mathrm{i}\boldsymbol{k}=-\left( \boldsymbol{\bar{%
\hat{R}}}_{n\boldsymbol{\ell }}^{\left( S_{1}\right) }\left( \boldsymbol{k}%
\right) -\boldsymbol{\bar{\hat{R}}}_{n\boldsymbol{\ell }}^{\left(
S_{2}\right) }\left( \boldsymbol{k}\right) \right) \,\mathrm{i}\boldsymbol{k}%
=\boldsymbol{\bar{R}}_{n\ell }^{\left( S_{1}\right) }\left( \boldsymbol{k}%
\right) -\boldsymbol{\bar{R}}_{n\ell }^{\left( S_{2}\right) }\left( 
\boldsymbol{k}\right) ,
\end{equation*}%
according (\ref{R-ovi-bar})$_{2}$ and (\ref{R-nl-bar-1-minus-2}). The
non-local displacement vector field in Fourier domain (\ref{u-bar-conserv-2})%
$_{2}$ takes the form%
\begin{equation}
\boldsymbol{\bar{u}}_{n\ell }^{\left( S_{3}\right) }\left( \boldsymbol{k}%
\right) =\boldsymbol{\bar{u}}_{n\ell }^{\left( S_{1}\right) }\left( 
\boldsymbol{k}\right) -\boldsymbol{\bar{u}}_{n\ell }^{\left( S_{2}\right)
}\left( \boldsymbol{k}\right) ,  \label{u-nl-bar-S3}
\end{equation}%
with $\boldsymbol{\bar{u}}_{n\ell }^{\left( S_{1}\right) }$ given by (\ref%
{u-nl-bar-S1}) and $\boldsymbol{\bar{u}}_{n\ell }^{\left( S_{2}\right) }$
given by (\ref{u-nl-bar-S2}), so that, after inverse Fourier transform, one
has%
\begin{align*}
\boldsymbol{u}_{n\ell }^{\left( S_{3}\right) }\left( \boldsymbol{r}\right) &
=\boldsymbol{u}_{n\ell }^{\left( S_{1}\right) }\left( \boldsymbol{r}\right) -%
\boldsymbol{u}_{n\ell }^{\left( S_{2}\right) }\left( \boldsymbol{r}\right) 
\\
& =-\frac{L^{2}}{4\pi }\frac{1}{\lambda +2\mu }\left( \frac{1}{r^{3}}\left(
1-3\left( \frac{\boldsymbol{L}}{L}\cdot \frac{\boldsymbol{r}}{r}\right)
^{2}\right) \right) \ast _{\boldsymbol{r}}\boldsymbol{f}_{b}\left( 
\boldsymbol{r}\right) .
\end{align*}

Alternatively, one can obtained the non-local displacement vector field
expressed trough the scalar function $\phi $ as%
\begin{align*}
\boldsymbol{u}_{n\ell }^{\left( S_{3}\right) }\left( \boldsymbol{r}\right) 
&=\boldsymbol{u}_{n\ell }^{\left( S_{1}\right) }\left( \boldsymbol{r}%
\right) -\boldsymbol{u}_{n\ell }^{\left( S_{2}\right) }\left( \boldsymbol{r}%
\right)  \\
&=-\frac{L^{2}}{\lambda +2\mu }\nabla \phi \left( \boldsymbol{r}\right) -%
\frac{1}{4\pi }\frac{1}{\lambda +2\mu }\left( \left( \left( \frac{%
\boldsymbol{L}}{L}\cdot \frac{\boldsymbol{r}}{r}\right) \frac{L^{2}}{r^{2}}%
\right) \ast \phi \left( \boldsymbol{r}\right) \right) \frac{\boldsymbol{L}}{%
L} \\
&\quad -\frac{1}{8\pi }\frac{1}{\lambda +2\mu }\left( \left( 1-3\left( \frac{%
\boldsymbol{L}}{L}\cdot \frac{\boldsymbol{r}}{r}\right) ^{2}\right) \frac{%
L^{2}}{r^{2}}\frac{\boldsymbol{r}}{r}\right) \ast \phi \left( \boldsymbol{r}%
\right) ,
\end{align*}
according to (\ref{u-nl-S1-T1-fi}) and (\ref{u-nl-S2-T3-fi}).

\subsubsection{Models $T_{2}$ and $T_{4}$}

In the case of models $T_{2}$ and $T_{4},$ the non-local resolvent vector in
Fourier domain becomes%
\begin{equation}
\boldsymbol{\bar{R}}_{n\ell }^{\left( T_{2}/T_{4}\right) }\left( \boldsymbol{%
k}\right) =-\frac{1}{\mu }\left( \boldsymbol{L}\otimes \boldsymbol{L}\right)
\,\mathrm{i}\boldsymbol{k}+\frac{\lambda +\mu }{\mu \left( \lambda +2\mu
\right) }\frac{\left( \boldsymbol{L}\cdot \boldsymbol{k}\right) ^{2}}{k^{2}}%
\,\mathrm{i}\boldsymbol{k},  \label{R-nl-bar-T2-T4}
\end{equation}%
according to (\ref{R-ovi-bar})$_{2}$ and respectively (\ref{R-nl-bar-T2})
and (\ref{R-nl-bar-T4}), since%
\begin{align*}
\boldsymbol{\bar{R}}_{n\ell }^{\left( T_{2}\right) }\left( \boldsymbol{k}%
\right) & =\boldsymbol{\bar{\hat{R}}}_{n\ell }^{\left( T_{2}\right) }\left( 
\boldsymbol{k}\right) \,\mathrm{i}\boldsymbol{k}=\frac{1}{\mu }\left( 
\boldsymbol{L}\otimes \boldsymbol{L}\right) \,\mathrm{i}\boldsymbol{k}-\frac{%
\lambda +\mu }{\mu \left( \lambda +2\mu \right) }\frac{\boldsymbol{k}\otimes 
\boldsymbol{k}}{k^{2}}\left( \boldsymbol{L}\otimes \boldsymbol{L}\right) \,%
\mathrm{i}\boldsymbol{k},\quad \text{and} \\
\boldsymbol{\bar{R}}_{n\ell }^{\left( T_{4}\right) }\left( \boldsymbol{k}%
\right) & =\boldsymbol{\bar{\hat{R}}}_{n\ell }^{\left( T_{4}\right) }\left( 
\boldsymbol{k}\right) \,\mathrm{i}\boldsymbol{k}=\frac{1}{\mu }\left( 
\boldsymbol{L}\otimes \boldsymbol{L}\right) \frac{\boldsymbol{k}\otimes 
\boldsymbol{k}}{k^{2}}\,\mathrm{i}\boldsymbol{k}-\frac{\lambda +\mu }{\mu
\left( \lambda +2\mu \right) }\left( \boldsymbol{L}\cdot \boldsymbol{k}%
\right) ^{2}\frac{\boldsymbol{k}\otimes \boldsymbol{k}}{k^{4}}\,\mathrm{i}%
\boldsymbol{k},
\end{align*}%
where $\left( \boldsymbol{k}\otimes \boldsymbol{k}\right) \left( \boldsymbol{%
L}\otimes \boldsymbol{L}\right) \mathrm{i}\boldsymbol{k}=\left( \boldsymbol{L%
}\cdot \boldsymbol{k}\right) ^{2}\mathrm{i}\boldsymbol{k}$ and $\left( 
\boldsymbol{k}\otimes \boldsymbol{k}\right) \mathrm{i}\boldsymbol{k}=k^{2}%
\mathrm{i}\boldsymbol{k}$ is used. The non-local displacement vector field
in Fourier domain (\ref{u-bar-conserv-2})$_{2}$, using the non-local
resolvent vector in Fourier domain (\ref{R-nl-bar-T2-T4}), takes the form%
\begin{align}
\boldsymbol{\bar{u}}_{n\ell }^{\left( T_{2}/T_{4}\right) }\left( \boldsymbol{%
k}\right) & =-\frac{1}{\mu }\left( \boldsymbol{L}\otimes \boldsymbol{L}%
\right) \,\mathrm{i}\boldsymbol{k\,}\bar{\phi}\left( \boldsymbol{k}\right) +%
\frac{\lambda +\mu }{\mu \left( \lambda +2\mu \right) }\frac{\left( 
\boldsymbol{L}\cdot \boldsymbol{k}\right) ^{2}}{k^{2}}\,\mathrm{i}%
\boldsymbol{k\,}\bar{\phi}\left( \boldsymbol{k}\right) 
\label{u-nl-bar-T2-1} \\
& =\frac{1}{\mu }\left( \boldsymbol{L}\otimes \boldsymbol{L}\right) 
\boldsymbol{\bar{f}}_{b}\left( \boldsymbol{k}\right) -\frac{\lambda +\mu }{%
\mu \left( \lambda +2\mu \right) }\frac{\left( \boldsymbol{L}\cdot 
\boldsymbol{k}\right) ^{2}}{k^{2}}\boldsymbol{\bar{f}}_{b}\left( \boldsymbol{%
k}\right)   \label{u-nl-bar-T2}
\end{align}%
and becomes%
\begin{equation*}
\boldsymbol{u}_{n\ell }^{\left( T_{2}/T_{4}\right) }\left( \boldsymbol{r}%
\right) =\frac{1}{\mu }\left( \boldsymbol{L}\otimes \boldsymbol{L}\right) 
\boldsymbol{f}_{b}\left( \boldsymbol{r}\right) -\frac{L^{2}}{4\pi }\frac{%
\lambda +\mu }{\mu \left( \lambda +2\mu \right) }\left( \frac{1}{r^{3}}%
\left( 1-3\left( \frac{\boldsymbol{L}}{L}\cdot \frac{\boldsymbol{r}}{r}%
\right) ^{2}\right) \right) \ast _{\boldsymbol{r}}\boldsymbol{f}_{b}\left( 
\boldsymbol{r}\right) ,
\end{equation*}%
after the Fourier transform inversion, where the inverse Fourier transform
of $\frac{\left( \boldsymbol{L}\cdot \boldsymbol{k}\right) ^{2}}{k^{2}},$
calculated in Appendix \ref{skalarni-kroz-kvadrat}, see (\ref%
{skalarni-kroz-kv}), is used.

Equivalently, one may express the non-local displacement vector field in
terms of scalar function $\phi $ by inverting the expression (\ref%
{u-bar-conserv-2})$_{2}$, yielding%
\begin{align}
\boldsymbol{u}_{n\ell }^{\left( T_{2}/T_{4}\right) }\left( \boldsymbol{r}%
\right)  &=\boldsymbol{R}_{n\boldsymbol{\ell }}^{\left( T_{2}/T_{4}\right)
}\left( \boldsymbol{r}\right) \ast \phi \left( \boldsymbol{r}\right)   \notag
\\
&=-\frac{1}{\mu }\left( \boldsymbol{L}\otimes \boldsymbol{L}\right)
\,\nabla \phi \left( \boldsymbol{r}\right) -\frac{1}{4\pi }\frac{\lambda
+\mu }{\mu \left( \lambda +2\mu \right) }\left( \left( \left( \frac{%
\boldsymbol{L}}{L}\cdot \frac{\boldsymbol{r}}{r}\right) \frac{L^{2}}{r^{2}}%
\right) \ast \phi \left( \boldsymbol{r}\right) \right) \frac{\boldsymbol{L}}{%
L} \notag \\ 
& \quad -\frac{1}{8\pi }\frac{\lambda +\mu }{\mu \left( \lambda +2\mu \right) }%
\left( \left( 1-3\left( \frac{\boldsymbol{L}}{L}\cdot \frac{\boldsymbol{r}}{r%
}\right) ^{2}\right) \frac{L^{2}}{r^{2}}\frac{\boldsymbol{r}}{r}\right) \ast
\phi \left( \boldsymbol{r}\right) ,  \label{u-nl-T2-T4-fi}
\end{align}%
where%
\begin{equation*}
\boldsymbol{R}_{n\boldsymbol{\ell }}^{\left( T_{2}/T_{4}\right) }\left( 
\boldsymbol{r}\right) =\mathcal{F}^{-1}\left[ \boldsymbol{\bar{R}}_{n\ell
}^{\left( T_{2}/T_{4}\right) }\left( \boldsymbol{k}\right) \right] \left( 
\boldsymbol{r}\right) =-\frac{1}{\mu }\left( \boldsymbol{L}\otimes 
\boldsymbol{L}\right) \mathcal{F}^{-1}\left[ \mathrm{i}\boldsymbol{k}\right]
\left( \boldsymbol{r}\right) +\frac{\lambda +\mu }{\mu \left( \lambda +2\mu
\right) }\mathcal{F}^{-1}\left[ \frac{\left( \boldsymbol{L}\cdot \boldsymbol{%
k}\right) ^{2}}{k^{2}}\,\mathrm{i}\boldsymbol{k}\right] \left( \boldsymbol{r}%
\right) ,
\end{equation*}%
since $\mathcal{F}^{-1}\left[ \mathrm{i}\boldsymbol{k}\right] \left( 
\boldsymbol{r}\right) =\nabla _{\boldsymbol{r}}$, as well as according to
calculations in Appendix \ref{skalarni-kroz-kvadrat-ik}, see (\ref{izraz}).
Note, the non-local displacement field, given by (\ref{u-nl-T2-T4-fi}), is
the Fourier transform inversion of expression (\ref{u-nl-bar-T2-1}).

\subsubsection{Model $T_{5}$}

In the case of model $T_{5},$ the tensor function $\boldsymbol{\bar{\hat{%
\mathcal{T}}}}_{5}$ is a zero tensor, i.e., $\boldsymbol{\bar{\hat{\mathcal{T%
}}}}_{5}=\boldsymbol{\hat{0}}$, see Table \ref{Modeli}, yielding that the
non-local resolvent tensor in Fourier domain $\boldsymbol{\bar{\hat{R}}}_{n%
\boldsymbol{\ell }}$, given by (\ref{R-nl-bar}), is also a zero tensor,
i.e., $\boldsymbol{\bar{\hat{R}}}_{n\boldsymbol{\ell }}^{\left( T_{5}\right)
}=\boldsymbol{\hat{0}}$, and therefore the non-local resolvent vector in
Fourier domain $\boldsymbol{\bar{R}}_{n\boldsymbol{\ell }}^{\left(
T_{5}\right) }$, given by (\ref{R-ovi-bar})$_{2}, $ is a zero vector,
implying that the non-local displacement vector field in Fourier domain,
given by (\ref{u-bar-conserv-2})$_{2},$ takes a zero value, i.e.,%
\begin{equation}
\boldsymbol{\bar{u}}_{n\ell }^{\left( T_{5}\right) }\left( \boldsymbol{k}%
\right) =\boldsymbol{0}.  \label{u-nl-bar-T5}
\end{equation}%
Finally, the non-local displacement vector field is a zero vector $%
\boldsymbol{u}_{n\ell }^{\left( T_{5}\right) }\left( \boldsymbol{r}\right) =%
\boldsymbol{0}$.

\section{Gravitational force as an example of conservative force}

In order to illustrate the displacement vector field, as a solution to a
problem of statics of an infinite non-local three-dimensional elastic body
under the existence of prescribed body force, the case of gravitational body
force%
\begin{equation}
\boldsymbol{f}_{b}\left( \boldsymbol{r}\right) =-\gamma \rho M\frac{%
\boldsymbol{r}}{r^{3}}  \label{grav-b-f}
\end{equation}%
is considered, where $\gamma $ is an universal gravitational constant, $\rho 
$ is a density of a non-local three-dimensional body, and mass $M$ is a
source of gravitational force field. The gravitational body force in Fourier
domain takes the form%
\begin{equation}
\boldsymbol{\bar{f}}_{b}\left( \boldsymbol{k}\right) =4\pi \gamma \rho M%
\frac{\mathrm{i}\boldsymbol{k}}{k^{2}},  \label{f-bar-grav}
\end{equation}%
according to expression (\ref{ik-kroz-kv}) calculated in Appendix \ref%
{ik-kroz-k-na-kv}.

\subsection{Displacement field for gravitational body force}

The local displacement vector field in Fourier domain (\ref{u-l-bar}) in the
presence of the gravitational body force takes the form%
\begin{equation*}
\boldsymbol{\bar{u}}_{\ell }\left( \boldsymbol{k}\right) =\frac{1}{\lambda
+2\mu }\frac{1}{k^{2}}\boldsymbol{\bar{f}}_{b}\left( \boldsymbol{k}\right)
=4\pi \gamma \rho M\frac{1}{\lambda +2\mu }\frac{\mathrm{i}\boldsymbol{k}}{%
k^{4}}=4\pi \frac{\gamma M}{c_{c}^{2}}\frac{\mathrm{i}\boldsymbol{k}}{k^{4}},
\end{equation*}%
where $c_{c}=\sqrt{\frac{\lambda +2\mu }{\rho }}$ is the compressive wave
propagation speed, becoming%
\begin{equation}
\boldsymbol{u}_{\ell }\left( \boldsymbol{r}\right) =-\frac{1}{2}\frac{\gamma
M}{c_{c}^{2}}\frac{\boldsymbol{r}}{r}  \label{u-loc-grav}
\end{equation}%
after inverse Fourier transform, according to expression (\ref%
{ik-kroz-k-na-4}) given in Appendix \ref{ik-kroz-k-na-kv}.

The non-local displacement vector field in Fourier domain in the presence of
the gravitational body force, corresponding to scalar-type non-locality
models $S_{1}$ - $S_{3}$ and tensor-type non-locality models $T_{1}$ - $%
T_{5},$ takes the following forms:%
\begin{align*}
\boldsymbol{\bar{u}}_{n\ell }^{\left( S_{1}/T_{1}\right) }\left( \boldsymbol{%
k}\right) &=\frac{L^{2}}{\lambda +2\mu }\boldsymbol{\bar{f}}_{b}\left( 
\boldsymbol{k}\right) =4\pi L^{2}\frac{\gamma M}{c_{c}^{2}}\frac{\mathrm{i}%
\boldsymbol{k}}{k^{2}}, \\
\boldsymbol{\bar{u}}_{n\ell }^{\left( S_{2}/T_{3}\right) }\left( \boldsymbol{%
k}\right) &=\frac{1}{\lambda +2\mu }\frac{\left( \boldsymbol{L}\cdot 
\boldsymbol{k}\right) ^{2}}{k^{2}}\boldsymbol{\bar{f}}_{b}\left( \boldsymbol{%
k}\right) =4\pi \frac{\gamma M}{c_{c}^{2}}\frac{\left( \boldsymbol{L}\cdot 
\boldsymbol{k}\right) ^{2}}{k^{4}}\mathrm{i}\boldsymbol{k}, \\
\boldsymbol{\bar{u}}_{n\ell }^{\left( S_{3}\right) }\left( \boldsymbol{k}%
\right) &=\boldsymbol{\bar{u}}_{n\ell }^{\left( S_{1}\right) }\left( 
\boldsymbol{k}\right) -\boldsymbol{\bar{u}}_{n\ell }^{\left( S_{2}\right)
}\left( \boldsymbol{k}\right), \\
\boldsymbol{\bar{u}}_{n\ell }^{\left( T_{2}/T_{4}\right) }\left( \boldsymbol{%
k}\right) &=\frac{1}{\mu }\left( \boldsymbol{L}\otimes \boldsymbol{L}\right) 
\boldsymbol{\bar{f}}_{b}\left( \boldsymbol{k}\right) -\frac{\lambda +\mu }{%
\mu \left( \lambda +2\mu \right) }\frac{\left( \boldsymbol{L}\cdot 
\boldsymbol{k}\right) ^{2}}{k^{2}}\boldsymbol{\bar{f}}_{b}\left( \boldsymbol{%
k}\right), \\
&=4\pi \frac{\gamma M}{c_{s}^{2}}\left( \boldsymbol{L}\otimes \boldsymbol{L}%
\right) \frac{\mathrm{i}\boldsymbol{k}}{k^{2}}-4\pi \gamma M\left( \frac{1}{%
c_{s}^{2}}-\frac{1}{c_{c}^{2}}\right) \frac{\left( \boldsymbol{L}\cdot 
\boldsymbol{k}\right) ^{2}}{k^{4}}\mathrm{i}\boldsymbol{k}, \\
\boldsymbol{\bar{u}}_{n\ell }^{\left( T_{5}\right) }\left( \boldsymbol{k}%
\right) &=\boldsymbol{0},
\end{align*}%
where $c_{s}=\sqrt{\frac{\mu }{\rho }}$ is the shear wave propagation speed,
according to (\ref{u-nl-bar-S1}), (\ref{u-nl-bar-S2}), (\ref{u-nl-bar-S3}), (%
\ref{u-nl-bar-T2}), and (\ref{u-nl-bar-T5}), as well as (\ref{f-bar-grav}),
since the gravitational body force is conservative, so that displacement
vector fields become 
\begin{align*}
\boldsymbol{u}_{n\ell }^{\left( S_{1}/T_{1}\right) }\left( \boldsymbol{r}%
\right) &=-\frac{\gamma M}{c_{c}^{2}}\frac{L^{2}}{r^{2}}\frac{\boldsymbol{r}%
}{r}, \\
\boldsymbol{u}_{n\ell }^{\left( S_{2}/T_{3}\right) }\left( \boldsymbol{r}%
\right) &=-\frac{1}{2}\frac{\gamma M}{c_{c}^{2}}\frac{L^{2}}{r^{2}}\left(
\left( 1-3\left( \frac{\boldsymbol{L}}{L}\cdot \frac{\boldsymbol{r}}{r}%
\right) ^{2}\right) \frac{\boldsymbol{r}}{r}+2\left( \frac{\boldsymbol{L}}{L}%
\cdot \frac{\boldsymbol{r}}{r}\right) \frac{\boldsymbol{L}}{L}\right), \\
\boldsymbol{u}_{n\ell }^{\left( S_{3}\right) }\left( \boldsymbol{r}\right) &=%
\boldsymbol{u}_{n\ell }^{\left( S_{1}\right) }\left( \boldsymbol{r}\right) -%
\boldsymbol{u}_{n\ell }^{\left( S_{2}\right) }\left( \boldsymbol{r}\right) \\
&=-\frac{1}{2}\frac{\gamma M}{c_{c}^{2}}\frac{L^{2}}{r^{2}}\left( \left(
1+3\left( \frac{\boldsymbol{L}}{L}\cdot \frac{\boldsymbol{r}}{r}\right)
^{2}\right) \frac{\boldsymbol{r}}{r}-2\left( \frac{\boldsymbol{L}}{L}\cdot 
\frac{\boldsymbol{r}}{r}\right) \frac{\boldsymbol{L}}{L}\right), \\
\boldsymbol{u}_{n\ell }^{\left( T_{2}/T_{4}\right) }\left( \boldsymbol{r}%
\right) &=-\frac{\gamma M}{c_{s}^{2}}\frac{L^{2}}{r^{2}}\frac{\boldsymbol{L}%
\otimes \boldsymbol{L}}{L^{2}}\frac{\boldsymbol{r}}{r}+\frac{1}{2}\gamma
M\left( \frac{1}{c_{s}^{2}}-\frac{1}{c_{c}^{2}}\right) \frac{L^{2}}{r^{2}}%
\left( \left( 1-3\left( \frac{\boldsymbol{L}}{L}\cdot \frac{\boldsymbol{r}}{r%
}\right) ^{2}\right) \frac{\boldsymbol{r}}{r}+2\left( \frac{\boldsymbol{L}}{L%
}\cdot \frac{\boldsymbol{r}}{r}\right) \frac{\boldsymbol{L}}{L}\right) \\
&=-\frac{\gamma M}{c_{c}^{2}}\left( \frac{\boldsymbol{L}}{L}\cdot \frac{%
\boldsymbol{r}}{r}\right) \frac{L^{2}}{r^{2}}\frac{\boldsymbol{L}}{L}+\frac{1%
}{2}\gamma M\left( \frac{1}{c_{s}^{2}}-\frac{1}{c_{c}^{2}}\right) \left(
1-3\left( \frac{\boldsymbol{L}}{L}\cdot \frac{\boldsymbol{r}}{r}\right)
^{2}\right) \frac{L^{2}}{r^{2}}\frac{\boldsymbol{r}}{r}, \\
\boldsymbol{u}_{n\ell }^{\left( T_{5}\right) }\left( \boldsymbol{r}\right) &=%
\boldsymbol{0},
\end{align*}
after the Fourier transform inversion.

Finally, the displacement vector field, corresponding to scalar-type
non-locality models $S_{1}$ - $S_{3}$ and tensor-type non-locality models $%
T_{1}$ - $T_{5}$ in the presence of the gravitational body force, is
represented as a superposition of local and non-local displacement vector
fields, see (\ref{u-loc-i-u-non-loc}) and also (\ref{u-loc-grav}) as well as
previous expressions, so that it takes the following forms:%
\begin{align}
\boldsymbol{u}^{\left( S_{1}/T_{1}\right) }\left( \boldsymbol{r}\right) & =%
\boldsymbol{u}_{\ell }\left( \boldsymbol{r}\right) +\boldsymbol{u}_{n\ell
}^{\left( S_{1}/T_{1}\right) }\left( \boldsymbol{r}\right)  \notag \\
& =-\frac{1}{2}\frac{\gamma M}{c_{c}^{2}}\left( 1+2\frac{L^{2}}{r^{2}}%
\right) \frac{\boldsymbol{r}}{r},  \label{u-S1-T1-grav} \\
\boldsymbol{u}^{\left( S_{2}/T_{3}\right) }\left( \boldsymbol{r}\right) & =%
\boldsymbol{u}_{\ell }\left( \boldsymbol{r}\right) +\boldsymbol{u}_{n\ell
}^{\left( S_{2}/T_{3}\right) }\left( \boldsymbol{r}\right)  \notag \\
& =-\frac{1}{2}\frac{\gamma M}{c_{c}^{2}}\left( \left( 1+\left( 1-3\left( 
\frac{\boldsymbol{L}}{L}\cdot \frac{\boldsymbol{r}}{r}\right) ^{2}\right) 
\frac{L^{2}}{r^{2}}\right) \frac{\boldsymbol{r}}{r}+2\left( \frac{%
\boldsymbol{L}}{L}\cdot \frac{\boldsymbol{r}}{r}\right) \frac{L^{2}}{r^{2}}%
\frac{\boldsymbol{L}}{L}\right),  \label{u-S2-T3-grav} \\
\boldsymbol{u}^{\left( S_{3}\right) }\left( \boldsymbol{r}\right) & =%
\boldsymbol{u}_{\ell }\left( \boldsymbol{r}\right) +\boldsymbol{u}_{n\ell
}^{\left( S_{3}\right) }\left( \boldsymbol{r}\right)  \notag \\
& =-\frac{1}{2}\frac{\gamma M}{c_{c}^{2}}\left( \left( 1+\left( 1+3\left( 
\frac{\boldsymbol{L}}{L}\cdot \frac{\boldsymbol{r}}{r}\right) ^{2}\right) 
\frac{L^{2}}{r^{2}}\right) \frac{\boldsymbol{r}}{r}-2\left( \frac{%
\boldsymbol{L}}{L}\cdot \frac{\boldsymbol{r}}{r}\right) \frac{L^{2}}{r^{2}}%
\frac{\boldsymbol{L}}{L}\right),  \label{u-S3-grav} \\
\boldsymbol{u}^{\left( T_{2}/T_{4}\right) }\left( \boldsymbol{r}\right) & =%
\boldsymbol{u}_{\ell }\left( \boldsymbol{r}\right) +\boldsymbol{u}_{n\ell
}^{\left( T_{2}/T_{4}\right) }\left( \boldsymbol{r}\right)  \notag \\
& =-\frac{1}{2}\frac{\gamma M}{c_{c}^{2}}\frac{\boldsymbol{r}}{r}-\frac{%
\gamma M}{c_{c}^{2}}\left( \frac{\boldsymbol{L}}{L}\cdot \frac{\boldsymbol{r}%
}{r}\right) \frac{L^{2}}{r^{2}}\frac{\boldsymbol{L}}{L}+\frac{1}{2}\gamma
M\left( \frac{1}{c_{s}^{2}}-\frac{1}{c_{c}^{2}}\right) \left( 1-3\left( 
\frac{\boldsymbol{L}}{L}\cdot \frac{\boldsymbol{r}}{r}\right) ^{2}\right) 
\frac{L^{2}}{r^{2}}\frac{\boldsymbol{r}}{r}  \notag \\
& =-\frac{\gamma M}{c_{c}^{2}}\left( \frac{\boldsymbol{L}}{L}\cdot \frac{%
\boldsymbol{r}}{r}\right) \frac{L^{2}}{r^{2}}\frac{\boldsymbol{L}}{L}-\frac{1%
}{2}\frac{\gamma M}{c_{c}^{2}}\left( \!1+\left( \!1-3\left( \frac{%
\boldsymbol{L}}{L}\cdot \frac{\boldsymbol{r}}{r}\right) ^{2}\right) \frac{%
L^{2}}{r^{2}}\right) \frac{\boldsymbol{r}}{r}+\frac{1}{2}\frac{\gamma M}{%
c_{s}^{2}}\left( \!1-3\left( \frac{\boldsymbol{L}}{L}\cdot \frac{\boldsymbol{%
r}}{r}\right) ^{2}\right) \!\frac{L^{2}}{r^{2}}\frac{\boldsymbol{r}}{r} 
\notag \\
& =-\frac{\gamma M}{c_{c}^{2}}\left( \frac{\boldsymbol{L}}{L}\cdot \frac{%
\boldsymbol{r}}{r}\right) \frac{L^{2}}{r^{2}}\frac{\boldsymbol{L}}{L}-\frac{1%
}{2}\frac{\gamma M}{c_{c}^{2}}\left( 1-\left( \frac{c_{c}^{2}}{c_{s}^{2}}%
-1\right) \left( 1-3\left( \frac{\boldsymbol{L}}{L}\cdot \frac{\boldsymbol{r}%
}{r}\right) ^{2}\right) \frac{L^{2}}{r^{2}}\right) \frac{\boldsymbol{r}}{r},
\label{u-T2-T4-grav} \\
\boldsymbol{u}^{\left( T_{5}\right) }\left( \boldsymbol{r}\right) & =%
\boldsymbol{u}_{\ell }\left( \boldsymbol{r}\right) +\boldsymbol{u}_{n\ell
}^{\left( T_{5}\right) }\left( \boldsymbol{r}\right) =\boldsymbol{u}_{\ell
}\left( \boldsymbol{r}\right) =-\frac{1}{2}\frac{\gamma M}{c_{c}^{2}}\frac{%
\boldsymbol{r}}{r}.  \label{u-T5-grav}
\end{align}

\subsection{Numerical examples}

\begin{sidewaysfigure}
\centering


\begin{subfigure}{0.95\columnwidth}

\begin{minipage}{0.18\columnwidth}
\includegraphics[width=\columnwidth]{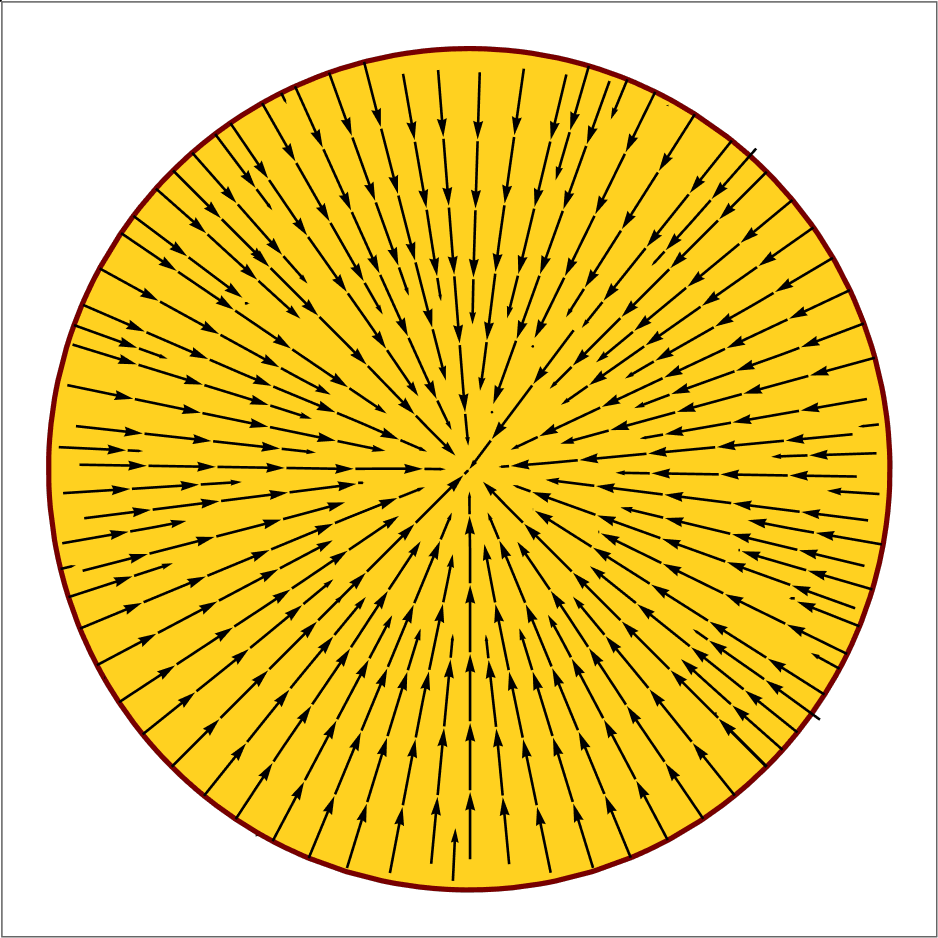}
\end{minipage}
\hfill
\begin{minipage}{0.18\columnwidth}
\includegraphics[width=\columnwidth]{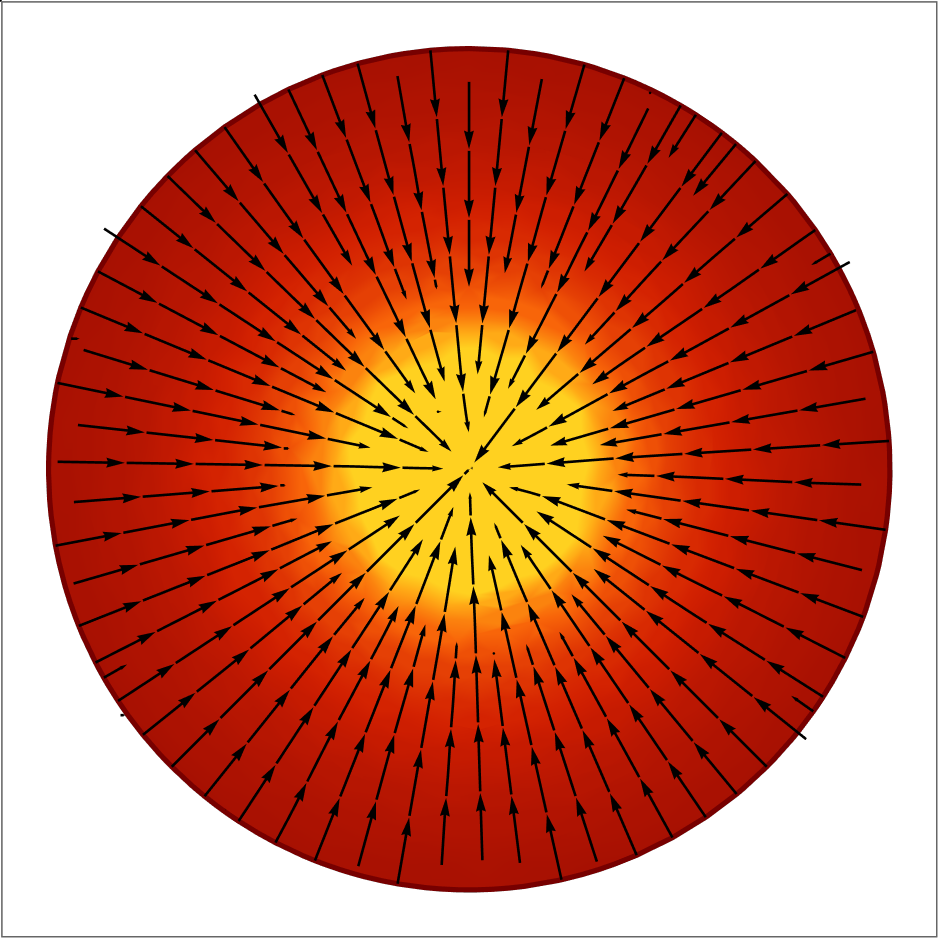}
\end{minipage}
\hfill
\begin{minipage}{0.18\columnwidth}
\includegraphics[width=\columnwidth]{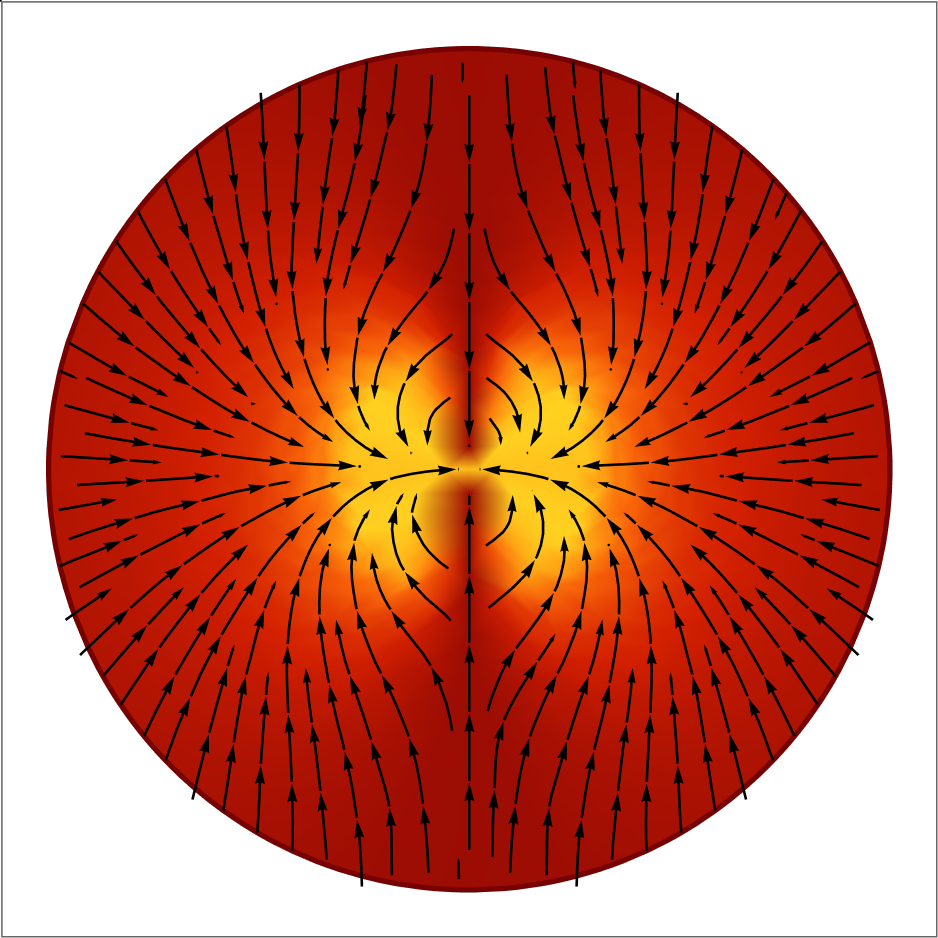}
\end{minipage}
\hfill
\begin{minipage}{0.18\columnwidth}
\includegraphics[width=\columnwidth]{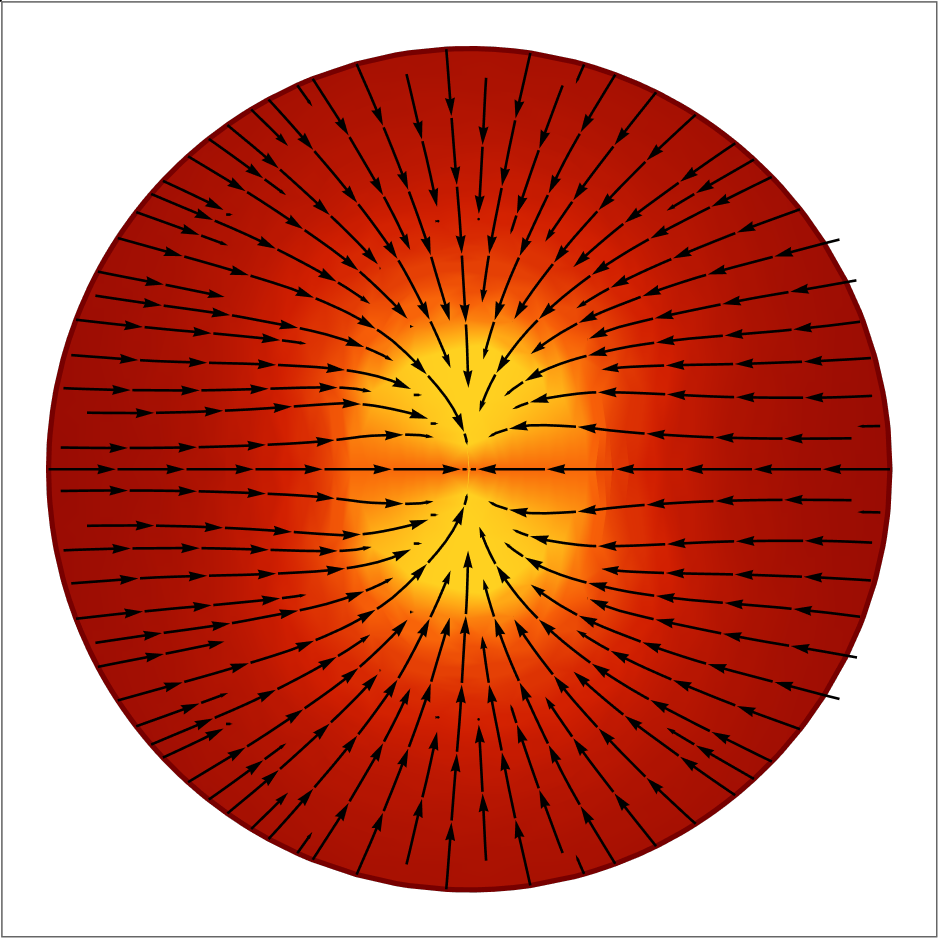}
\end{minipage}
\hfill
\begin{minipage}{0.18\columnwidth}
\includegraphics[width=\columnwidth]{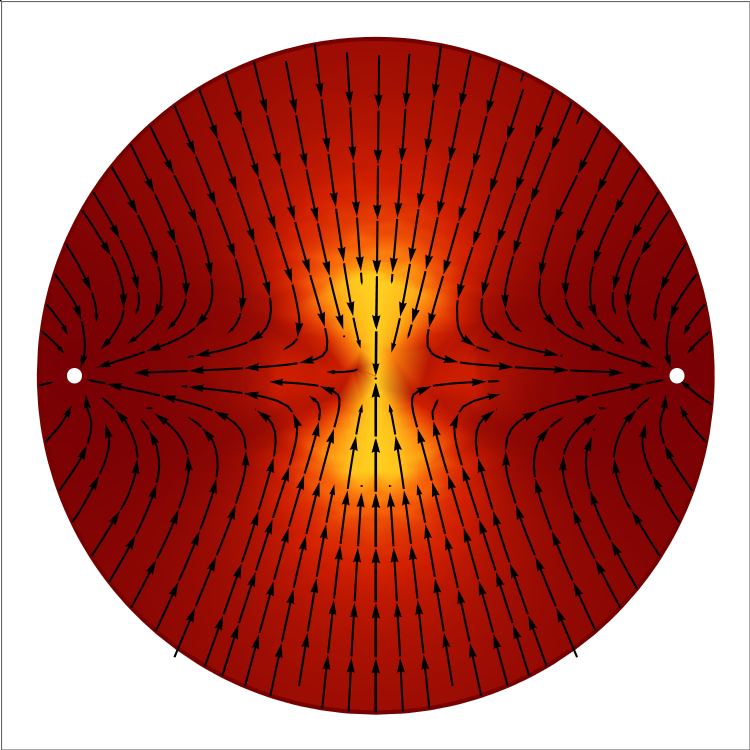}
\end{minipage}

\vspace{2mm}

\begin{minipage}{0.18\columnwidth}
		\centering
		Hooke's law and model $T_5$
	\end{minipage}
\hfil
	\begin{minipage}{0.18\columnwidth}
		\centering
		models $S_1$ and $T_1$
	\end{minipage}
\hfil
	\begin{minipage}{0.18\columnwidth}
		\centering
		models $S_2$ and $T_3$
	\end{minipage}
\hfil
	\begin{minipage}{0.18\columnwidth}
		\centering
		model $S_3$
	\end{minipage}
\hfil
	\begin{minipage}{0.18\columnwidth}
		\centering
		\quad models $T_2$ and $T_4$, $\beta=2.25$
	\end{minipage}

\vspace{2mm}

\caption{Displacement vector fields corresponding to Hooke's law and models $S_1$ -  $S_3$
and $T_1$ - $T_5$.\medskip}
\label{svi-modeli-S1-T5}
\end{subfigure}

\vspace{8mm}


\begin{subfigure}{0.95\columnwidth}

\begin{minipage}{0.18\columnwidth}
\includegraphics[width=\columnwidth]{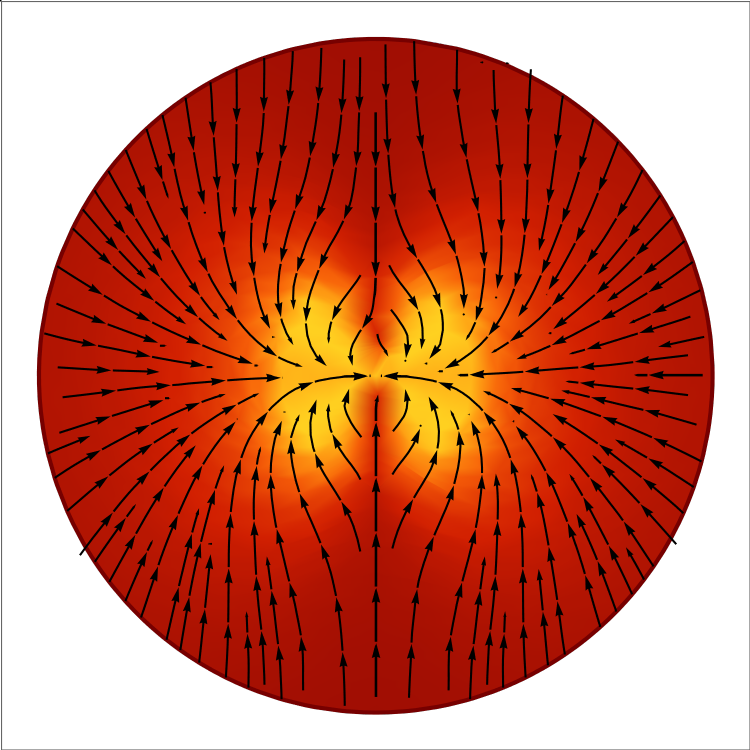}
\end{minipage}
\hfill
\begin{minipage}{0.18\columnwidth}
\includegraphics[width=\columnwidth]{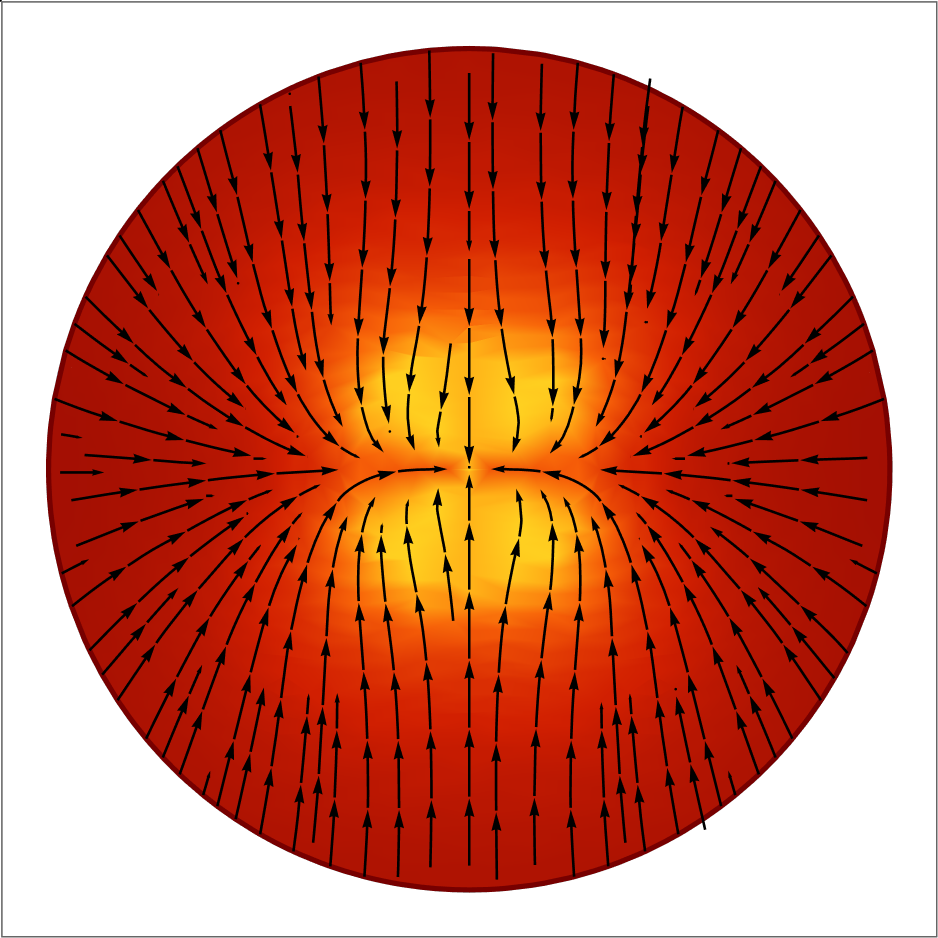}
\end{minipage}
\hfill
\begin{minipage}{0.18\columnwidth}
\includegraphics[width=\columnwidth]{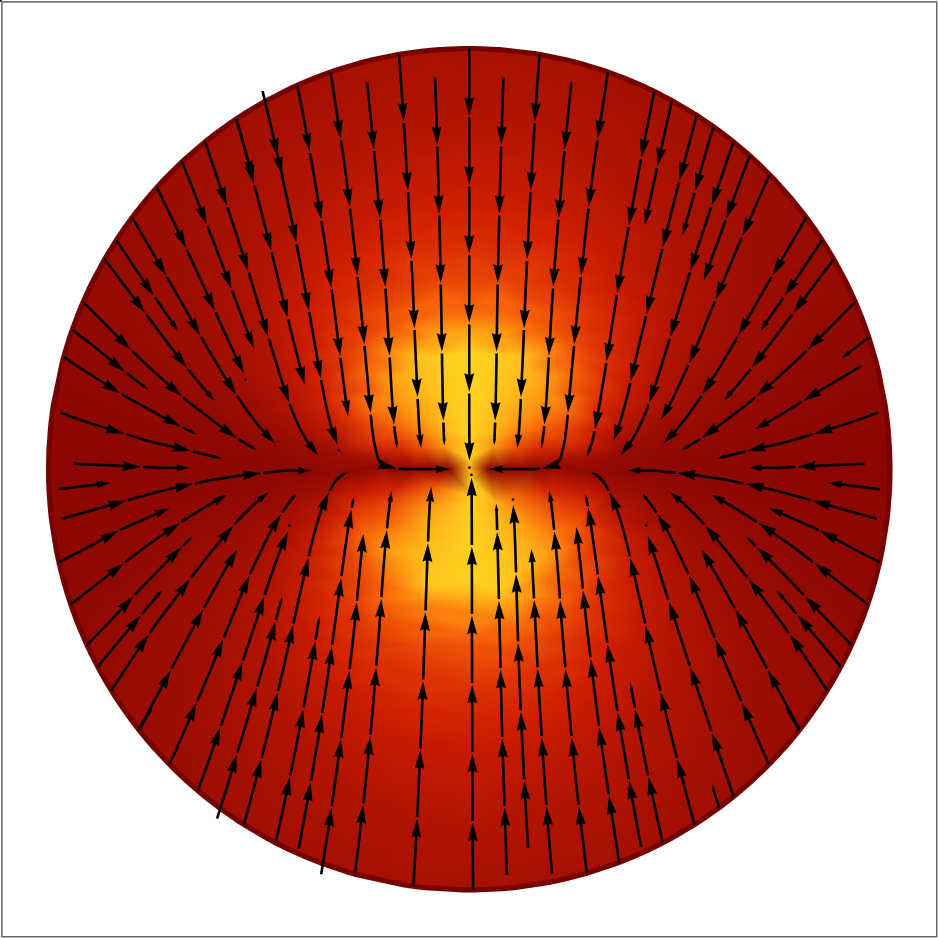}
\end{minipage}
\hfill
\begin{minipage}{0.18\columnwidth}
\includegraphics[width=\columnwidth]{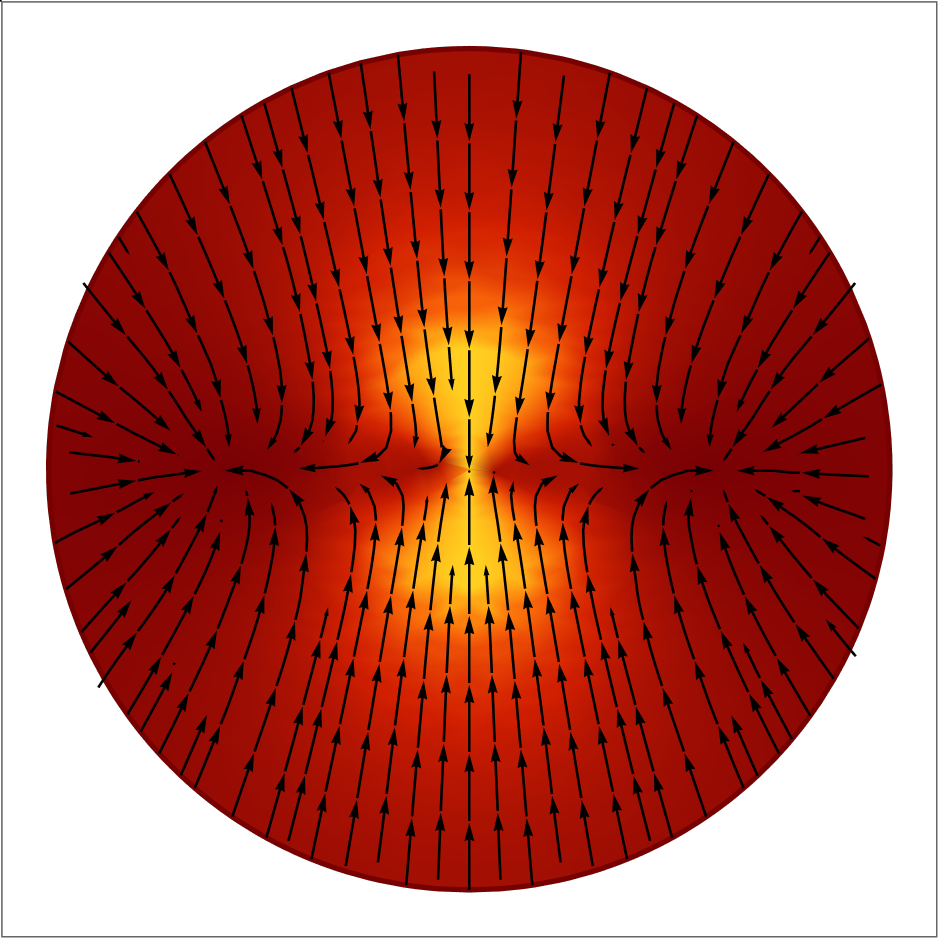}
\end{minipage}
\hfill
\begin{minipage}{0.18\columnwidth}
\includegraphics[width=\columnwidth]{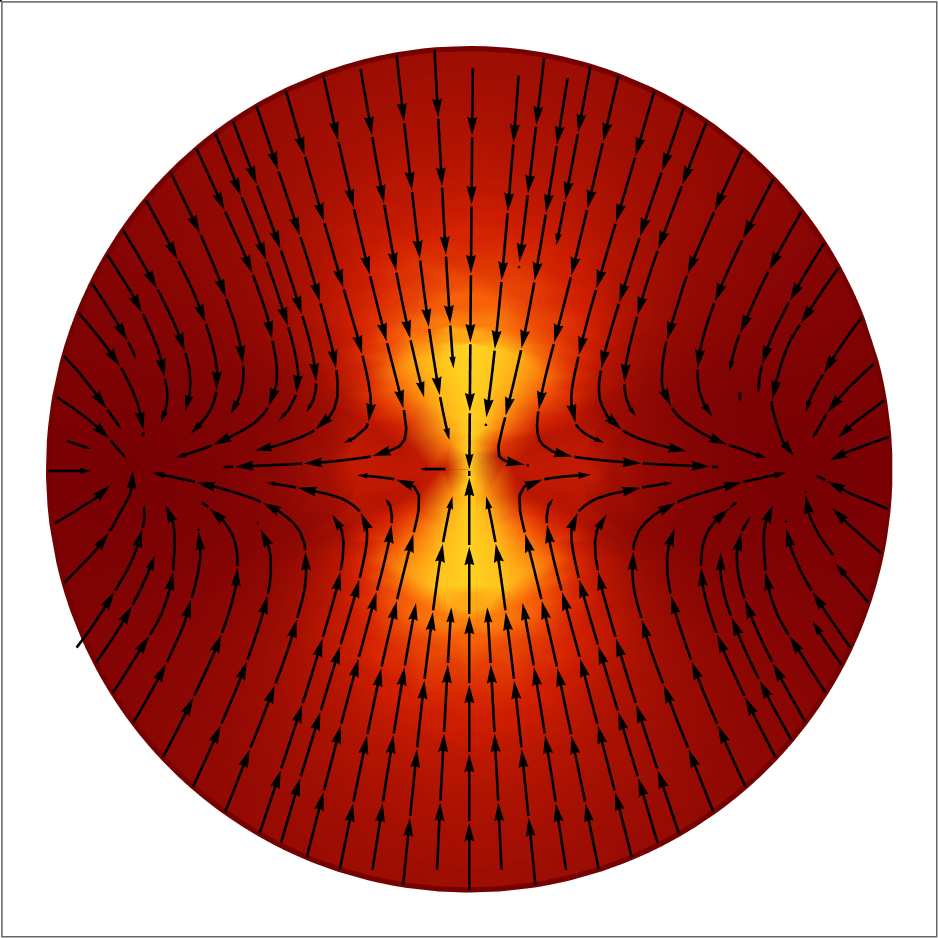}
\end{minipage}

\vspace{2mm}

\begin{minipage}{0.18\columnwidth}
		\centering
		$\beta=0.1$
	\end{minipage}
\hfil
	\begin{minipage}{0.18\columnwidth}
		\centering
		$\beta=0.5$
	\end{minipage}
\hfil
	\begin{minipage}{0.18\columnwidth}
		\centering
		$\beta=1$
	\end{minipage}
\hfil
	\begin{minipage}{0.18\columnwidth}
		\centering
		$\beta=1.5$
	\end{minipage}
\hfil
	\begin{minipage}{0.18\columnwidth}
		\centering
		$\beta=2$
	\end{minipage}
\vspace{2mm}

\caption{Displacement vector field corresponding to both tensor-type non-locality models $T_2$ and $T_4$ for various values of $\beta= \frac{c_{c}^{2}}{c_{s}^{2}}$.\medskip}
\label{samo-T2-i-T4}
\end{subfigure}
\vspace{5mm}

\caption{Displacement vector fields corresponding to Hooke's law and scalar-type non-locality models $S_1$ -  $S_3$
and tensor-type non-locality models $T_1$ -  $T_5$ in the presence of the gravitational body force, depicted in $\rho O z$-plane on domain $\frac{r}{L}\in \lbrack 0,1.25]$ and $\theta \in \lbrack 0,2\pi ]$, when the non-locality vector $\boldsymbol{L}$ is parallel to the $z$-axis.}
\label{non-loc-displ-field-grav}

\end{sidewaysfigure}

Setting the non-locality vector $\boldsymbol{L}$ to be parallel with the $z$%
-axis, i.e., $\boldsymbol{L}=L\boldsymbol{e}_{z},$ and by writing the radius
vector $\boldsymbol{r}$ in cylindrical coordinates, i.e., $\boldsymbol{r}%
=\rho \boldsymbol{e}_{\rho }+z\boldsymbol{e}_{z}$, the expressions (\ref%
{u-S1-T1-grav}) - (\ref{u-T5-grav}) are transformed into 
\begin{align}
\boldsymbol{u}^{\left( S_{1}/T_{1}\right) }\left( r,\theta \right) & =-\frac{%
1}{2}\frac{\gamma M}{c_{c}^{2}}\left( 1+2\frac{L^{2}}{r^{2}}\right) \left(
\sin \theta \boldsymbol{e}_{\rho }+\cos \theta \boldsymbol{e}_{z}\right) ,
\label{u-S1-T1-grav-1} \\
\boldsymbol{u}^{\left( S_{2}/T_{3}\right) }\left( r,\theta \right) & =-\frac{%
1}{2}\frac{\gamma M}{c_{c}^{2}}\left( \left( 1+\left( 1-3\cos ^{2}\theta
\right) \frac{L^{2}}{r^{2}}\right) \sin \theta \boldsymbol{e}_{\rho }+\left(
1+3\sin ^{2}\theta \,\frac{L^{2}}{r^{2}}\right) \cos \theta \boldsymbol{e}%
_{z}\right) ,  \label{u-S2-T3-grav-1} \\
\boldsymbol{u}^{\left( S_{3}\right) }\left( r,\theta \right) & =-\frac{1}{2}%
\frac{\gamma M}{c_{c}^{2}}\left( \left( 1+\left( 1+3\cos ^{2}\theta \right) 
\frac{L^{2}}{r^{2}}\right) \sin \theta \boldsymbol{e}_{\rho }+\left(
1-\left( 1-3\cos ^{2}\theta \right) \frac{L^{2}}{r^{2}}\right) \cos \theta 
\boldsymbol{e}_{z}\right) ,  \label{u-S3-grav-1} \\
\boldsymbol{u}^{\left( T_{2}/T_{4}\right) }\left( r,\theta \right) & =-\frac{%
1}{2}\frac{\gamma M}{c_{c}^{2}}\left( \left( 1-\left( \frac{c_{c}^{2}}{%
c_{s}^{2}}-1\right) \left( 1-3\cos ^{2}\theta \right) \frac{L^{2}}{r^{2}}%
\right) \sin \theta \boldsymbol{e}_{\rho }\right.  \notag \\
& \quad \quad \quad \quad \quad \left. +\left( 1+2\frac{L^{2}}{r^{2}}-\left( 
\frac{c_{c}^{2}}{c_{s}^{2}}-1\right) \left( 1-3\cos ^{2}\theta \right) \frac{%
L^{2}}{r^{2}}\right) \cos \theta \boldsymbol{e}_{z}\right) ,
\label{u-T2-T4-grav-1} \\
\boldsymbol{u}^{\left( T_{5}\right) }\left( r,\theta \right) & =\boldsymbol{u%
}_{\ell }\left( r,\theta \right) =-\frac{1}{2}\frac{\gamma M}{c_{c}^{2}}%
\left( \sin \theta \boldsymbol{e}_{\rho }+\cos \theta \boldsymbol{e}%
_{z}\right) ,  \label{u-T5-grav-1}
\end{align}%
where $\theta $ is an angle between $z$-axis and radius vector $\boldsymbol{r%
}$, having intensity $r$, and where the ratio between compressive and shear
propagation speeds $\frac{c_{c}^{2}}{c_{s}^{2}}$ is denoted by $\beta $, so
that the expressions (\ref{u-S1-T1-grav-1}) - (\ref{u-T5-grav-1}) are used
in order to graphically represent displacement vector fields 
\begin{equation*}
\boldsymbol{u}^{\left( S_{i}/T_{j}\right) }\left( r,\theta \right) =u_{\rho
}^{\left( S_{i}/T_{j}\right) }\left( r,\theta \right) \,\boldsymbol{e}_{\rho
}+u_{z}^{\left( S_{i}/T_{j}\right) }\left( r,\theta \right) \,\boldsymbol{e}%
_{z}
\end{equation*}%
in Figure \ref{non-loc-displ-field-grav} and depict them in $\rho Oz$-plane
on domain $\frac{r}{L}\in \lbrack 0,1.25]$ and $\theta \in \lbrack 0,2\pi ]$%
, since, apart from displacement vector fields from expressions (\ref%
{u-S1-T1-grav}) - (\ref{u-T5-grav}) depending on the non-locality vector $%
\boldsymbol{L}=L\boldsymbol{e}_{z},$ they also depend on the radius vector $%
\boldsymbol{r}=\rho \boldsymbol{e}_{\rho }+z\boldsymbol{e}_{z}$. Therefore,
one has the same picture for any polar angle, implying that the
three-dimensional image of the displacement vector field can be
reconstructed by rotating the two-dimensional cross-section pictures in
Figure \ref{non-loc-displ-field-grav} about the vertical $z$-axis. Moreover,
due to the term $\frac{\boldsymbol{L}}{L}\cdot \frac{\boldsymbol{r}}{r}=\cos
\theta $, appearing in expressions (\ref{u-S1-T1-grav}) - (\ref{u-T5-grav}),
i.e., (\ref{u-S1-T1-grav-1}) - (\ref{u-T5-grav-1}), the displacement vector
fields are also symmetrical with respect to the $z$-axis. Since the
magnitude of displacement vector fields is scaled by the factor $\frac{1}{2}%
\frac{\gamma M}{c_{c}^{2}}$, Figure \ref{non-loc-displ-field-grav} actually
presents plots of scaled displacement vector fields. Finally, arrows
represent the direction of the scaled vector field, while colors represent
its intensity, so the brighter the color is, the greater the intensity is.

The scalar-type non-locality kernels ${\mathcal{K}}_{1}$ - ${\mathcal{K}}%
_{3} $, as well as tensor-type non-locality kernels $\boldsymbol{\hat{{%
\mathcal{K}}}}_{1}$ - $\boldsymbol{\hat{{\mathcal{K}}}}_{5}$, see Table \ref%
{Modeli-new}, define the way in which information about the non-locality
field is transferred between spatial points of an infinite non-local
three-dimensional body, modeled by scalar- and tensor-type stress-gradient
Eringen three-dimensional constitutive equations (\ref{Hook-non-loc}) and (%
\ref{Hook-non-loc-t}). Therefore, the characteristics of non-local kernels
are reflected in the non-locality character of displacement fields (\ref%
{u-S1-T1-grav}) - (\ref{u-T5-grav}), corresponding to the mentioned
constitutive models and presented in Figure \ref{non-loc-displ-field-grav}
for the case of gravitational body force, which is more clearly observed in
the case of scalar- than tensor-type models, since the scalar-type
non-locality kernels ${\mathcal{K}}_{1}$ - ${\mathcal{K}}_{3}$ are
represented only by scalar functions and the Dirac delta distribution, while
the non-local character of tensor-type non-locality models is affected by
the scalar part of the non-locality kernel $\boldsymbol{\hat{{\mathcal{K}}}}%
_{1}$ - $\boldsymbol{\hat{{\mathcal{K}}}}_{5}$ together with the Dirac delta
distribution, as well as the tensors that give these kernels a tensor
character.

The second figure in Figure \ref{svi-modeli-S1-T5} clearly demonstrates the
radial character of the displacement vector field's non-locality for the
scalar-type model $S_{1} $, which is consistent with the corresponding
non-locality kernel ${\mathcal{K}}_{1} (r) = \frac{1}{4\pi rL^{2}}\mathrm{e}%
^{-\frac{r}{L}}$, since the non-locality vector $\boldsymbol{L}$ acts solely
as a scaling parameter that controls the magnitude of the non-local effect.
Furthermore, the consistency between non-local character of the kernel and
displacement field is also evident for the scalar-type models $S_{2}$ and $%
S_{3}$. Namely, the displacement vector fields from third and fourth figures
in Figure \ref{svi-modeli-S1-T5} exhibit a preferential direction of
non-locality, which is parallel with the direction of the non-locality
vector $\boldsymbol{L}$ for the scalar-type model $S_{2}$ and perpendicular
to the direction of the same vector in the case of model $S_{3}$, reflecting
the non-local anisotropy of kernels 
\begin{equation*}
\mathcal{K} _{2} (r) = \frac{1}{2L}\delta \left( \boldsymbol{r}_{\perp
}\right) \mathrm{e}^{-\frac{\left\vert r_{\parallel }\right\vert }{L}} \quad 
\text{and} \quad \mathcal{K} _{3} (r) = \frac{1}{2\pi L^{2}}\delta \left(
r_{\parallel }\right) K_{0}\left( \frac{r_{\perp }}{L}\right),
\end{equation*}
see Table \ref{Modeli-new}, which account for non-locality effects either in
the direction parallel or perpendicular to the direction of non-locality
vector $\boldsymbol{L}$, such that the effects are weighted either by
exponential or Macdonald function. In the vicinity of $\rho$-, respectively $%
z$-axis, in the case of scalar-type model $S_{2}$, respectively model $S_{3}$%
, the direction of displacement field seems to be radial, which is in
accordance with the kernels ${\mathcal{K}}_{2}$ and ${\mathcal{K}}_{3}$,
which do not exhibit non-locality effects in the direction perpendicular,
respectively parallel, to the direction of non-locality vector $\boldsymbol{L%
}$. Overall, the non-local character of displacement field vanishes for
large values of the radius vector and the radial character of local-type
displacement vector field remains, compare expressions (\ref{u-S1-T1-grav-1}%
), (\ref{u-S2-T3-grav-1}), and (\ref{u-S3-grav-1}) for large values of
radius vector with (\ref{u-T5-grav-1}).

Less evidently, the non-locality kernel 
\begin{equation*}
\boldsymbol{\hat{{\mathcal{K}}}}_{1} \left( \boldsymbol{r}\right) =2\delta
\left( \boldsymbol{r}\right) \left( \boldsymbol{\hat{I}}-2\frac{\boldsymbol{r%
}\otimes \boldsymbol{r}}{r^{2}}\right) -\frac{1}{4\pi r^{3}}\mathrm{e}^{-%
\frac{r}{L}}\left( \left( 1+\frac{r}{L}\right) \boldsymbol{\hat{I}}+\left(
3\left( 1+\frac{r}{L}\right) +\frac{r^{2}}{L^{2}}\right) \frac{\boldsymbol{r}%
\otimes \boldsymbol{r}}{r^{2}}\right),
\end{equation*}
see Table \ref{Modeli-new}, corresponding to the tensor-type model $T_{1}$
suggests a radial character of the displacement vector field's non-locality,
evident form expression (\ref{u-S1-T1-grav-1}) and shown in the second
figure in Figure \ref{svi-modeli-S1-T5}, which can be attributed to the
second term of the kernel, in which both the scalar function and the tensor
indicate such a radial character, whereas the first term, due to the
presence of Dirac delta distribution, does not contribute to the non-local
character. It is worth noting that the non-locality kernel ${\mathcal{K}}%
_{1} $ and the non-locality kernel $\boldsymbol{\hat{{\mathcal{K}}}}_{1}$
are the scalar and tensor counterparts that yield not only qualitatively
same non-locality character of displacement vector field, but also
quantitatively same character in the case of a conservative body force, see (%
\ref{u-non-loc-S1-konz}), as shown in the second figure in Figure \ref%
{svi-modeli-S1-T5}.

Nevertheless, the non-local character of displacement field cannot be
readily anticipated from the form of non-locality kernel alone in the case
of tensor-type models $T_{2}$, $T_{3}$, and $T_{4}$, since in expressions 
\begin{align}
\boldsymbol{\hat{{\mathcal{K}}}}_{2}\left( \boldsymbol{r}\right) & =\delta
\left( \boldsymbol{r}\right) \left( \boldsymbol{\hat{I}}-\frac{\boldsymbol{L}%
\otimes \boldsymbol{L}}{L^{2}}\right) +\frac{1}{4\pi L^{2}r}\mathrm{e}^{-%
\frac{r}{L}}\frac{\boldsymbol{L}\otimes \boldsymbol{L}}{L^{2}},  \notag \\
\boldsymbol{\hat{{\mathcal{K}}}}_{3}\left( \boldsymbol{r}\right) & =\delta
\left( \boldsymbol{r}\right) \left( \boldsymbol{\hat{I}}-\frac{\boldsymbol{L}%
\otimes \boldsymbol{L}}{L^{2}}\right) +\frac{1}{4L}\delta \left( \boldsymbol{%
r}_{\perp }\right) \mathrm{e}^{-\frac{\left\vert r_{\parallel }\right\vert }{%
L}}\frac{\boldsymbol{L}\otimes \boldsymbol{L}}{L^{2}}+\frac{\delta \left( 
\boldsymbol{r}_{\perp }\right) }{r_{\perp }}\mathrm{e}^{-\frac{\left\vert
r_{\parallel }\right\vert }{L}}\left( \frac{\boldsymbol{L}}{L}\otimes \frac{%
\boldsymbol{r}_{\perp }}{r_{\perp }}\right),  \notag \\
\boldsymbol{\hat{{\mathcal{K}}}}_{4}\left( \boldsymbol{r}\right) & =\delta
\left( \boldsymbol{r}\right) \left( \boldsymbol{\hat{I}}-\frac{\boldsymbol{L}%
\otimes \boldsymbol{L}}{L^{2}}\right) +\frac{1}{4L}\delta \left( \boldsymbol{%
r}_{\perp }\right) \mathrm{e}^{-\frac{\left\vert r_{\parallel }\right\vert }{%
L}}\frac{\boldsymbol{L}\otimes \boldsymbol{L}}{L^{2}}+\frac{\delta \left( 
\boldsymbol{r}_{\perp }\right) }{r_{\perp }}\mathrm{e}^{-\frac{\left\vert
r_{\parallel }\right\vert }{L}}\left( \frac{\boldsymbol{r}_{\perp }}{%
r_{\perp }}\otimes \frac{\boldsymbol{L}}{L}\right),  \notag
\end{align}%
see Table \ref{Modeli-new}, the non-local anisotropy originates either only
from tensor $\boldsymbol{L}\otimes \boldsymbol{L}$ in the case of
non-locality kernel $\boldsymbol{\hat{{\mathcal{K}}}}_{2}$, or from both
tensor, constructed from non-locality vector $\boldsymbol{L}$, and
exponential function singling out preferential direction of the vector $%
\boldsymbol{L}$, in the case of non-locality kernels $\boldsymbol{\hat{{%
\mathcal{K}}}}_{3}$ and $\boldsymbol{\hat{{\mathcal{K}}}}_{4}$. It is worth
noting that models $T_{2}$ and $T_{4}$ yield the same displacement field in
the case of a gravitational body force, see (\ref{u-T2-T4-grav-1}) and the
last figure in Figure \ref{svi-modeli-S1-T5}, even though models $T_{3}$ and 
$T_{4}$ exhibit the smallest difference in their non-locality kernels $%
\boldsymbol{\hat{{\mathcal{K}}}}_{3}$ and $\boldsymbol{\hat{{\mathcal{K}}}}%
_{4}$, while the models $S_{2}$ and $T_{3}$ represent scalar and tensor
counterparts yielding the same displacement vector field, see (\ref%
{u-S2-T3-grav-1}) and the third figure in Figure \ref{svi-modeli-S1-T5}, in
the case of a gravitational body force.

The peculiarity of displacement vector field corresponding to models $T_{2}$
and $T_{4}$ is the occurrence of points representing the sinks of vector
field lines located on the $\rho $-axis if $\beta >1$, which become further
apart as the value of the parameter $\beta $ increases, as observed from the
last figure in Figure \ref{svi-modeli-S1-T5} and the last two figures in
Figure \ref{samo-T2-i-T4}, while there are no sink points when $\beta <1$ as
evident from the first two figures in Figure \ref{samo-T2-i-T4}. Actually,
these points are intersection points of a $\rho Oz$-plane and a circle of
radius $\rho ^{\ast }$ located in the plane $z=0$, along which the
displacement field is a zero vector. Moreover, the location $\rho ^{\ast }$
of sink points on $\rho $-axis can be analytically calculated, by
considering the expression of displacement vector field corresponding to
models $T_{2}$ and $T_{4}$, given by (\ref{u-T2-T4-grav-1}), which reduces to%
\begin{align}
u_{\rho }^{\left( T_{2}/T_{4}\right) }\left( r,\theta \right) \,& =-\frac{1}{%
2}\frac{\gamma M}{c_{c}^{2}}\left( 1-\left( \beta -1\right) \left( 1-3\cos
^{2}\theta \right) \frac{L^{2}}{r^{2}}\right) \sin \theta ,
\label{u-T-2-rho} \\
u_{z}^{\left( T_{2}/T_{4}\right) }\left( r,\theta \right) & =-\frac{1}{2}%
\frac{\gamma M}{c_{c}^{2}}\left( 1+2\frac{L^{2}}{r^{2}}-\left( \beta
-1\right) \left( 1-3\cos ^{2}\theta \right) \frac{L^{2}}{r^{2}}\right) \cos
\theta .  \label{u-T-2-z}
\end{align}%
By considering the zeros of displacement vector field $\boldsymbol{u}%
^{\left( T_{2}/T_{4}\right) }$, namely $u_{\rho }^{\left( T_{2}/T_{4}\right)
}\left( r,\theta \right) =0$ and $u_{z}^{\left( T_{2}/T_{4}\right) }\left(
r,\theta \right) =0$ , the case in which the terms in parentheses vanish
simultaneously can be ruled out, since they differ by the factor $2\frac{%
L^{2}}{r^{2}}>0$. The case $\sin \theta =0$ can also be excluded, due to $%
u_{z}^{\left( T_{2}/T_{4}\right) }\left( r,0\right) =u_{z}^{\left(
T_{2}/T_{4}\right) }\left( r,\pi \right) =-\left( 1+2\beta \frac{L^{2}}{r^{2}%
}\right) \neq 0$, since $\beta =\frac{c_{c}^{2}}{c_{s}^{2}}\geqslant 0$.
However, the assumption $\theta =\frac{\pi }{2}$ or $\theta =\frac{3\pi }{2}$%
, implies $u_{z}^{\left( T_{2}/T_{4}\right) }\left( r,\frac{\pi }{2}\right)
=u_{z}^{\left( T_{2}/T_{4}\right) }\left( r,\frac{3\pi }{2}\right) =0$,
while $u_{\rho }^{\left( T_{2}/T_{4}\right) }\left( r,\frac{\pi }{2}\right)
=-u_{\rho }^{\left( T_{2}/T_{4}\right) }\left( r,\frac{3\pi }{2}\right)
=-\left( 1-\left( \beta -1\right) \frac{L^{2}}{r^{2}}\right) =0$ yields $%
\frac{\rho ^{\ast }}{L}=\sqrt{\beta -1}$ as a solution, so that if $\beta <1$
there are no sink points, since $\rho ^{\ast }$ is imaginary, while if $%
\beta =1$ there is a sink point at the origin, and position $\rho ^{\ast }$
of sink points attains a larger value as $\beta >1$ increases. The fact that
points located at $\rho ^{\ast }$ are sinks for the field lines of the
displacement vector field can be observed on the $\rho $-axis by plugging $%
\frac{\rho ^{\ast }}{L}=\sqrt{\beta -1}$ into (\ref{u-T-2-rho}) for $\theta =%
\frac{\pi }{2}$ and $\theta =\frac{3\pi }{2}$ yielding%
\begin{gather*}
u_{\rho }^{\left( T_{2}/T_{4}\right) }\left( \rho ,\frac{\pi }{2}\right)
=-u_{\rho }^{\left( T_{2}/T_{4}\right) }\left( \rho ,\frac{3\pi }{2}\right)
=-\frac{1}{2}\frac{\gamma M}{c_{c}^{2}}\left( 1-\frac{\rho ^{\ast 2}}{\rho
^{2}}\right) ,\quad \text{implying} \\
u_{\rho }^{\left( T_{2}/T_{4}\right) }\left( \rho ,\frac{\pi }{2}\right)
=-u_{\rho }^{\left( T_{2}/T_{4}\right) }\left( \rho ,\frac{3\pi }{2}\right)
>0\quad \text{for}\quad \rho <\rho ^{\ast }\quad \text{while} \\
u_{\rho }^{\left( T_{2}/T_{4}\right) }\left( \rho ,\frac{\pi }{2}\right)
=-u_{\rho }^{\left( T_{2}/T_{4}\right) }\left( \rho ,\frac{3\pi }{2}\right)
<0\quad \text{for}\quad \rho >\rho ^{\ast },
\end{gather*}%
i.e., field's component $u_{\rho }^{\left( T_{2}/T_{4}\right) }$ implies
vector field's direction being the direction of unit vector $\boldsymbol{e}%
_{\rho }$ when $\rho <\rho ^{\ast }$ and opposite direction when $\rho >\rho
^{\ast }$. Moreover, by rewriting the expression (\ref{u-T-2-z}) for $%
u_{z}^{\left( T_{2}/T_{4}\right) }$ as%
\begin{equation*}
u_{z}^{\left( T_{2}/T_{4}\right) }\left( \rho ,z\right) =-\frac{1}{2}\frac{%
\gamma M}{c_{c}^{2}}\left( 1+2\frac{L^{2}}{\rho ^{2}+z^{2}}-\left( \beta
-1\right) \left( 1-3\frac{z^{2}}{\rho ^{2}+z^{2}}\right) \frac{L^{2}}{\rho
^{2}+z^{2}}\right) \frac{z}{\sqrt{\rho ^{2}+z^{2}}}
\end{equation*}%
and by considering its asymptotic behavior at $\rho =\rho ^{\ast }$ as $%
z\rightarrow 0$, with $\frac{\rho ^{\ast }}{L}=\sqrt{\beta -1}$ and $\cos
\theta =\frac{z}{\sqrt{\rho ^{\ast 2}+z^{2}}}\sim \frac{z}{\rho ^{\ast }}$,
one has%
\begin{align*}
u_{z}^{\left( T_{2}/T_{4}\right) }\left( \rho ^{\ast },z\right) & \sim -%
\frac{1}{2}\frac{\gamma M}{c_{c}^{2}}\left( 1+2\frac{L^{2}}{\rho ^{\ast 2}}%
-\left( \beta -1\right) \frac{L^{2}}{\rho ^{\ast 2}}\right) \frac{z}{\rho
^{\ast }}=-\frac{\gamma M}{c_{c}^{2}}\frac{L^{2}}{\rho ^{\ast 2}}\frac{z}{%
\rho ^{\ast }}\quad \text{as\quad }z\rightarrow 0\text{,}\quad \text{%
implying,} \\
\quad u_{z}^{\left( T_{2}/T_{4}\right) }\left( \rho ^{\ast },z\right) &
>0\quad \text{when\quad }z\rightarrow 0^{-}\quad \text{and}\quad
u_{z}^{\left( T_{2}/T_{4}\right) }\left( \rho ^{\ast },z\right) <0\quad 
\text{when\quad }z\rightarrow 0^{+},
\end{align*}%
i.e., field's component $u_{z}^{\left( T_{2}/T_{4}\right) }$ implies vector
field's direction being the direction of unit vector $\boldsymbol{e}_{z}$
when $z\rightarrow 0^{-}$ and opposite direction when $z\rightarrow 0^{+}$.

Note, the displacement vector field corresponding to models $T_{2}$ and $%
T_{4}$ for small values of the parameter $\beta $ qualitatively resembles
the displacement vector field corresponding to models $S_{2}$ and $T_{3}$,
compare the first two figures in Figure \ref{samo-T2-i-T4} with the third
figure in Figure \ref{svi-modeli-S1-T5}, while if $\beta =0$ all of these
models yield the same displacement vector field, presented in the third
figure in Figure \ref{svi-modeli-S1-T5}, since the expression (\ref%
{u-T2-T4-grav-1}) corresponding to models $T_{2}$ and $T_{4}$ reduces to
expression (\ref{u-S2-T3-grav-1}) corresponding to models $S_{2}$ and $T_{3}$%
, i.e., the expression (\ref{u-T2-T4-grav}) reduces to expression (\ref%
{u-S2-T3-grav}).

Contrary to all other scalar- and tensor-type stress-gradient Eringen three
dimensional constitutive models (\ref{non-loc-scalar-type}) and (\ref%
{non-loc-tensor-type}), models $T_{2}$ and $T_{4}$ yield the displacement
vector field $\boldsymbol{u}^{\left( T_{2}/T_{4}\right) }$ that has sinks of
the field lines, different that possible sink at the origin, which are
actually located on the circle. The fact that other models do not yield
sinks of the field lines is obvious from expressions (\ref{u-S1-T1-grav-1}),
(\ref{u-S2-T3-grav-1}), (\ref{u-S3-grav-1}), and (\ref{u-T5-grav-1}).
Namely, the displacement vector fields $\boldsymbol{u}^{\left(
S_{1}/T_{1}\right) }$ and $\boldsymbol{u}^{\left( T_{5}\right) }$, see (\ref%
{u-S1-T1-grav-1}) and (\ref{u-T5-grav-1}), have no zeros, which is obvious
since sine and cosine functions cannot be zero simultaneously. Moreover, the
displacement vector fields $\boldsymbol{u}^{\left( S_{2}/T_{3}\right) }$ and 
$\boldsymbol{u}^{\left( S_{3}\right) }$, see (\ref{u-S2-T3-grav-1}) and (\ref%
{u-S3-grav-1}), also have no zeros, since the projections $u_{z }^{\left(
S_{2}/T_{3}\right) }$ and $u_{\rho}^{\left( S_{3}\right) }$ are zero if $%
\theta \in \{\frac{\pi}{2},\frac{3\pi}{2}\}$, respectively $\theta \in
\{0,\pi\}$, however $u_{\rho }^{\left( S_{2}/T_{3}\right) }$ and $%
u_{z}^{\left( S_{3}\right) }$ do not admit zeros for specified values of
angle $\theta$.

The displacement vector field $\boldsymbol{u}^{\left( T_{5}\right) }$,
corresponding to model $T_{5}$, has only the local contribution, as
presented in the first figure in Figure \ref{svi-modeli-S1-T5}, since in the
framework of static elasticity problem the corresponding non-local
contribution is zero. %

\section{Conclusion}

Unbounded three-dimensional medium of non-local character is modeled by
scalar- and tensor-type stress-gradient Eringen three-dimensional
constitutive equations (\ref{non-loc-scalar-type}) and (\ref%
{non-loc-tensor-type}), proposed in \cite{SD-6}, see also Table \ref%
{Modeli-new}, grouped into two classes: three scalar-type models (models $%
S_{1}$ - $S_{3}$) and five tensor-type models (models $T_{1}$ - $T_{5}$),
which represent a three-dimensional generalization of the one-dimensional
Eringen model (\ref{1D-Eringen}). A problem of statics of an unbounded solid
medium under the influence of prescribed body force is defined in the
framework of three-dimensional non-local elasticity and it is represented by
a system of equations consisting of equilibrium equation (\ref%
{equilibrium-eq}), infinitesimal strain (\ref{str}), and constitutive
equations either (\ref{Hook-non-loc}) or (\ref{Hook-non-loc-t}) which
incorporate the elastic character of a non-local medium through Hooke's law.
In parallel, the problem of statics of classical three-dimensional elastic
body, subject to prescribed body force, is analyzed by considering a system
of equations which, in addition to the equilibrium equation (\ref%
{equilibrium-eq}) and infinitesimal strain (\ref{str}), contains Hooke's law
(\ref{Hook-loc}). By applying the Fourier integral transform method, the
system of equations of these two problems is solved for arbitrary body
force, giving the displacement vector field in integral form (\ref{u-loc})
for the case when the medium is of a local character and by expressions (\ref%
{u-non-loc-S1}) - (\ref{u-non-loc-T5}) for the case when the medium is of
non-local character, modeled by scalar- and tensor-type stress-gradient
Eringen three-dimensional constitutive equations (\ref{Hook-non-loc}) and (%
\ref{Hook-non-loc-t}). Further, due to the linearity of governing system of
equations, for each of the eight non-local models the displacement field is
obtained in the form of a superposition of the local displacement vector
field (\ref{u-loc}), common to each model, and the non-local displacement
vector field, specific to the non-local model. It turns out that models $%
S_{1}$ and $T_{1}$ are scalar- and tensor-type counterparts and give the
same non-local displacement field (\ref{nltdf-S1}), which is also the case
with models $S_{2}$ and $T_{3}$, which correspond to a non-local
displacement field (\ref{nltdf-S2}), while expression (\ref{nltdf-S3}) is
specific to a scalar-type model $S_{3}$ and following expressions (\ref%
{nltdf-T2}), (\ref{nltdf-T4}), and (\ref{nltdf-T5}) are specific to
individual tensor-type models $T_{2}$, $T_{4}$, and $T_{5}$, respectively.

The governing system of equations describing mentioned two static problems
is solved for conservative body force given by (\ref{f-konz}), as well as
even specifically for the case when this force is a gravitational body force
(\ref{grav-b-f}), also providing analytical solutions (\ref%
{u-loc-i-u-non-loc}) for the displacement vector fields, each represented by
a superposition of local and non-local displacement vector field,
corresponding to three-dimensional elastic bodies modeled by constitutive
equations (\ref{Hook-non-loc}) and (\ref{Hook-non-loc-t}). Compared to the
displacement vector fields (\ref{u-non-loc-S1}) - (\ref{u-non-loc-T5}),
obtained for arbitrary body force, the displacement vector fields (\ref%
{u-non-loc-S1-konz}) - (\ref{u-non-loc-T5-konz}) for conservative body force
take significantly simplified forms, whereby the conservative body force
additionally implies the same solution (\ref{u-non-loc-T2-konz}) for the
displacement field in the case of tensor-type models $T_{2}$ and $T_{4}$.
Even simpler expressions (\ref{u-S1-T1-grav}) - (\ref{u-T5-grav}) for the
displacement fields are obtained in the case of three-dimensional non-local
bodies exposed to the gravitational force. Since the gravitational body
force is central, these displacement field solutions are considered within
the framework of a spherical coordinate system. It turns out that they are
independent of the polar angle and can therefore be represented graphically
in the $\rho O z$-plane, as shown in Figure \ref{non-loc-displ-field-grav}
using arrows to present the direction of the vector field and colors for the
intensity of the vector field, where the non-locality vector $\boldsymbol{L}$
is chosen to be parallel with the $z$-axis.

The scalar-type non-locality kernels ${\mathcal{K}}_{1}$ - ${\mathcal{K}}%
_{3} $, and tensor-type non-locality kernels $\boldsymbol{\hat{{\mathcal{K}}}%
}_{1} $ - $\boldsymbol{\hat{{\mathcal{K}}}}_{5}$, given in Table \ref%
{Modeli-new}, contain information about the non-locality characteristics of
the models and it should also be manifested in the displacement field, which
is observed in Figure \ref{non-loc-displ-field-grav}. Namely, the
displacement vector field (\ref{u-S1-T1-grav}), corresponding to models $%
S_{1}$ and $T_{1}$, exhibits spherically symmetric non-local character,
while the displacement vector field (\ref{u-S2-T3-grav}), corresponding to
models $S_{2}$ and $T_{3}$, exhibits preferential direction of non-locality
parallel to the non-locality vector $\boldsymbol{L}$ and finally the
displacement vector field (\ref{u-S3-grav}), corresponding to model $S_{3}$,
exhibits preferential direction of non-locality perpendicular to the
non-locality vector $\boldsymbol{L}$. Further, displacement field
corresponding to model $T_{5}$ and displacement field corresponding to
models $T_{2}$ and $T_{4}$ display more pronounced specificities, such that
for the former displacement field (\ref{u-T5-grav}) is spherically symmetric
without non-local contribution, while for the latter displacement field (\ref%
{u-T2-T4-grav}) exhibits parametric dependence on the ratio $\beta $ of
compressive and shear wave propagation speeds. If the value of the parameter 
$\beta $ is greater than one, then the force lines of the displacement
vector field corresponding to models $T_{2}$ and $T_{4}$ have a sink, unlike
the cases when the value of the parameter $\beta $ is less than one. Based
on the non-locality kernel, the non-locality character of the displacement
field can be foreseen more easily in the case of scalar- than in the case of
tensor-type models, since in the case of the former kernel's, non-locality
is influenced only by the scalar function and Dirac delta distribution,
while in the case of the latter, the more complex form of tensor-type
kernels in which tensors affect the non-local character in addition to the
scalar function and Dirac delta distribution, makes it less obvious.

\appendix

\section{Calculation of $\mathcal{F}^{-1}\left[ \frac{\mathrm{i}\boldsymbol{k%
}}{k^{2}}\right] \left( \boldsymbol{r}\right) $ and $\mathcal{F}^{-1}\left[ 
\frac{\mathrm{i}\boldsymbol{k}}{k^{4}}\right] \left( \boldsymbol{r}\right) $ 
\label{ik-kroz-k-na-kv}}

The Fourier transform inversion of vector functions $\frac{\mathrm{i}%
\boldsymbol{k}}{k^{2}}$ and $\frac{\mathrm{i}\boldsymbol{k}}{k^{4}}$ takes
the respective forms%
\begin{equation}
\mathcal{F}^{-1}\left[ \frac{\mathrm{i}\boldsymbol{k}}{k^{2}}\right] \left( 
\boldsymbol{r}\right) =-\frac{1}{4\pi r^{2}}\frac{\boldsymbol{r}}{r}\quad 
\text{and\quad }\mathcal{F}^{-1}\left[ \frac{\mathrm{i}\boldsymbol{k}}{k^{4}}%
\right] \left( \boldsymbol{r}\right) =-\frac{1}{8\pi }\frac{\boldsymbol{r}}{r%
}.  \label{ik-kroz-kv}
\end{equation}

Namely, using the Cartesian coordinate system for vector representation and
expressing vector components in spherical coordinate system, such that the
radius vector $\boldsymbol{r}$ is parallel to the $z$-axis, according to the
definition of inverse Fourier transform (\ref{IFT}), one has%
\begin{align*}
\mathcal{F}^{-1}\left[ \frac{\mathrm{i}k_{x}}{k^{2}}\right] \left( 
\boldsymbol{r}\right) & =\frac{1}{\left( 2\pi \right) ^{3}}\int_{%
\mathbb{R}
^{3}}\frac{\mathrm{i}k_{x}}{k^{2}}\mathrm{e}^{\mathrm{i}\boldsymbol{k}\cdot 
\boldsymbol{r}}\mathrm{d}_{\boldsymbol{k}}V \\
& =\frac{\mathrm{i}}{8\pi ^{3}}\int_{0}^{2\pi }\!\!\!\int_{0}^{\infty
}\!\!\!\int_{0}^{\pi }\frac{k\sin \theta \cos \varphi }{k^{2}}\mathrm{e}^{%
\mathrm{i}kr\cos \theta }k^{2}\sin \theta \mathrm{d}\theta \mathrm{d}k%
\mathrm{d}\varphi =0,
\end{align*}%
and%
\begin{align*}
\mathcal{F}^{-1}\left[ \frac{\mathrm{i}k_{y}}{k^{2}}\right] \left( 
\boldsymbol{r}\right) & =\frac{1}{\left( 2\pi \right) ^{3}}\int_{%
\mathbb{R}
^{3}}\frac{\mathrm{i}k_{y}}{k^{2}}\mathrm{e}^{\mathrm{i}\boldsymbol{k}\cdot 
\boldsymbol{r}}\mathrm{d}_{\boldsymbol{k}}V \\
& =\frac{\mathrm{i}}{8\pi ^{3}}\int_{0}^{2\pi }\!\!\!\int_{0}^{\infty
}\!\!\!\int_{0}^{\pi }\frac{k\sin \theta \sin \varphi }{k^{2}}\mathrm{e}^{%
\mathrm{i}kr\cos \theta }k^{2}\sin \theta \mathrm{d}\theta \mathrm{d}k%
\mathrm{d}\varphi =0,
\end{align*}%
since $\int_{0}^{2\pi }\cos \varphi \,\mathrm{d}\varphi =\int_{0}^{2\pi
}\sin \varphi \,\mathrm{d}\varphi =0,$ as well as%
\begin{align*}
\mathcal{F}^{-1}\left[ \frac{\mathrm{i}k_{z}}{k^{2}}\right] \left( 
\boldsymbol{r}\right) & =\frac{1}{\left( 2\pi \right) ^{3}}\int_{%
\mathbb{R}
^{3}}\frac{\mathrm{i}k_{z}}{k^{2}}\mathrm{e}^{\mathrm{i}\boldsymbol{k}\cdot 
\boldsymbol{r}}\mathrm{d}_{\boldsymbol{k}}V \\
& =\frac{\mathrm{i}}{8\pi ^{3}}\int_{0}^{2\pi }\!\!\!\int_{0}^{\infty
}\!\!\!\int_{0}^{\pi }\frac{k\cos \theta }{k^{2}}\mathrm{e}^{\mathrm{i}%
kr\cos \theta }k^{2}\sin \theta \mathrm{d}\theta \mathrm{d}k\mathrm{d}\varphi
\\
& =\frac{\mathrm{i}}{4\pi ^{2}}\int_{0}^{\infty }\!\!\!\int_{0}^{\pi }k\cos
\theta \mathrm{e}^{\mathrm{i}kr\cos \theta }\sin \theta \mathrm{d}\theta 
\mathrm{d}k \\
& =\frac{\mathrm{i}}{4\pi ^{2}}\int_{0}^{\infty }k\int_{-\mathrm{i}kr}^{%
\mathrm{i}kr}\frac{u}{\mathrm{i}kr}\mathrm{e}^{u}\frac{\mathrm{d}u}{\mathrm{i%
}kr}\mathrm{d}k \\
& =\frac{1}{2\pi ^{2}r^{2}}\int_{0}^{\infty }\frac{1}{k}\int_{-\mathrm{i}%
kr}^{\mathrm{i}kr}u\frac{\mathrm{e}^{u}}{2\mathrm{i}}\mathrm{d}u\mathrm{d}k
\\
& =\frac{1}{2\pi ^{2}r^{2}}\int_{0}^{\infty }\frac{kr\cos \left( kr\right) -%
\mathrm{\sin }\left( kr\right) }{k}\mathrm{d}k \\
& =\frac{1}{2\pi r}\left( \frac{1}{\pi }\int_{0}^{\infty }\cos \left(
kr\right) \mathrm{d}k\right) -\frac{1}{2\pi ^{2}r^{2}}\int_{0}^{\infty }%
\frac{\mathrm{\sin }\left( kr\right) }{k}\mathrm{d}k \\
& =\frac{1}{2\pi r}\delta \left( r\right) -\frac{1}{2\pi ^{2}r^{2}}\frac{\pi 
}{2}\mathrm{sgn\,}r \\
& =r\delta \left( \boldsymbol{r}\right) -\frac{1}{4\pi r^{2}}\mathrm{sgn\,}r,
\end{align*}%
since $\int u\mathrm{e}^{u}\mathrm{d}u=\left( u-1\right) \mathrm{e}^{u}+%
\mathrm{const}$, as well as%
\begin{equation}
\frac{1}{\pi }\int_{0}^{\infty }\cos \left( kr\right) \mathrm{d}k=\delta
\left( r\right) ,\quad \int_{0}^{\infty }\frac{\sin \left( kr\right) }{k}%
\mathrm{d}k=\frac{\pi }{2}\mathrm{sgn\,}r,\quad \text{and\quad }\delta
\left( \boldsymbol{r}\right) =\frac{\delta \left( r\right) }{2\pi r^{2}}.
\label{signum-i-mignum}
\end{equation}

Therefore, by collecting the previous results, one has 
\begin{equation*}
\mathcal{F}^{-1}\left[ \frac{\mathrm{i}\boldsymbol{k}}{k^{2}}\right] \left( 
\boldsymbol{r}\right) =\boldsymbol{r}\delta \left( \boldsymbol{r}\right) -%
\frac{\mathrm{sgn\,}r}{4\pi r^{2}}\frac{\boldsymbol{r}}{r},
\end{equation*}%
since the unit vector of $z$-axis is written as $\boldsymbol{e}_{z}=\frac{%
\boldsymbol{r}}{r}$, thus obtaining (\ref{ik-kroz-kv})$_{1}$, due to $%
\boldsymbol{r}\delta \left( \boldsymbol{r}\right) =\boldsymbol{0}$.

Additionally, the inversion $\mathcal{F}^{-1}\left[ \frac{1}{k^{2}}\right]
\left( \boldsymbol{r}\right) $ gives%
\begin{align}
\mathcal{F}^{-1}\left[ \frac{1}{k^{2}}\right] \left( \boldsymbol{r}\right) &
=\frac{1}{\left( 2\pi \right) ^{3}}\int_{%
\mathbb{R}
^{3}}\frac{1}{k^{2}}\mathrm{e}^{\mathrm{i}\boldsymbol{k}\cdot \boldsymbol{r}}%
\mathrm{d}_{\boldsymbol{k}}V  \notag \\
& =\frac{1}{8\pi ^{3}}\int_{0}^{2\pi }\!\!\!\int_{0}^{\infty
}\!\!\!\int_{0}^{\pi }\frac{1}{k^{2}}\mathrm{e}^{\mathrm{i}kr\cos \theta
}k^{2}\sin \theta \mathrm{d}\theta \mathrm{d}k\mathrm{d}\varphi  \notag \\
& =\frac{1}{4\pi ^{2}}\int_{0}^{\infty }\!\!\!\int_{-\mathrm{i}kr}^{\mathrm{i%
}kr}\mathrm{e}^{u}\frac{\mathrm{d}u}{\mathrm{i}kr}\mathrm{d}k  \notag \\
& =\frac{1}{2\pi ^{2}r}\int_{0}^{\infty }\frac{1}{k}\int_{-\mathrm{i}kr}^{%
\mathrm{i}kr}\frac{\mathrm{e}^{u}}{2\mathrm{i}}\mathrm{d}u\mathrm{d}k  \notag
\\
& =\frac{1}{2\pi ^{2}r}\int_{0}^{\infty }\frac{\mathrm{\sin }\left(
kr\right) }{k}\mathrm{d}k  \notag \\
& =\frac{1}{2\pi ^{2}r}\frac{\pi }{2}\mathrm{sgn\,}r  \notag \\
& =\frac{1}{4\pi r}\mathrm{sgn\,}r,  \label{1-kroz-4-pi-r}
\end{align}%
where (\ref{signum-i-mignum})$_{2}$ is used, so that one obtains (\ref%
{ik-kroz-kv}) by exploiting%
\begin{equation*}
\func{grad}\frac{\mathrm{sgn\,}r}{r}=\frac{\mathrm{d}}{\mathrm{d}r}\left( 
\frac{\mathrm{sgn\,}r}{r}\right) \frac{\boldsymbol{r}}{r}=-\mathrm{sgn\,}r%
\frac{\boldsymbol{r}}{r^{3}}+2\delta \left( r\right) \frac{\boldsymbol{r}}{%
r^{2}}=-\mathrm{sgn\,}r\frac{\boldsymbol{r}}{r^{3}}+4\pi \delta \left( 
\boldsymbol{r}\right) \boldsymbol{r},
\end{equation*}%
according (\ref{signum-i-mignum})$_{3}$, since$\mathcal{F}^{-1}\left[ \frac{%
\mathrm{i}\boldsymbol{k}}{k^{2}}\right] \left( \boldsymbol{r}\right) =\func{%
grad}\left( \mathcal{F}^{-1}\left[ \frac{1}{k^{2}}\right] \left( \boldsymbol{%
r}\right) \right) $.

In order to obtain the Fourier transform inversion of vector function $\frac{%
\mathrm{i}\boldsymbol{k}}{k^{4}},$ i.e., $\mathcal{F}^{-1}\left[ \frac{%
\mathrm{i}\boldsymbol{k}}{k^{4}}\right] \left( \boldsymbol{r}\right) $, the
convolution property of Fourier transform is used, so that, starting from
inversions $\mathcal{F}^{-1}\left[ \frac{\mathrm{i}\boldsymbol{k}}{k^{2}}%
\right] \left( \boldsymbol{r}\right) $ and $\mathcal{F}^{-1}\left[ \frac{1}{%
k^{2}}\right] \left( \boldsymbol{r}\right) $, given by (\ref{ik-kroz-kv})
and (\ref{1-kroz-4-pi-r}) respectively, by the use of Cartesian coordinate
system for radius vector representation and expressing the coordinates in
spherical coordinate system, as well as by setting the radius vector $%
\boldsymbol{r}$ to be parallel to the $z$-axis, one has%
\begin{align}
\mathcal{F}^{-1}\left[ \frac{\mathrm{i}\boldsymbol{k}}{k^{4}}\right] \left( 
\boldsymbol{r}\right) & =\mathcal{F}^{-1}\left[ \frac{\mathrm{i}\boldsymbol{k%
}}{k^{2}}\right] \left( \boldsymbol{r}\right) \ast _{\boldsymbol{r}}\mathcal{%
F}^{-1}\left[ \frac{1}{k^{2}}\right] \left( \boldsymbol{r}\right)  \notag \\
& =-\frac{\boldsymbol{r}}{4\pi r^{3}}\ast _{\boldsymbol{r}}\frac{1}{4\pi r} 
\notag \\
& =-\frac{1}{16\pi ^{2}}\int_{%
\mathbb{R}
^{3}}\frac{\boldsymbol{r}^{\prime }}{r^{\prime 3}}\frac{1}{\left\vert 
\boldsymbol{r}-\boldsymbol{r}^{\prime }\right\vert }\mathrm{d}_{\boldsymbol{r%
}}V  \notag \\
& =-\frac{1}{16\pi ^{2}}\int_{0}^{2\pi }\!\!\!\int_{0}^{\infty
}\!\!\!\int_{0}^{\pi }\frac{r^{\prime }\sin \theta \cos \varphi \,%
\boldsymbol{e}_{x}+r^{\prime }\sin \theta \sin \varphi \,\boldsymbol{e}%
_{y}+r^{\prime }\cos \theta \,\boldsymbol{e}_{z}}{r^{\prime 3}\sqrt{%
r^{2}-2rr^{\prime }\cos \theta +r^{\prime 2}}}r^{\prime 2}\sin \theta 
\mathrm{d}\theta \mathrm{d}r^{\prime }\mathrm{d}\varphi  \notag \\
& =-\frac{\boldsymbol{e}_{z}}{8\pi }\int_{0}^{\infty }\!\!\!\int_{0}^{\pi }%
\frac{\cos \theta }{\sqrt{r^{2}-2rr^{\prime }\cos \theta +r^{\prime 2}}}\sin
\theta \mathrm{d}\theta \mathrm{d}r^{\prime }  \notag \\
& =-\frac{\boldsymbol{e}_{z}}{32\pi r^{2}}\int_{0}^{\infty }\frac{1}{%
r^{\prime 2}}\int_{\left( r-r^{\prime }\right) ^{2}}^{\left( r+r^{\prime
}\right) ^{2}}\left( \frac{r^{2}+r^{\prime 2}}{\sqrt{u}}-\sqrt{u}\right) 
\mathrm{d}u\mathrm{d}r^{\prime }  \notag \\
& =-\frac{\boldsymbol{e}_{z}}{16\pi r^{2}}\int_{0}^{\infty }\frac{1}{%
r^{\prime 2}}\left( \left( r^{2}+r^{\prime 2}\right) \left( \left\vert
r+r^{\prime }\right\vert -\left\vert r-r^{\prime }\right\vert \right) -\frac{%
1}{3}\left( \left\vert r+r^{\prime }\right\vert ^{3}-\left\vert r-r^{\prime
}\right\vert ^{3}\right) \right) \mathrm{d}r^{\prime }  \notag \\
& =-\frac{\boldsymbol{e}_{z}}{16\pi r^{2}}\left( \int_{0}^{r}\frac{4}{3}%
r^{\prime }\mathrm{d}r^{\prime }-\int_{r}^{\infty }\frac{4}{3}\frac{r^{3}}{%
r^{\prime 2}}\mathrm{d}r^{\prime }\right)  \notag \\
& =-\frac{1}{8\pi }\frac{\boldsymbol{r}}{r},  \label{ik-kroz-k-na-4}
\end{align}%
since $\int_{0}^{2\pi }\sin \varphi \mathrm{d}\varphi =\int_{0}^{2\pi }\cos
\varphi \mathrm{d}\varphi =0$, since the unit vector of $z$-axis is written
as $\boldsymbol{e}_{z}=\frac{\boldsymbol{r}}{r}$, becoming (\ref{ik-kroz-kv})%
$_{2}.$

\section{Calculation of $\mathcal{F}^{-1}\left[ \frac{\boldsymbol{k}\otimes 
\boldsymbol{k}}{k^{4}}\right] \left( \boldsymbol{r}\right) $ \label%
{dijadski-kroz-cetvrti}}

The Fourier transform inversion of tensor function $\frac{\boldsymbol{k}%
\otimes \boldsymbol{k}}{k^{4}}$ is of the form%
\begin{equation}
\mathcal{F}^{-1}\left[ \frac{\boldsymbol{k}\otimes \boldsymbol{k}}{k^{4}}%
\right] \left( \boldsymbol{r}\right) =\frac{1}{8\pi r}\left( \boldsymbol{%
\hat{I}}-\frac{\boldsymbol{r}\otimes \boldsymbol{r}}{r^{2}}\right) .
\label{k-dij-k-kroz-cetvrti}
\end{equation}

Namely, the tensor $\frac{\boldsymbol{k}\otimes \boldsymbol{k}}{k^{4}}$ can
be written in the matrix form as%
\begin{equation}
\frac{\boldsymbol{k}\otimes \boldsymbol{k}}{k^{4}}=\frac{1}{k^{4}}\left[ 
\begin{array}{ccc}
k_{x}^{2} & k_{x}k_{y} & k_{x}k_{z} \\ 
k_{x}k_{y} & k_{y}^{2} & k_{y}k_{z} \\ 
k_{x}k_{z} & k_{y}k_{z} & k_{z}^{2}%
\end{array}%
\right] =\frac{1}{k^{2}}\left[ 
\begin{array}{ccc}
\sin ^{2}\theta \cos ^{2}\varphi & \sin ^{2}\theta \sin \varphi \cos \varphi
& \sin \theta \cos \theta \cos \varphi \\ 
\sin ^{2}\theta \sin \varphi \cos \varphi & \sin ^{2}\theta \sin ^{2}\varphi
& \sin \theta \cos \theta \sin \varphi \\ 
\sin \theta \cos \theta \cos \varphi & \sin \theta \cos \theta \sin \varphi
& \cos ^{2}\theta%
\end{array}%
\right] ,  \label{k-dij-k-kroz-cetvrti-matricno}
\end{equation}%
using Cartesian coordinate system for tensor representation and expressing
tensor components in spherical coordinate system, as well as by setting
radius vector $\boldsymbol{r}$ to be parallel with $z$-axis, so that, its
Fourier transform inversion is achieved by inverting each component of the
matrix, according to the definition of inverse Fourier transform (\ref{IFT}%
), implying%
\begin{align*}
\mathcal{F}^{-1}\left[ \frac{k_{x}^{2}}{k^{4}}\right] \left( \boldsymbol{r}%
\right) & =\frac{1}{\left( 2\pi \right) ^{3}}\int_{%
\mathbb{R}
^{3}}\frac{k_{x}^{2}}{k^{4}}\mathrm{e}^{\mathrm{i}\boldsymbol{k}\cdot 
\boldsymbol{r}}\mathrm{d}_{\boldsymbol{k}}V \\
& =\frac{1}{8\pi ^{3}}\int_{0}^{2\pi }\!\!\!\int_{0}^{\infty
}\!\!\!\int_{0}^{\pi }\frac{\sin ^{2}\theta \cos ^{2}\varphi }{k^{2}}\mathrm{%
e}^{\mathrm{i}kr\cos \theta }k^{2}\sin \theta \mathrm{d}\theta \mathrm{d}k%
\mathrm{d}\varphi \\
& =\frac{1}{8\pi ^{2}}\int_{0}^{\infty }\!\!\!\int_{0}^{\pi }\left( 1-\cos
^{2}\theta \right) \mathrm{e}^{\mathrm{i}kr\cos \theta }\sin \theta \mathrm{d%
}\theta \mathrm{d}k \\
& =\frac{1}{8\pi ^{2}}\int_{0}^{\infty }\!\!\!\int_{-\mathrm{i}kr}^{\mathrm{i%
}kr}\left( 1+\frac{u^{2}}{k^{2}r^{2}}\right) \mathrm{e}^{u}\frac{\mathrm{d}u%
}{\mathrm{i}kr}\mathrm{d}k \\
& =\frac{1}{4\pi ^{2}r^{3}}\int_{0}^{\infty }\frac{1}{k^{3}}\int_{-\mathrm{i}%
kr}^{\mathrm{i}kr}\left( k^{2}r^{2}+u^{2}\right) \frac{\mathrm{e}^{u}}{2%
\mathrm{i}}\mathrm{d}u\mathrm{d}k \\
& =\frac{1}{4\pi ^{2}r^{3}}\int_{0}^{\infty }\frac{-2kr\cos \left( kr\right)
+2\sin \left( kr\right) }{k^{3}}\mathrm{d}k \\
& =\frac{1}{2\pi ^{2}r^{3}}\frac{1}{4}\pi r^{2}\mathrm{sgn\,}r \\
& =\frac{1}{8\pi r},
\end{align*}%
since $\int u^{2}\mathrm{e}^{u}\mathrm{d}u=\left( u^{2}-2u+2\right) \mathrm{e%
}^{u}+\mathrm{const}$ and%
\begin{equation}
\int_{0}^{\infty }\left( \frac{\sin \left( kr\right) }{k^{3}}-\frac{r\cos
\left( kr\right) }{k^{2}}\right) \mathrm{d}k=\frac{1}{4}\pi r^{2}\mathrm{%
sgn\,}r.  \label{inace-divergira-ali-zajedno-ne}
\end{equation}%
Note, when considered separately, both integrals have non-integrable
singularities at zero which cancel when integrals are combined. Similarly,
it can be shown that%
\begin{equation*}
\mathcal{F}^{-1}\left[ \frac{k_{y}^{2}}{k^{4}}\right] \left( \boldsymbol{r}%
\right) =\frac{1}{8\pi r}.
\end{equation*}%
Further, one has%
\begin{align*}
\mathcal{F}^{-1}\left[ \frac{k_{z}^{2}}{k^{4}}\right] \left( \boldsymbol{r}%
\right) & =\frac{1}{\left( 2\pi \right) ^{3}}\int_{%
\mathbb{R}
^{3}}\frac{k_{z}^{2}}{k^{4}}\mathrm{e}^{\mathrm{i}\boldsymbol{k}\cdot 
\boldsymbol{r}}\mathrm{d}_{\boldsymbol{k}}V \\
& =\frac{1}{8\pi ^{3}}\int_{0}^{2\pi }\!\!\!\int_{0}^{\infty
}\!\!\!\int_{0}^{\pi }\frac{\cos ^{2}\theta }{k^{2}}\mathrm{e}^{\mathrm{i}%
kr\cos \theta }k^{2}\sin \theta \mathrm{d}\theta \mathrm{d}k\mathrm{d}\varphi
\\
& =\frac{1}{4\pi ^{2}}\int_{0}^{\infty }\!\!\!\int_{0}^{\pi }\cos ^{2}\theta 
\mathrm{e}^{\mathrm{i}kr\cos \theta }\sin \theta \mathrm{d}\theta \mathrm{d}k
\\
& =-\frac{1}{4\pi ^{2}}\int_{0}^{\infty }\!\!\!\int_{-\mathrm{i}kr}^{\mathrm{%
i}kr}\frac{u^{2}}{k^{2}r^{2}}\mathrm{e}^{u}\frac{\mathrm{d}u}{\mathrm{i}kr}%
\mathrm{d}k \\
& =-\frac{1}{2\pi ^{2}r^{3}}\int_{0}^{\infty }\frac{1}{k^{3}}\int_{-\mathrm{i%
}kr}^{\mathrm{i}kr}u^{2}\frac{\mathrm{e}^{u}}{2\mathrm{i}}\mathrm{d}u\mathrm{%
d}k \\
& =\frac{1}{2\pi ^{2}r^{3}}\int_{0}^{\infty }\frac{-k^{2}r^{2}\sin \left(
kr\right) -2kr\cos \left( kr\right) +2\sin \left( kr\right) }{k^{3}}\mathrm{d%
}k \\
& =-\frac{1}{2\pi ^{2}r}\int_{0}^{\infty }\frac{\sin \left( kr\right) }{k}%
\mathrm{d}k+\frac{1}{\pi ^{2}r^{3}}\int_{0}^{\infty }\frac{kr\cos \left(
kr\right) -\sin \left( kr\right) }{k^{3}}\mathrm{d}k \\
& =-\frac{1}{2\pi ^{2}r}\frac{\pi }{2}\mathrm{sgn\,}r+\frac{1}{\pi ^{2}r^{3}}%
\frac{1}{4}\pi r^{2}\mathrm{sgn\,}r \\
& =0,
\end{align*}%
since $\int u^{2}\mathrm{e}^{u}\mathrm{d}u=\left( u^{2}-2u+2\right) \mathrm{e%
}^{u}+\mathrm{const}$ and according to (\ref{signum-i-mignum})$_{2}$ and (%
\ref{inace-divergira-ali-zajedno-ne}). The inversion of all other tensor
components also results in their zero contribution, i.e.,%
\begin{align*}
\mathcal{F}^{-1}\left[ \frac{k_{x}k_{y}}{k^{4}}\right] \left( \boldsymbol{r}%
\right) & =\frac{1}{\left( 2\pi \right) ^{3}}\int_{%
\mathbb{R}
^{3}}\frac{k_{x}k_{y}}{k^{4}}\mathrm{e}^{\mathrm{i}\boldsymbol{k}\cdot 
\boldsymbol{r}}\mathrm{d}_{\boldsymbol{k}}V \\
& =\frac{1}{8\pi ^{3}}\int_{0}^{2\pi }\!\!\!\int_{0}^{\infty
}\!\!\!\int_{0}^{\pi }\frac{\sin ^{2}\theta \sin \varphi \cos \varphi }{k^{2}%
}\mathrm{e}^{\mathrm{i}kr\cos \theta }k^{2}\sin \theta \mathrm{d}\theta 
\mathrm{d}k\mathrm{d}\varphi \\
& =0,
\end{align*}%
due to $\int_{0}^{2\pi }\sin \varphi \cos \varphi \mathrm{d}\varphi =0$, and
similarly, it can be shown that%
\begin{equation*}
\mathcal{F}^{-1}\left[ \frac{k_{x}k_{z}}{k^{4}}\right] \left( \boldsymbol{r}%
\right) =\mathcal{F}^{-1}\left[ \frac{k_{y}k_{z}}{k^{4}}\right] \left( 
\boldsymbol{r}\right) =0.
\end{equation*}

Therefore, by substituting all previously calculated inversions into (\ref%
{k-dij-k-kroz-cetvrti-matricno}), one has%
\begin{equation*}
\mathcal{F}^{-1}\left[ \frac{\boldsymbol{k}\otimes \boldsymbol{k}}{k^{4}}%
\right] \left( \boldsymbol{r}\right) =\frac{1}{8\pi r}\left[ 
\begin{array}{ccc}
1 & 0 & 0 \\ 
0 & 1 & 0 \\ 
0 & 0 & 0%
\end{array}%
\right] =\frac{1}{8\pi r}\left( \boldsymbol{\hat{I}}-\frac{\boldsymbol{r}%
\otimes \boldsymbol{r}}{r^{2}}\right) ,
\end{equation*}%
since the unit vector of $z$-axis is written as $\boldsymbol{e}_{z}=\frac{%
\boldsymbol{r}}{r}$, thus obtaining (\ref{k-dij-k-kroz-cetvrti}).

\section{Calculation of $\mathcal{F}^{-1}\left[ \frac{\boldsymbol{k}\otimes 
\boldsymbol{k}}{k^{2}}\right] \left( \boldsymbol{r}\right)$ \label%
{dijadski-kroz-kvadrat}}

The Fourier transform inversion of tensor function $\frac{\boldsymbol{k}%
\otimes \boldsymbol{k}}{k^{2}}$ is of the form%
\begin{equation}
\mathcal{F}^{-1}\left[ \frac{\boldsymbol{k}\otimes \boldsymbol{k}}{k^{2}}%
\right] \left( \boldsymbol{r}\right) =-\delta \left( \boldsymbol{r}\right)
\left( \boldsymbol{\hat{I}}-4\frac{\boldsymbol{r}\otimes \boldsymbol{r}}{%
r^{2}}\right) +\frac{1}{4\pi r^{3}}\left( \boldsymbol{\hat{I}}-3\frac{%
\boldsymbol{r}\otimes \boldsymbol{r}}{r^{2}}\right) .
\label{k-dij-k-kroz-drugi}
\end{equation}

Namely, the tensor $\frac{\boldsymbol{k}\otimes \boldsymbol{k}}{k^{2}}$ can
be written in the matrix form as%
\begin{equation}
\frac{\boldsymbol{k}\otimes \boldsymbol{k}}{k^{2}}=\frac{1}{k^{2}}\left[ 
\begin{array}{ccc}
k_{x}^{2} & k_{x}k_{y} & k_{x}k_{z} \\ 
k_{x}k_{y} & k_{y}^{2} & k_{y}k_{z} \\ 
k_{x}k_{z} & k_{y}k_{z} & k_{z}^{2}%
\end{array}%
\right] =\left[ 
\begin{array}{ccc}
\sin ^{2}\theta \cos ^{2}\varphi & \sin ^{2}\theta \sin \varphi \cos \varphi
& \sin \theta \cos \theta \cos \varphi \\ 
\sin ^{2}\theta \sin \varphi \cos \varphi & \sin ^{2}\theta \sin ^{2}\varphi
& \sin \theta \cos \theta \sin \varphi \\ 
\sin \theta \cos \theta \cos \varphi & \sin \theta \cos \theta \sin \varphi
& \cos ^{2}\theta%
\end{array}%
\right] ,  \label{k-dij-k-kroz-drugii-matricno}
\end{equation}%
using Cartesian coordinate system for tensor representation and expressing
tensor components in spherical coordinate system, as well as by setting
radius vector $\boldsymbol{r}$ to be parallel with $z$-axis, so that, its
Fourier transform inversion is achieved by inverting each component of the
matrix, according to the definition of inverse Fourier transform (\ref{IFT}%
), implying%
\begin{align*}
\mathcal{F}^{-1}\left[ \frac{k_{x}^{2}}{k^{2}}\right] \left( \boldsymbol{r}%
\right) & =\frac{1}{\left( 2\pi \right) ^{3}}\int_{%
\mathbb{R}
^{3}}\frac{k_{x}^{2}}{k^{2}}\mathrm{e}^{\mathrm{i}\boldsymbol{k}\cdot 
\boldsymbol{r}}\mathrm{d}_{\boldsymbol{k}}V \\
& =\frac{1}{8\pi ^{3}}\int_{0}^{2\pi }\!\!\!\int_{0}^{\infty
}\!\!\!\int_{0}^{\pi }\sin ^{2}\theta \cos ^{2}\varphi \mathrm{e}^{\mathrm{i}%
kr\cos \theta }k^{2}\sin \theta \mathrm{d}\theta \mathrm{d}k\mathrm{d}\varphi
\\
& =\frac{1}{8\pi ^{2}}\int_{0}^{\infty }\!\!\!\int_{0}^{\pi }k^{2}\left(
1-\cos ^{2}\theta \right) \mathrm{e}^{\mathrm{i}kr\cos \theta }\sin \theta 
\mathrm{d}\theta \mathrm{d}k \\
& =\frac{1}{8\pi ^{2}}\int_{0}^{\infty }\!\!\!\int_{-\mathrm{i}kr}^{\mathrm{i%
}kr}k^{2}\left( 1+\frac{u^{2}}{k^{2}r^{2}}\right) \mathrm{e}^{u}\frac{%
\mathrm{d}u}{\mathrm{i}kr}\mathrm{d}k \\
& =\frac{1}{4\pi ^{2}r^{3}}\int_{0}^{\infty }\frac{1}{k}\int_{-\mathrm{i}%
kr}^{\mathrm{i}kr}\left( k^{2}r^{2}+u^{2}\right) \frac{\mathrm{e}^{u}}{2%
\mathrm{i}}\mathrm{d}u\mathrm{d}k \\
& =\frac{1}{4\pi ^{2}r^{3}}\int_{0}^{\infty }\frac{-2kr\cos \left( kr\right)
+2\sin \left( kr\right) }{k}\mathrm{d}k \\
& =-\frac{1}{2\pi r^{2}}\frac{1}{\pi }\int_{0}^{\infty }\cos \left(
kr\right) \mathrm{d}k+\frac{1}{2\pi ^{2}r^{3}}\int_{0}^{\infty }\frac{\sin
\left( kr\right) }{k}\mathrm{d}k \\
& =-\frac{1}{2\pi r^{2}}\delta \left( r\right) +\frac{1}{2\pi ^{2}r^{3}}%
\frac{\pi }{2}\mathrm{sgn\,}r \\
& =-\delta \left( \boldsymbol{r}\right) +\frac{1}{4\pi r^{3}}\mathrm{sgn\,}r,
\end{align*}%
since $\int u^{2}\mathrm{e}^{u}\mathrm{d}u=\left( u^{2}-2u+2\right) \mathrm{e%
}^{u}+\mathrm{const}$ and according to (\ref{signum-i-mignum}). Similarly,
it can be shown that%
\begin{equation*}
\mathcal{F}^{-1}\left[ \frac{k_{y}^{2}}{k^{2}}\right] \left( \boldsymbol{r}%
\right) =-\delta \left( \boldsymbol{r}\right) +\frac{1}{4\pi r^{3}}\mathrm{%
sgn\,}r.
\end{equation*}%
Further, one has%
\begin{align*}
\mathcal{F}^{-1}\left[ \frac{k_{z}^{2}}{k^{2}}\right] \left( \boldsymbol{r}%
\right) & =\frac{1}{\left( 2\pi \right) ^{3}}\int_{%
\mathbb{R}
^{3}}\frac{k_{z}^{2}}{k^{2}}\mathrm{e}^{\mathrm{i}\boldsymbol{k}\cdot 
\boldsymbol{r}}\mathrm{d}_{\boldsymbol{k}}V \\
& =\frac{1}{8\pi ^{3}}\int_{0}^{2\pi }\!\!\!\int_{0}^{\infty
}\!\!\!\int_{0}^{\pi }\cos ^{2}\theta \mathrm{e}^{\mathrm{i}kr\cos \theta
}k^{2}\sin \theta \mathrm{d}\theta \mathrm{d}k\mathrm{d}\varphi \\
& =\frac{1}{4\pi ^{2}}\int_{0}^{\infty }\!\!\!\int_{0}^{\pi }k^{2}\cos
^{2}\theta \mathrm{e}^{\mathrm{i}kr\cos \theta }\sin \theta \mathrm{d}\theta 
\mathrm{d}k \\
& =-\frac{1}{4\pi ^{2}}\int_{0}^{\infty }\!\!\!\int_{-\mathrm{i}kr}^{\mathrm{%
i}kr}k^{2}\frac{u^{2}}{k^{2}r^{2}}\mathrm{e}^{u}\frac{\mathrm{d}u}{\mathrm{i}%
kr}\mathrm{d}k \\
& =-\frac{1}{2\pi ^{2}r^{3}}\int_{0}^{\infty }\frac{1}{k}\int_{-\mathrm{i}%
kr}^{\mathrm{i}kr}u^{2}\frac{\mathrm{e}^{u}}{2\mathrm{i}}\mathrm{d}u\mathrm{d%
}k \\
& =\frac{1}{2\pi ^{2}r^{3}}\int_{0}^{\infty }\frac{k^{2}r^{2}\sin \left(
kr\right) +2kr\cos \left( kr\right) -2}{k}\mathrm{d}k \\
& =\frac{1}{2\pi ^{2}r}\int_{0}^{\infty }k\sin \left( kr\right) \mathrm{d}k+%
\frac{1}{\pi r^{2}}\frac{1}{\pi }\int_{0}^{\infty }\cos \left( kr\right) 
\mathrm{d}k-\frac{1}{\pi ^{2}r^{3}}\int_{0}^{\infty }\frac{\sin \left(
kr\right) }{k}\mathrm{d}k \\
& =-\frac{1}{2\pi r}\frac{\mathrm{d}}{\mathrm{d}r}\left( \frac{1}{\pi }%
\int_{0}^{\infty }\cos \left( kr\right) \mathrm{d}k\right) +\frac{1}{\pi
r^{2}}\frac{1}{\pi }\int_{0}^{\infty }\cos \left( kr\right) \mathrm{d}k-%
\frac{1}{\pi ^{2}r^{3}}\int_{0}^{\infty }\frac{\sin \left( kr\right) }{k}%
\mathrm{d}k \\
& =-\frac{1}{2\pi r}\frac{\mathrm{d}}{\mathrm{d}r}\delta \left( r\right) +%
\frac{1}{\pi r^{2}}\delta \left( r\right) -\frac{1}{\pi ^{2}r^{3}}\frac{\pi 
}{2}\mathrm{sgn\,}r \\
& =\frac{1}{2\pi r^{2}}\delta \left( r\right) +\frac{1}{\pi r^{2}}\delta
\left( r\right) -\frac{1}{2\pi r^{3}}\mathrm{sgn\,}r \\
& =\delta \left( \boldsymbol{r}\right) +2\delta \left( \boldsymbol{r}\right)
-\frac{1}{2\pi r^{3}}\mathrm{sgn\,}r \\
& =3\delta \left( \boldsymbol{r}\right) -\frac{1}{2\pi r^{3}}\mathrm{sgn\,}r,
\end{align*}%
since $\int u^{2}\mathrm{e}^{u}\mathrm{d}u=\left( u^{2}-2u+2\right) \mathrm{e%
}^{u}+\mathrm{const}$ and according to (\ref{signum-i-mignum}), with $r\frac{%
\mathrm{d}}{\mathrm{d}r}\delta \left( r\right) =-\delta \left( r\right) .$
On the other hand, the inversion of all other tensor components results in
their zero contributions, i.e.,%
\begin{align*}
\mathcal{F}^{-1}\left[ \frac{k_{x}k_{y}}{k^{2}}\right] \left( \boldsymbol{r}%
\right) & =\frac{1}{\left( 2\pi \right) ^{3}}\int_{%
\mathbb{R}
^{3}}\frac{k_{x}k_{y}}{k^{2}}\mathrm{e}^{\mathrm{i}\boldsymbol{k}\cdot 
\boldsymbol{r}}\mathrm{d}_{\boldsymbol{k}}V \\
& =\frac{1}{8\pi ^{3}}\int_{0}^{2\pi }\!\!\!\int_{0}^{\infty
}\!\!\!\int_{0}^{\pi }\sin ^{2}\theta \sin \varphi \cos \varphi \mathrm{e}^{%
\mathrm{i}kr\cos \theta }k^{2}\sin \theta \mathrm{d}\theta \mathrm{d}k%
\mathrm{d}\varphi \\
& =0,
\end{align*}%
due to $\int_{0}^{2\pi }\sin \varphi \cos \varphi \mathrm{d}\varphi =0$ and
similarly, it can be shown that%
\begin{equation*}
\mathcal{F}^{-1}\left[ \frac{k_{x}k_{z}}{k^{2}}\right] \left( \boldsymbol{r}%
\right) =\mathcal{F}^{-1}\left[ \frac{k_{y}k_{z}}{k^{2}}\right] \left( 
\boldsymbol{r}\right) =0.
\end{equation*}

Substitution of all previously calculated inversions into (\ref%
{k-dij-k-kroz-drugii-matricno}) yields%
\begin{align*}
\mathcal{F}^{-1}\left[ \frac{\boldsymbol{k}\otimes \boldsymbol{k}}{k^{2}}%
\right] \left( \boldsymbol{r}\right) & =\left[ 
\begin{array}{ccc}
-\delta \left( \boldsymbol{r}\right) +\frac{1}{4\pi r^{3}} & 0 & 0 \\ 
0 & -\delta \left( \boldsymbol{r}\right) +\frac{1}{4\pi r^{3}} & 0 \\ 
0 & 0 & 3\delta \left( \boldsymbol{r}\right) -\frac{1}{2\pi r^{3}}%
\end{array}%
\right] \\
& =\left( -\delta \left( \boldsymbol{r}\right) +\frac{1}{4\pi r^{3}}\right) 
\boldsymbol{\hat{I}}+\left( 4\delta \left( \boldsymbol{r}\right) -\frac{3}{%
4\pi r^{3}}\right) \frac{\boldsymbol{r}\otimes \boldsymbol{r}}{r^{2}} \\
& =-\delta \left( \boldsymbol{r}\right) \left( \boldsymbol{\hat{I}}-4\frac{%
\boldsymbol{r}\otimes \boldsymbol{r}}{r^{2}}\right) +\frac{1}{4\pi r^{3}}%
\left( \boldsymbol{\hat{I}}-3\frac{\boldsymbol{r}\otimes \boldsymbol{r}}{%
r^{2}}\right) ,
\end{align*}%
since the unit vector of $z$-axis is written as $\boldsymbol{e}_{z}=\frac{%
\boldsymbol{r}}{r},$ taking the form (\ref{k-dij-k-kroz-drugi}).

\section{Calculation of $\mathcal{F}^{-1}\left[ \frac{\left( \boldsymbol{L}%
\cdot \boldsymbol{k}\right) ^{2}}{k^{2}}\right] \left( \boldsymbol{r}\right) 
$ \label{skalarni-kroz-kvadrat}}

The Fourier transform inversion of scalar function $\frac{\left( \boldsymbol{%
L}\cdot \boldsymbol{k}\right) ^{2}}{k^{2}}$ is of the form%
\begin{equation}
\mathcal{F}^{-1}\left[ \frac{\left( \boldsymbol{L}\cdot \boldsymbol{k}%
\right) ^{2}}{k^{2}}\right] \left( \boldsymbol{r}\right) =L^{2}\delta \left( 
\boldsymbol{r}\right) +\frac{L^{2}}{4\pi r^{3}}\left( 1-3\left( \frac{%
\boldsymbol{L}}{L}\cdot \frac{\boldsymbol{r}}{r}\right) ^{2}\right) .
\label{skalarni-kroz-kv}
\end{equation}

Namely, by writing $\boldsymbol{k}=k_{\parallel }\frac{\boldsymbol{L}}{L}+%
\boldsymbol{k}_{\perp }=k_{\parallel }\frac{\boldsymbol{L}}{L}+k_{\perp }%
\boldsymbol{e}_{k_{\perp }}$ and $\boldsymbol{r}=r_{\parallel }\frac{%
\boldsymbol{L}}{L}+\boldsymbol{r}_{\perp }$, where $k_{\parallel
},r_{\parallel }\in \mathbb{R}$ are coordinates, while $k_{\perp }=|{%
\boldsymbol{k}_{\perp }}|\geqslant 0$ is a vector intensity, so that $%
\boldsymbol{k}\cdot \boldsymbol{r}=k_{\parallel }r_{\parallel }+\boldsymbol{k%
}_{\perp }\cdot \boldsymbol{r}_{\perp }=k_{\parallel }r_{\parallel
}+k_{\perp }r_{\perp }\cos \varphi $, where $r_{\perp }=|{\boldsymbol{r}%
_{\perp }}|\geqslant 0$ is the vector intensity and where the angle $\varphi 
$, taken to increase from the line determined by the direction of vector ${%
\boldsymbol{r}_{\perp }}$, is a coordinate in the polar coordinate system
determined by the unit vectors $\boldsymbol{e}_{k_{\perp }}$ and $%
\boldsymbol{e}_{\varphi }$, so that according to the definition of inverse
Fourier transform (\ref{IFT}), one has%
\begin{align}
\mathcal{F}^{-1}& \left[ \frac{\left( \boldsymbol{L}\cdot \boldsymbol{k}%
\right) ^{2}}{k^{2}}\right] \left( \boldsymbol{r}\right) =\frac{1}{\left(
2\pi \right) ^{3}}\int_{%
\mathbb{R}
^{3}}\frac{\left( \boldsymbol{L}\cdot \boldsymbol{k}\right) ^{2}}{k^{2}}%
\mathrm{e}^{\mathrm{i}\boldsymbol{k}\cdot \boldsymbol{r}}\mathrm{d}_{%
\boldsymbol{k}}V  \notag \\
& =\frac{L^{2}}{\left( 2\pi \right) ^{3}}\int_{%
\mathbb{R}
^{2}}\int_{%
\mathbb{R}
}\frac{k_{\parallel }^{2}}{k_{\parallel }^{2}+\left\vert \boldsymbol{k}%
_{\perp }\right\vert ^{2}}\mathrm{e}^{\mathrm{i}k_{\boldsymbol{\parallel }%
}r_{\boldsymbol{\parallel }}}\mathrm{e}^{\mathrm{i}\boldsymbol{k}_{\perp
}\cdot \boldsymbol{r}_{\perp }}\mathrm{d}k_{\parallel }\,\mathrm{d}_{%
\boldsymbol{k}_{\perp }}S  \notag \\
& =\frac{L^{2}}{\left( 2\pi \right) ^{3}}\int_{%
\mathbb{R}
^{3}}\mathrm{e}^{\mathrm{i}\boldsymbol{k}\cdot \boldsymbol{r}}\mathrm{d}_{%
\boldsymbol{k}}V-\frac{L^{2}}{4\pi ^{2}}\int_{%
\mathbb{R}
^{2}}\left\vert \boldsymbol{k}_{\perp }\right\vert ^{2}\left( \frac{1}{2\pi }%
\int_{%
\mathbb{R}
}\frac{1}{k_{\parallel }^{2}+\left\vert \boldsymbol{k}_{\perp }\right\vert
^{2}}\mathrm{e}^{\mathrm{i}k_{\boldsymbol{\parallel }}r_{\boldsymbol{%
\parallel }}}\mathrm{d}k_{\parallel }\right) \mathrm{e}^{\mathrm{i}%
\boldsymbol{k}_{\perp }\cdot \boldsymbol{r}_{\perp }}\,\mathrm{d}_{%
\boldsymbol{k}_{\perp }}S  \label{C-1} \\
& =L^{2}\delta \left( \boldsymbol{r}\right) -\frac{L^{2}}{8\pi ^{2}}\int_{%
\mathbb{R}
^{2}}\left\vert \boldsymbol{k}_{\perp }\right\vert \mathrm{e}^{-\left\vert
r_{\boldsymbol{\parallel }}\right\vert \sqrt{\left\vert \boldsymbol{k}%
_{\perp }\right\vert ^{2}}}\mathrm{e}^{\mathrm{i}\boldsymbol{k}_{\perp
}\cdot \boldsymbol{r}_{\perp }}\,\mathrm{d}_{\boldsymbol{k}_{\perp }}S 
\notag \\
& =L^{2}\delta \left( \boldsymbol{r}\right) -\frac{L^{2}}{8\pi ^{2}}%
\int_{0}^{2\pi }\!\!\!\int_{0}^{\infty }k_{\perp }\mathrm{e}^{-k_{\perp
}\left( \left\vert r_{\boldsymbol{\parallel }}\right\vert -\mathrm{i}%
r_{\perp }\cos \varphi \right) }\,k_{\perp }\mathrm{d}k_{\perp }\mathrm{d}%
\varphi  \label{C-2} \\
& =L^{2}\delta \left( \boldsymbol{r}\right) -\frac{L^{2}}{4\pi ^{2}}%
\int_{0}^{2\pi }\frac{1}{\left( \left\vert r_{\boldsymbol{\parallel }%
}\right\vert -\mathrm{i}r_{\perp }\cos \varphi \right) ^{3}}\mathrm{d}\varphi
\notag \\
& =L^{2}\delta \left( \boldsymbol{r}\right) -\frac{L^{2}}{4\pi ^{2}}\left( 
\frac{1}{2}\frac{\partial ^{2}}{\partial \left\vert r_{\boldsymbol{\parallel 
}}\right\vert ^{2}}\int_{0}^{2\pi }\frac{1}{\left\vert r_{\boldsymbol{%
\parallel }}\right\vert -\mathrm{i}r_{\perp }\cos \varphi }\mathrm{d}\varphi
\right)  \label{C-3} \\
& =L^{2}\delta \left( \boldsymbol{r}\right) -\frac{L^{2}}{4\pi ^{2}}\left(
-\pi \frac{r_{\perp }^{2}-2r_{\boldsymbol{\parallel }}^{2}}{\sqrt{r_{%
\boldsymbol{\parallel }}^{2}+r_{\perp }^{2}}^{5}}\right)  \notag \\
& =L^{2}\delta \left( \boldsymbol{r}\right) +\frac{L^{2}}{4\pi }\frac{%
r^{2}-3r_{\boldsymbol{\parallel }}^{2}}{r^{5}}  \notag \\
& =L^{2}\delta \left( \boldsymbol{r}\right) +\frac{L^{2}}{4\pi r^{3}}\left(
1-3\left( \frac{\boldsymbol{L}}{L}\cdot \frac{\boldsymbol{r}}{r}\right)
^{2}\right) ,  \notag
\end{align}%
being (\ref{skalarni-kroz-kv}), due to $r_{\boldsymbol{\parallel }}=%
\boldsymbol{r}\cdot \frac{\boldsymbol{L}}{L}$, since%
\begin{gather}
\mathcal{F}^{-1}\left[ \frac{1}{k^{2}+\lambda }\right] \left( x\right) =%
\frac{1}{2\pi }\int_{%
\mathbb{R}
}\frac{1}{k^{2}+\lambda }\mathrm{e}^{\mathrm{i}kx}\mathrm{d}k=\frac{1}{2%
\sqrt{\lambda }}\mathrm{e}^{-\left\vert x\right\vert \sqrt{\lambda }},\quad 
\text{for\quad }\func{Re}\lambda >0,  \label{C-4} \\
\int_{0}^{\infty }u^{2}\mathrm{e}^{-\alpha u}\,\mathrm{d}u=\frac{2}{\alpha
^{3}},\quad \text{for\quad }\func{Re}\alpha >0,  \label{C-5} \\
\int_{0}^{2\pi }\frac{1}{\alpha +\mathrm{i}\beta \cos \varphi }\mathrm{d}%
\varphi =2\pi \frac{1}{\sqrt{\alpha ^{2}+\beta ^{2}}},\quad \text{for\quad }%
\alpha >0\quad \text{and\quad }\beta \in 
\mathbb{R}
\label{C-6}
\end{gather}%
is used in (\ref{C-1}), (\ref{C-2}), and (\ref{C-3}), respectively. Note,
the integral in (\ref{C-4}) is the standard Fourier transform inversion,
that can be calculated by the integration in complex plane along the
suitable contour, while the integral in (\ref{C-6}) is calculated by the
integration in complex plane along the unit circle.

\section{Calculation of $\mathcal{F}^{-1}\left[ \frac{\left( \boldsymbol{L}%
\cdot \boldsymbol{k}\right) ^{2}}{k^{4}}\mathrm{i}\boldsymbol{k}\right]
\left( \boldsymbol{r}\right) $ \label{skalarni-kroz-kvadrat-ik}}

The Fourier transform inversion of vector function $\frac{\left( \boldsymbol{%
L}\cdot \boldsymbol{k}\right) ^{2}}{k^{2}}\mathrm{i}\boldsymbol{k}$ is of
the form%
\begin{equation}
\mathcal{F}^{-1}\left[ \frac{\left( \boldsymbol{L}\cdot \boldsymbol{k}%
\right) ^{2}}{k^{4}}\mathrm{i}\boldsymbol{k}\right] \left( \boldsymbol{r}%
\right) =-\frac{1}{4\pi }\left( \frac{\boldsymbol{L}}{L}\cdot \frac{%
\boldsymbol{r}}{r}\right) \frac{L^{2}}{r^{2}}\frac{\boldsymbol{L}}{L}-\frac{1%
}{8\pi }\left( 1-3\left( \frac{\boldsymbol{L}}{L}\cdot \frac{\boldsymbol{r}}{%
r}\right) ^{2}\right) \frac{L^{2}}{r^{2}}\frac{\boldsymbol{r}}{r}.
\label{izraz}
\end{equation}

Namely, by choosing the Cartesian coordinate system such that its $z$-axis
coincides with the non-locality vector $\boldsymbol{L},$ i.e., $\boldsymbol{L%
}=L\boldsymbol{e}_{z},$ and by choosing, for the sake of simplicity,
decomposition of the radius vector as $\boldsymbol{r}=r_{\parallel }%
\boldsymbol{e}_{z}+r_{\perp }\boldsymbol{e}_{x},$ for the vector $%
\boldsymbol{k}$ one has $\boldsymbol{k}=k_{\parallel }\boldsymbol{e}_{z}+%
\boldsymbol{k}_{\perp }=k_{\perp }\cos \varphi \boldsymbol{e}_{x}+k_{\perp
}\sin \varphi \boldsymbol{e}_{y}+k_{\parallel }\boldsymbol{e}_{z}$, i.e., $%
\left( k_{x},k_{y},k_{z}\right) =\left( k_{\perp }\cos \varphi ,k_{\perp
}\sin \varphi ,k_{\parallel }\right) $, where $k_{\parallel }\in 
\mathbb{R}
$ is a coordinate and $k_{\perp }=|{\boldsymbol{k}_{\perp }}|\geqslant 0$ is
the vector intensity, thus the scalar product $\boldsymbol{k}\cdot 
\boldsymbol{r}=k_{\parallel }r_{\parallel }+k_{\perp }r_{\perp }\cos \varphi 
$, so that, according to the definition of inverse Fourier transform (\ref%
{IFT}), one has 
\begin{align}
\mathcal{F}^{-1}& \left[ \frac{\left( \boldsymbol{L}\cdot \boldsymbol{k}%
\right) ^{2}}{k^{4}}\mathrm{i}k_{x}\right] \left( \boldsymbol{r}\right) =%
\frac{1}{\left( 2\pi \right) ^{3}}\int_{%
\mathbb{R}
^{3}}\frac{\left( \boldsymbol{L}\cdot \boldsymbol{k}\right) ^{2}}{k^{4}}%
\mathrm{i}k_{x}\mathrm{e}^{\mathrm{i}\boldsymbol{k}\cdot \boldsymbol{r}}%
\mathrm{d}_{\boldsymbol{k}}V  \notag \\
& =\frac{\mathrm{i}L^{2}}{8\pi ^{3}}\int_{0}^{2\pi }\!\!\!\int_{0}^{\infty
}\!\!\!\int_{-\infty }^{\infty }\frac{k_{\parallel }^{2}k_{\perp }\cos
\varphi }{\left( k_{\parallel }^{2}+k_{\perp }^{2}\right) ^{2}}\mathrm{e}^{%
\mathrm{i}k_{\boldsymbol{\parallel }}r_{\boldsymbol{\parallel }}}\mathrm{e}^{%
\mathrm{i}k_{\perp }r_{\perp }\cos \varphi }\mathrm{d}k_{\parallel
}\,k_{\perp }\mathrm{d}k_{\perp }\mathrm{d}\varphi  \label{E-1} \\
& =\frac{\mathrm{i}L^{2}}{16\pi ^{2}}\int_{0}^{2\pi }\!\!\!\int_{0}^{\infty
}k_{\perp }\mathrm{e}^{-k_{\perp }\left( \left\vert r_{\boldsymbol{\parallel 
}}\right\vert -\mathrm{i}r_{\perp }\cos \varphi \right) }\cos \varphi \,%
\mathrm{d}k_{\perp }\mathrm{d}\varphi -\frac{\mathrm{i}L^{2}}{16\pi ^{2}}%
\int_{0}^{2\pi }\!\!\!\int_{0}^{\infty }\left\vert r_{\boldsymbol{\parallel }%
}\right\vert k_{\perp }^{2}\mathrm{e}^{-k_{\perp }\left( \left\vert r_{%
\boldsymbol{\parallel }}\right\vert -\mathrm{i}r_{\perp }\cos \varphi
\right) }\cos \varphi \,\mathrm{d}k_{\perp }\mathrm{d}\varphi  \label{E-3} \\
& =\frac{\mathrm{i}L^{2}}{16\pi ^{2}}\int_{0}^{2\pi }\frac{1}{\left(
\left\vert r_{\boldsymbol{\parallel }}\right\vert -\mathrm{i}r_{\perp }\cos
\varphi \right) ^{2}}\cos \varphi \,\mathrm{d}\varphi -\frac{\mathrm{i}L^{2}%
}{16\pi ^{2}}\int_{0}^{2\pi }\frac{2\left\vert r_{\boldsymbol{\parallel }%
}\right\vert }{\left( \left\vert r_{\boldsymbol{\parallel }}\right\vert -%
\mathrm{i}r_{\perp }\cos \varphi \right) ^{3}}\cos \varphi \,\mathrm{d}%
\varphi  \notag \\
& =-\frac{L^{2}}{16\pi ^{2}}\frac{\partial }{\partial r_{\perp }}%
\int_{0}^{2\pi }\frac{1}{\left\vert r_{\boldsymbol{\parallel }}\right\vert -%
\mathrm{i}r_{\perp }\cos \varphi }\mathrm{d}\varphi -\frac{L^{2}}{16\pi ^{2}}%
\left\vert r_{\boldsymbol{\parallel }}\right\vert \frac{\partial ^{2}}{%
\partial r_{\perp }\partial \left\vert r_{\boldsymbol{\parallel }%
}\right\vert }\int_{0}^{2\pi }\frac{1}{\left\vert r_{\boldsymbol{\parallel }%
}\right\vert -\mathrm{i}r_{\perp }\cos \varphi }\mathrm{d}\varphi
\label{E-5} \\
& =-\frac{L^{2}}{16\pi ^{2}}\left( -2\pi \frac{r_{\perp }}{\sqrt{r_{%
\boldsymbol{\parallel }}^{2}-r_{\perp }^{2}}^{3}}+6\pi \left\vert r_{%
\boldsymbol{\parallel }}\right\vert \frac{\left\vert r_{\boldsymbol{%
\parallel }}\right\vert r_{\perp }}{\sqrt{r_{\boldsymbol{\parallel }%
}^{2}-r_{\perp }^{2}}^{5}}\right)  \notag \\
& =-\frac{L^{2}}{8\pi r^{3}}r_{\perp }\left( 1-3\frac{r_{\boldsymbol{%
\parallel }}^{2}}{r^{2}}\right) ,  \notag
\end{align}%
where in (\ref{E-1}) the integral%
\begin{equation}
\int_{-\infty }^{\infty }\frac{x^{2}}{\left( x^{2}+\lambda ^{2}\right) ^{2}}%
\mathrm{e}^{\mathrm{i}\alpha x}\mathrm{d}x=2\int_{0}^{\infty }\frac{%
x^{2}\cos \left( \alpha x\right) }{\left( x^{2}+\lambda ^{2}\right) ^{2}}%
\mathrm{d}x=-2\frac{\partial ^{2}}{\partial \left\vert \alpha \right\vert
^{2}}\int_{0}^{\infty }\frac{\cos \left( \alpha x\right) }{\left(
x^{2}+\lambda ^{2}\right) ^{2}}\mathrm{d}x=\frac{\pi }{2}\frac{1}{\lambda }%
\left( 1-\left\vert \alpha \right\vert \lambda \right) \mathrm{e}%
^{-\left\vert \alpha \right\vert \lambda }  \label{D-9}
\end{equation}%
is used, which relies on the following integral%
\begin{equation}
\int_{0}^{\infty }\frac{\cos \left( \alpha x\right) }{\left( x^{2}+\lambda
^{2}\right) ^{2}}\mathrm{d}x=\frac{\pi }{4\lambda ^{3}}\left( 1+\left\vert
\alpha \right\vert \lambda \right) \mathrm{e}^{-\left\vert \alpha
\right\vert \lambda },\quad \text{for\quad }\lambda >0\quad \text{and\quad }%
\alpha \in 
\mathbb{R}
,  \label{D-8}
\end{equation}%
further in (\ref{E-3}) integrals 
\begin{equation}
\int_{0}^{\infty }u\,\mathrm{e}^{-\alpha u}\,\mathrm{d}u=\frac{1}{\alpha ^{2}%
},\quad \text{for\quad }\func{Re}\alpha >0,  \label{DD-10}
\end{equation}%
and (\ref{C-5}) are used, and finally in (\ref{E-5}) the integral (\ref{C-6}%
) is used. Note, the integral (\ref{D-8}) is calculated using Cauchy
residues theorem $\oint_{\Gamma }\frac{\mathrm{e}^{\mathrm{i}\alpha z}}{%
\left( z^{2}+\lambda ^{2}\right) ^{2}}\mathrm{d}z=2\pi \mathrm{i}\func{Res}%
\left[ \frac{\mathrm{e}^{\mathrm{i}\alpha z}}{\left( z^{2}+\lambda
^{2}\right) ^{2}},\mathrm{i}\lambda \right] $, where contour $\Gamma $ is a
semi-circle, lying in the upper complex half-plane, with the diameter
coinciding with the axis $\func{Re}z$.

Inversion of another vector component results in a zero contribution, i.e.,%
\begin{align}
\mathcal{F}^{-1} & \left[ \frac{\left( \boldsymbol{L}\cdot \boldsymbol{k}%
\right) ^{2}}{k^{4}}\mathrm{i}k_{y}\right] \left( \boldsymbol{r}\right) =%
\frac{1}{\left( 2\pi \right) ^{3}}\int_{%
\mathbb{R}
^{3}}\frac{\left( \boldsymbol{L}\cdot \boldsymbol{k}\right) ^{2}}{k^{4}}%
\mathrm{i}k_{y}\mathrm{e}^{\mathrm{i}\boldsymbol{k}\cdot \boldsymbol{r}}%
\mathrm{d}_{\boldsymbol{k}}V  \notag \\
& =\frac{\mathrm{i}L^{2}}{8\pi ^{3}}\int_{0}^{2\pi }\!\!\!\int_{0}^{\infty
}\!\!\!\int_{-\infty }^{\infty }\frac{k_{\parallel }^{2}k_{\perp }\sin
\varphi }{\left( k_{\parallel }^{2}+k_{\perp }^{2}\right) ^{2}}\mathrm{e}^{%
\mathrm{i}k_{\boldsymbol{\parallel }}r_{\boldsymbol{\parallel }}}\mathrm{e}^{%
\mathrm{i}k_{\perp }r_{\perp }\cos \varphi }\mathrm{d}k_{\parallel
}\,k_{\perp }\mathrm{d}k_{\perp }\mathrm{d}\varphi  \label{E-1y} \\
& =\frac{\mathrm{i}L^{2}}{16\pi ^{2}}\int_{0}^{2\pi }\!\!\!\int_{0}^{\infty
}k_{\perp }\mathrm{e}^{-k_{\perp }\left( \left\vert r_{\boldsymbol{\parallel 
}}\right\vert -\mathrm{i}r_{\perp }\cos \varphi \right) }\sin \varphi \,%
\mathrm{d}k_{\perp }\mathrm{d}\varphi -\frac{\mathrm{i}L^{2}}{16\pi ^{2}}%
\int_{0}^{2\pi }\!\!\!\int_{0}^{\infty }\left\vert r_{\boldsymbol{\parallel }%
}\right\vert k_{\perp }^{2}\mathrm{e}^{-k_{\perp }\left( \left\vert r_{%
\boldsymbol{\parallel }}\right\vert -\mathrm{i}r_{\perp }\cos \varphi
\right) }\sin \varphi \,\mathrm{d}k_{\perp }\mathrm{d}\varphi  \label{E-3y}
\\
& =\frac{\mathrm{i}L^{2}}{16\pi ^{2}}\int_{0}^{2\pi }\frac{1}{\left(
\left\vert r_{\boldsymbol{\parallel }}\right\vert -\mathrm{i}r_{\perp }\cos
\varphi \right) ^{2}}\sin \varphi \,\mathrm{d}\varphi -\frac{\mathrm{i}L^{2}%
}{16\pi ^{2}}\int_{0}^{2\pi }\frac{2\left\vert r_{\boldsymbol{\parallel }%
}\right\vert }{\left( \left\vert r_{\boldsymbol{\parallel }}\right\vert -%
\mathrm{i}r_{\perp }\cos \varphi \right) ^{3}}\sin \varphi \,\mathrm{d}%
\varphi  \notag \\
& =0,  \notag
\end{align}%
since both integrands are odd functions and the limits of integration are
symmetric with respect to point $\varphi=\pi$, where in (\ref{E-1y}) the
integrals (\ref{D-9}) and (\ref{DD-10}) are used, while in (\ref{E-3y}) the
integral (\ref{C-5}) is used.

Inverting the last vector component one has%
\begin{align}
\mathcal{F}^{-1} & \left[ \frac{\left( \boldsymbol{L}\cdot \boldsymbol{k}%
\right) ^{2}}{k^{4}}\mathrm{i}k_{z}\right] \left( \boldsymbol{r}\right) =%
\frac{1}{\left( 2\pi \right) ^{3}}\int_{%
\mathbb{R}
^{3}}\frac{\left( \boldsymbol{L}\cdot \boldsymbol{k}\right) ^{2}}{k^{4}}%
\mathrm{i}k_{z}\mathrm{e}^{\mathrm{i}\boldsymbol{k}\cdot \boldsymbol{r}}%
\mathrm{d}_{\boldsymbol{k}}V  \notag \\
& =\frac{\mathrm{i}L^{2}}{8\pi ^{3}}\int_{0}^{2\pi }\!\!\!\int_{0}^{\infty
}\!\!\!\int_{-\infty }^{\infty }\frac{k_{\parallel }^{3}}{\left(
k_{\parallel }^{2}+k_{\perp }^{2}\right) ^{2}}\mathrm{e}^{\mathrm{i}k_{%
\boldsymbol{\parallel }}r_{\boldsymbol{\parallel }}}\mathrm{e}^{\mathrm{i}%
k_{\perp }r_{\perp }\cos \varphi }\mathrm{d}k_{\parallel }\,k_{\perp }%
\mathrm{d}k_{\perp }\mathrm{d}\varphi  \label{E-1z} \\
& =-\frac{L^{2}}{8\pi ^{2}}\mathrm{sgn}r_{\boldsymbol{\parallel }%
}\int_{0}^{2\pi }\!\!\!\int_{0}^{\infty }k_{\perp }\mathrm{e}^{-k_{\perp
}\left( \left\vert r_{\boldsymbol{\parallel }}\right\vert -\mathrm{i}%
r_{\perp }\cos \varphi \right) }\mathrm{d}k_{\perp }\mathrm{d}\varphi +\frac{%
L^{2}}{16\pi ^{2}}\mathrm{sgn}r_{\boldsymbol{\parallel }}\int_{0}^{2\pi
}\!\!\!\int_{0}^{\infty }\left\vert r_{\boldsymbol{\parallel }}\right\vert
k_{\perp }^{2}\mathrm{e}^{-k_{\perp }\left( \left\vert r_{\boldsymbol{%
\parallel }}\right\vert -\mathrm{i}r_{\perp }\cos \varphi \right) }\mathrm{d}%
k_{\perp }\mathrm{d}\varphi  \label{E-3z} \\
& =-\frac{L^{2}}{8\pi ^{2}}\mathrm{sgn}r_{\boldsymbol{\parallel }%
}\int_{0}^{2\pi }\frac{1}{\left( \left\vert r_{\boldsymbol{\parallel }%
}\right\vert -\mathrm{i}r_{\perp }\cos \varphi \right) ^{2}}\mathrm{d}%
\varphi +\frac{L^{2}}{16\pi ^{2}}\mathrm{sgn}r_{\boldsymbol{\parallel }%
}\int_{0}^{2\pi }\frac{2\left\vert r_{\boldsymbol{\parallel }}\right\vert }{%
\left( \left\vert r_{\boldsymbol{\parallel }}\right\vert -\mathrm{i}r_{\perp
}\cos \varphi \right) ^{3}}\mathrm{d}\varphi  \notag \\
& =\frac{L^{2}}{8\pi ^{2}}\mathrm{sgn}r_{\boldsymbol{\parallel }}\frac{%
\partial }{\partial \left\vert r_{\boldsymbol{\parallel }}\right\vert }%
\int_{0}^{2\pi }\frac{1}{\left\vert r_{\boldsymbol{\parallel }}\right\vert -%
\mathrm{i}r_{\perp }\cos \varphi }\mathrm{d}\varphi +\frac{L^{2}}{16\pi ^{2}}%
r_{\boldsymbol{\parallel }}\frac{\partial ^{2}}{\partial r_{\perp }\partial
\left\vert r_{\boldsymbol{\parallel }}\right\vert }\int_{0}^{2\pi }\frac{1}{%
\left\vert r_{\boldsymbol{\parallel }}\right\vert -\mathrm{i}r_{\perp }\cos
\varphi }\mathrm{d}\varphi  \label{E-5z} \\
& =-\frac{L^{2}}{8\pi ^{2}}\mathrm{sgn}r_{\boldsymbol{\parallel }}\frac{2\pi
\left\vert r_{\boldsymbol{\parallel }}\right\vert }{\sqrt{r_{\boldsymbol{%
\parallel }}^{2}+r_{\perp }^{2}}^{3}}-\frac{L^{2}}{16\pi ^{2}}\frac{2\pi r_{%
\boldsymbol{\parallel }}}{\sqrt{r_{\boldsymbol{\parallel }}^{2}+r_{\perp
}^{2}}^{3}}\left( 1-3\frac{r_{\boldsymbol{\parallel }}^{2}}{\sqrt{r_{%
\boldsymbol{\parallel }}^{2}+r_{\perp }^{2}}^{2}}\right)  \notag \\
& =-\frac{3L^{2}}{8\pi }\frac{r_{\boldsymbol{\parallel }}}{r^{3}}\left( 1-%
\frac{r_{\boldsymbol{\parallel }}^{2}}{r^{2}}\right),  \notag
\end{align}%
where in (\ref{E-1z}) the integral%
\begin{equation*}
\int_{-\infty }^{\infty }\frac{x^{3}}{\left( x^{2}+\lambda ^{2}\right) ^{2}}%
\mathrm{e}^{\mathrm{i}\alpha x}\mathrm{d}x=2\mathrm{i}\int_{0}^{\infty }%
\frac{x^{3}\sin \left( \alpha x\right) }{\left( x^{2}+\lambda ^{2}\right)
^{2}}\mathrm{d}x=2\mathrm{i}\frac{\partial ^{3}}{\partial \left\vert \alpha
\right\vert ^{3}}\int_{0}^{\infty }\frac{\cos \left( \alpha x\right) }{%
\left( x^{2}+\lambda ^{2}\right) ^{2}}\mathrm{d}x=\mathrm{i}\frac{\pi }{2}%
\mathrm{sgn}\alpha \,\left( 2-\left\vert \alpha \right\vert \lambda \right) 
\mathrm{e}^{-\left\vert \alpha \right\vert \lambda },
\end{equation*}%
is used, which relies on the integral (\ref{D-8}), in (\ref{E-3z}) the
integrals (\ref{DD-10}) and (\ref{C-5}) are used, while in (\ref{E-5z}) the
integral (\ref{C-6}) is used.

Therefore, by substituting all previously calculated inversions one has%
\begin{align*}
\mathcal{F}^{-1}\left[ \frac{\left( \boldsymbol{L}\cdot \boldsymbol{k}%
\right) ^{2}}{k^{4}}\mathrm{i}\boldsymbol{k}\right] \left( \boldsymbol{r}%
\right) & =-\frac{L^{2}}{8\pi }\left( 1-3\frac{r_{\boldsymbol{\parallel }%
}^{2}}{r^{2}}\right) \frac{r_{\perp }}{r^{3}}\boldsymbol{e}_{x}-\frac{3L^{2}%
}{8\pi }\left( 1-\frac{r_{\boldsymbol{\parallel }}^{2}}{r^{2}}\right) \frac{%
r_{\boldsymbol{\parallel }}}{r^{3}}\boldsymbol{e}_{z} \\
& =-\frac{1}{4\pi }\frac{L^{2}}{r^{2}}\frac{r_{\boldsymbol{\parallel }}}{r}%
\boldsymbol{e}_{z}-\frac{1}{8\pi }\left( 1-3\frac{r_{\boldsymbol{\parallel }%
}^{2}}{r^{2}}\right) \frac{L^{2}}{r^{3}}\left( r_{\perp }\boldsymbol{e}%
_{x}+r_{\boldsymbol{\parallel }}\boldsymbol{e}_{z}\right) \\
& =-\frac{1}{4\pi }\left( \frac{\boldsymbol{L}}{L}\cdot \frac{\boldsymbol{r}%
}{r}\right) \frac{L^{2}}{r^{2}}\frac{\boldsymbol{L}}{L}-\frac{1}{8\pi }%
\left( 1-3\left( \frac{\boldsymbol{L}}{L}\cdot \frac{\boldsymbol{r}}{r}%
\right) ^{2}\right) \frac{L^{2}}{r^{2}}\frac{\boldsymbol{r}}{r},
\end{align*}%
being (\ref{izraz}) due to $r_{\boldsymbol{\parallel }}=\boldsymbol{r}\cdot 
\frac{\boldsymbol{L}}{L}$, $\boldsymbol{e}_{z}=\frac{\boldsymbol{L}}{L}$,
and $\boldsymbol{r}=r_{\perp }\boldsymbol{e}_{x}+r_{\boldsymbol{\parallel }}%
\boldsymbol{e}_{z}$.

\section{Calculation of $\mathcal{F}^{-1}\left[ \left( \boldsymbol{L}\cdot 
\boldsymbol{k}\right) ^{2}\frac{\boldsymbol{k}\otimes \boldsymbol{k}}{k^{4}}%
\right] \left( \boldsymbol{r}\right) $ \label{dijadski-skalarni-kroz-cetvrti}%
}

The Fourier transform inversion of tensor function $\left( \boldsymbol{L}%
\cdot \boldsymbol{k}\right) ^{2}\frac{\boldsymbol{k}\otimes \boldsymbol{k}}{%
k^{4}}$ is of the form%
\begin{align}
\mathcal{F}^{-1}& \left[ \left( \boldsymbol{L}\cdot \boldsymbol{k}\right)
^{2}\frac{\boldsymbol{k}\otimes \boldsymbol{k}}{k^{4}}\right] \left( 
\boldsymbol{r}\right)  \notag \\
& =\delta \left( \boldsymbol{r}\right) \left( \boldsymbol{L}\otimes 
\boldsymbol{L}\right) +\frac{L^{2}}{8\pi r^{3}}\left( \left( 1-3\left( \frac{%
\boldsymbol{L}}{L}\cdot \frac{\boldsymbol{r}}{r}\right) ^{2}\right) 
\boldsymbol{\hat{I}}+2\frac{\boldsymbol{L}\otimes \boldsymbol{L}}{L^{2}}%
\right.  \notag \\
& \quad \left. -6\left( \frac{\boldsymbol{L}}{L}\cdot \frac{\boldsymbol{r}}{r%
}\right) \left( \frac{\boldsymbol{L}}{L}\otimes \frac{\boldsymbol{r}}{r}+%
\frac{\boldsymbol{r}}{r}\otimes \frac{\boldsymbol{L}}{L}\right) -3\left(
1-5\left( \frac{\boldsymbol{L}}{L}\cdot \frac{\boldsymbol{r}}{r}\right)
^{2}\right) \frac{\boldsymbol{r}\otimes \boldsymbol{r}}{r^{2}}\right) .
\label{operator-D}
\end{align}

Namely, the tensor $\left( \boldsymbol{L}\cdot \boldsymbol{k}\right) ^{2}%
\frac{\boldsymbol{k}\otimes \boldsymbol{k}}{k^{4}}$ can be written in the
matrix form as%
\begin{align}
\left( \boldsymbol{L}\cdot \boldsymbol{k}\right) ^{2}\frac{\boldsymbol{k}%
\otimes \boldsymbol{k}}{k^{4}}& =L^{2}\frac{k_{z}^{2}}{k^{4}}\left[ 
\begin{array}{ccc}
k_{x}^{2} & k_{x}k_{y} & k_{x}k_{z} \\ 
k_{x}k_{y} & k_{y}^{2} & k_{y}k_{z} \\ 
k_{x}k_{z} & k_{y}k_{z} & k_{z}^{2}%
\end{array}%
\right]  \notag \\
& =L^{2}\frac{k_{\parallel }^{2}}{\left( k_{\parallel }^{2}+k_{\perp
}^{2}\right) ^{2}}\left[ 
\begin{array}{ccc}
k_{\perp }^{2}\cos ^{2}\varphi & k_{\perp }^{2}\sin \varphi \cos \varphi & 
k_{\parallel }k_{\perp }\cos \varphi \\ 
k_{\perp }^{2}\sin \varphi \cos \varphi & k_{\perp }^{2}\sin ^{2}\varphi & 
k_{\parallel }k_{\perp }\sin \varphi \\ 
k_{\parallel }k_{\perp }\cos \varphi & k_{\parallel }k_{\perp }\sin \varphi
& k_{\parallel }^{2}%
\end{array}%
\right] ,  \label{operator-D-fd}
\end{align}%
using Cartesian coordinate system for tensor representation, expressing
tensor components in cylindrical coordinates, and assuming that the $z$-axis
of coordinate system coincides with the non-locality vector $\boldsymbol{L}$%
, i.e., $\boldsymbol{L}=L\boldsymbol{e}_{z}$, implying that $\left(
k_{x},k_{y},k_{z}\right) =\left( k_{\perp }\cos \varphi ,k_{\perp }\sin
\varphi ,k_{\parallel }\right) $. Moreover, by writing $\boldsymbol{r}%
=r_{\parallel }\frac{\boldsymbol{L}}{L}+\boldsymbol{r}_{\perp }=r_{\parallel
}\boldsymbol{e}_{z}+\boldsymbol{r}_{\perp }$ the scalar product $\boldsymbol{%
k}\cdot \boldsymbol{r}=k_{\parallel }r_{\parallel }+\boldsymbol{k}_{\perp
}\cdot \boldsymbol{r}_{\perp }=k_{\parallel }r_{\parallel }+k_{\perp
}r_{\perp }\cos \varphi $, for the sake of simplicity, implies $\boldsymbol{r%
}_{\perp }=r_{\perp }\boldsymbol{e}_{x},$ thus $\boldsymbol{r}=r_{\perp }%
\boldsymbol{e}_{x}+r_{\parallel }\boldsymbol{e}_{z}$, where $r_{\perp }=|{%
\boldsymbol{r}_{\perp }}|\geqslant 0$ is a vector intensity, while $%
r_{\parallel }\in \mathbb{R}$ is a coordinate. Therefore, the Fourier
transform inversion of tensor $\left( \boldsymbol{L}\cdot \boldsymbol{k}%
\right) ^{2}\frac{\boldsymbol{k}\otimes \boldsymbol{k}}{k^{4}}$ is achieved
by inverting each component of the matrix (\ref{operator-D-fd}) according to
the definition of the inverse Fourier transform (\ref{IFT}), yielding for
the first element on the main diagonal%
\begin{align}
\mathcal{F}^{-1}& \left[ \left( \boldsymbol{L}\cdot \boldsymbol{k}\right)
^{2}\frac{k_{x}^{2}}{k^{4}}\right] \left( \boldsymbol{r}\right) =\frac{1}{%
\left( 2\pi \right) ^{3}}\int_{%
\mathbb{R}
^{3}}\left( \boldsymbol{L}\cdot \boldsymbol{k}\right) ^{2}\frac{k_{x}^{2}}{%
k^{4}}\mathrm{e}^{\mathrm{i}\boldsymbol{k}\cdot \boldsymbol{r}}\mathrm{d}_{%
\boldsymbol{k}}V  \notag \\
& =\frac{L^{2}}{\left( 2\pi \right) ^{3}}\int_{0}^{2\pi
}\!\!\!\int_{0}^{\infty }\!\!\!\int_{-\infty }^{\infty }\frac{k_{\parallel
}^{2}k_{\perp }^{2}}{\left( k_{\parallel }^{2}+k_{\perp }^{2}\right) ^{2}}%
\cos ^{2}\varphi \,\mathrm{e}^{\mathrm{i}k_{\boldsymbol{\parallel }}r_{%
\boldsymbol{\parallel }}}\mathrm{e}^{\mathrm{i}k_{\perp }r_{\perp }\cos
\varphi }k_{\perp }\mathrm{d}k_{\parallel }\mathrm{d}k_{\perp }\mathrm{d}%
\varphi  \notag \\
& =\frac{L^{2}}{4\pi ^{3}}\int_{0}^{2\pi }\cos ^{2}\varphi \int_{0}^{\infty
}k_{\perp }^{3}\mathrm{e}^{\mathrm{i}k_{\perp }r_{\perp }\cos \varphi
}\int_{0}^{\infty }\frac{k_{\parallel }^{2}\cos \left( k_{\parallel
}r_{\parallel }\right) }{\left( k_{\parallel }^{2}+k_{\perp }^{2}\right) ^{2}%
}\mathrm{d}k_{\parallel }\mathrm{d}k_{\perp }\mathrm{d}\varphi  \label{D-6}
\\
& =\frac{L^{2}}{4\pi ^{3}}\int_{0}^{2\pi }I\left( \boldsymbol{r},\varphi
\right) \cos ^{2}\varphi \,\mathrm{d}\varphi \;\;\text{with}\;\;I\left( 
\boldsymbol{r},\varphi \right) =\int_{0}^{\infty }k_{\perp }^{3}\mathrm{e}^{%
\mathrm{i}k_{\perp }r_{\perp }\cos \varphi }\int_{0}^{\infty }\frac{%
k_{\parallel }^{2}\cos \left( k_{\parallel }r_{\parallel }\right) }{\left(
k_{\parallel }^{2}+k_{\perp }^{2}\right) ^{2}}\mathrm{d}k_{\parallel }%
\mathrm{d}k_{\perp }  \label{D-1} \\
& =\frac{L^{2}}{4\pi ^{3}}\int_{0}^{2\pi }\left( -\frac{\pi }{2}\frac{%
2\left\vert r_{\parallel }\right\vert +\mathrm{i}r_{\perp }\cos \varphi }{%
\left( \left\vert r_{\parallel }\right\vert -\mathrm{i}r_{\perp }\cos
\varphi \right) ^{4}}\right) \cos ^{2}\varphi \,\mathrm{d}\varphi  \notag \\
& =-\frac{L^{2}}{4\pi ^{2}}\left\vert r_{\parallel }\right\vert
\int_{0}^{2\pi }\frac{\cos ^{2}\varphi }{\left( \left\vert r_{\parallel
}\right\vert -\mathrm{i}r_{\perp }\cos \varphi \right) ^{4}}\mathrm{d}%
\varphi -\mathrm{i}\frac{L^{2}}{8\pi ^{2}}r_{\perp }\int_{0}^{2\pi }\frac{%
\cos ^{3}\varphi }{\left( \left\vert r_{\parallel }\right\vert -\mathrm{i}%
r_{\perp }\cos \varphi \right) ^{4}}\mathrm{d}\varphi  \notag \\
& =-\frac{L^{2}}{4\pi ^{2}}\left\vert r_{\parallel }\right\vert \left( \frac{%
1}{6}\frac{\partial ^{3}}{\partial \left\vert r_{\parallel }\right\vert
\partial r_{\perp }^{2}}\int_{0}^{2\pi }\frac{1}{\left\vert r_{\parallel
}\right\vert -\mathrm{i}r_{\perp }\cos \varphi }\mathrm{d}\varphi \right) -%
\mathrm{i}\frac{L^{2}}{8\pi ^{2}}r_{\perp }\left( -\frac{1}{6\mathrm{i}}%
\frac{\partial ^{3}}{\partial r_{\perp }^{3}}\int_{0}^{2\pi }\frac{1}{%
\left\vert r_{\parallel }\right\vert -\mathrm{i}r_{\perp }\cos \varphi }%
\mathrm{d}\varphi \right)  \label{D-2} \\
& =-\frac{L^{2}}{4\pi ^{2}}\left\vert r_{\parallel }\right\vert \left( \pi
\left\vert r_{\parallel }\right\vert \frac{r_{\parallel }^{2}-4r_{\perp }^{2}%
}{\sqrt{r_{\parallel }^{2}+r_{\perp }^{2}}^{7}}\right) +\frac{L^{2}}{8\pi
^{2}}r_{\perp }\left( \pi r_{\perp }\frac{3r_{\parallel }^{2}-2r_{\perp }^{2}%
}{\sqrt{r_{\parallel }^{2}+r_{\perp }^{2}}^{7}}\right)  \notag \\
& =\frac{L^{2}}{8\pi r^{3}}\left( -2+15\frac{r_{\parallel }^{2}}{r^{2}}-15%
\frac{r_{\parallel }^{4}}{r^{4}}\right) ,  \label{DD-1}
\end{align}%
where the integral (\ref{C-6}) is used in (\ref{D-2}), while the integral $I$
from (\ref{D-1}) is calculated as%
\begin{align}
I\left( \boldsymbol{r},\varphi \right) & =\int_{0}^{\infty }k_{\perp }^{3}%
\mathrm{e}^{\mathrm{i}k_{\perp }r_{\perp }\cos \varphi }\int_{0}^{\infty }%
\frac{k_{\parallel }^{2}\cos \left( k_{\parallel }r_{\parallel }\right) }{%
\left( k_{\parallel }^{2}+k_{\perp }^{2}\right) ^{2}}\mathrm{d}k_{\parallel }%
\mathrm{d}k_{\perp }  \notag \\
& =\int_{0}^{\infty }k_{\perp }^{3}\mathrm{e}^{\mathrm{i}k_{\perp }r_{\perp
}\cos \varphi }\left( -\frac{\partial ^{2}}{\partial r_{\parallel }^{2}}%
\int_{0}^{\infty }\frac{\cos \left( k_{\parallel }r_{\parallel }\right) }{%
\left( k_{\parallel }^{2}+k_{\perp }^{2}\right) ^{2}}\mathrm{d}k_{\parallel
}\right) \mathrm{d}k_{\perp }  \label{D-5} \\
& =\int_{0}^{\infty }k_{\perp }^{3}\mathrm{e}^{\mathrm{i}k_{\perp }r_{\perp
}\cos \varphi }\left( \frac{\pi }{4k_{\perp }^{3}}k_{\perp }^{2}\left(
1-\left\vert r_{\parallel }\right\vert k_{\perp }\right) \mathrm{e}%
^{-\left\vert r_{\parallel }\right\vert k_{\perp }}\right) \mathrm{d}%
k_{\perp }  \notag \\
& =\frac{\pi }{4}\int_{0}^{\infty }k_{\perp }^{2}\left( 1-\left\vert
r_{\parallel }\right\vert k_{\perp }\right) \mathrm{e}^{-k_{\perp }\left(
\left\vert r_{\parallel }\right\vert -\mathrm{i}r_{\perp }\cos \varphi
\right) }\mathrm{d}k_{\perp }  \label{D-4} \\
& =\frac{\pi }{4}\left( \frac{2}{\left( \left\vert r_{\parallel }\right\vert
-\mathrm{i}r_{\perp }\cos \varphi \right) ^{3}}-\left\vert r_{\parallel
}\right\vert \frac{6}{\left( \left\vert r_{\parallel }\right\vert -\mathrm{i}%
r_{\perp }\cos \varphi \right) ^{4}}\right)  \label{D-23} \\
& =-\frac{\pi }{2}\frac{2\left\vert r_{\parallel }\right\vert +\mathrm{i}%
r_{\perp }\cos \varphi }{\left( \left\vert r_{\parallel }\right\vert -%
\mathrm{i}r_{\perp }\cos \varphi \right) ^{4}},  \label{D-3}
\end{align}%
using the integral (\ref{D-8}) in (\ref{D-5}), as well as integrals (\ref%
{C-5}) and%
\begin{equation}
\int_{0}^{\infty }u^{3}\mathrm{e}^{-\alpha u}\,\mathrm{d}u=\frac{6}{\alpha
^{4}},\quad \text{for\quad }\func{Re}\alpha >0,  \label{D-15}
\end{equation}%
in (\ref{D-4}).

Further, for the second element on the main diagonal of matrix (\ref%
{operator-D-fd}), one has%
\begin{align*}
\mathcal{F}^{-1}& \left[ \left( \boldsymbol{L}\cdot \boldsymbol{k}\right)
^{2}\frac{k_{y}^{2}}{k^{4}}\right] \left( \boldsymbol{r}\right) =\frac{1}{%
\left( 2\pi \right) ^{3}}\int_{%
\mathbb{R}
^{3}}\left( \boldsymbol{L}\cdot \boldsymbol{k}\right) ^{2}\frac{k_{y}^{2}}{%
k^{4}}\mathrm{e}^{\mathrm{i}\boldsymbol{k}\cdot \boldsymbol{r}}\mathrm{d}_{%
\boldsymbol{k}}V \\
& =\frac{L^{2}}{\left( 2\pi \right) ^{3}}\int_{0}^{2\pi
}\!\!\!\int_{0}^{\infty }\!\!\!\int_{-\infty }^{\infty }\frac{k_{\parallel
}^{2}k_{\perp }^{2}}{\left( k_{\parallel }^{2}+k_{\perp }^{2}\right) ^{2}}%
\sin ^{2}\varphi \,\mathrm{e}^{\mathrm{i}k_{\boldsymbol{\parallel }}r_{%
\boldsymbol{\parallel }}}\mathrm{e}^{\mathrm{i}k_{\perp }r_{\perp }\cos
\varphi }k_{\perp }\mathrm{d}k_{\parallel }\mathrm{d}k_{\perp }\mathrm{d}%
\varphi \\
& =\frac{L^{2}}{4\pi ^{3}}\int_{0}^{2\pi }\left( 1-\cos ^{2}\varphi \right)
\int_{0}^{\infty }k_{\perp }^{3}\mathrm{e}^{\mathrm{i}k_{\perp }r_{\perp
}\cos \varphi }\int_{-\infty }^{\infty }\frac{k_{\parallel }^{2}\cos \left(
k_{\parallel }r_{\parallel }\right) }{\left( k_{\parallel }^{2}+k_{\perp
}^{2}\right) ^{2}}\mathrm{d}k_{\parallel }\mathrm{d}k_{\perp }\mathrm{d}%
\varphi \\
& =\frac{L^{2}}{4\pi ^{3}}\int_{0}^{2\pi }I\left( \boldsymbol{r},\varphi
\right) \mathrm{d}\varphi -\mathcal{F}^{-1}\left[ \left( \boldsymbol{L}\cdot 
\boldsymbol{k}\right) ^{2}\frac{k_{x}^{2}}{k^{4}}\right] \left( \boldsymbol{r%
}\right),
\end{align*}%
with the previously found inversion $\mathcal{F}^{-1}\left[ \left( 
\boldsymbol{L}\cdot \boldsymbol{k}\right) ^{2}\frac{k_{x}^{2}}{k^{4}}\right] 
$, see (\ref{D-6}), and with integral $I$ defined by (\ref{D-1})$_{2}$ and
solved as in (\ref{D-3}), implying that the calculation of $\mathcal{F}^{-1}%
\left[ \left( \boldsymbol{L}\cdot \boldsymbol{k}\right) ^{2}\frac{k_{y}^{2}}{%
k^{4}}\right] \boldsymbol{\ }$reduces to computing the following integral%
\begin{align}
\int_{0}^{2\pi }I\left( \boldsymbol{r},\varphi \right) \mathrm{d}\varphi & =-%
\frac{\pi }{2}\int_{0}^{2\pi }\frac{2\left\vert r_{\parallel }\right\vert +%
\mathrm{i}r_{\perp }\cos \varphi }{\left( \left\vert r_{\parallel
}\right\vert -\mathrm{i}r_{\perp }\cos \varphi \right) ^{4}}\mathrm{d}\varphi
\notag \\
& =-\pi \left\vert r_{\parallel }\right\vert \int_{0}^{2\pi }\frac{1}{\left(
\left\vert r_{\parallel }\right\vert -\mathrm{i}r_{\perp }\cos \varphi
\right) ^{4}}\mathrm{d}\varphi -\mathrm{i}\frac{\pi }{2}r_{\perp
}\int_{0}^{2\pi }\frac{\cos \varphi }{\left( \left\vert r_{\parallel
}\right\vert -\mathrm{i}r_{\perp }\cos \varphi \right) ^{4}}\mathrm{d}\varphi
\notag \\
& =-\pi \left\vert r_{\parallel }\right\vert \left( -\frac{1}{6}\frac{%
\partial ^{3}}{\partial \left\vert r_{\parallel }\right\vert ^{3}}%
\int_{0}^{2\pi }\!\!\!\frac{1}{\left\vert r_{\parallel }\right\vert -\mathrm{%
i}r_{\perp }\cos \varphi }\mathrm{d}\varphi \right) -\mathrm{i}\frac{\pi }{2}%
r_{\perp }\left( \frac{1}{6\mathrm{i}}\frac{\partial ^{3}}{\partial r_{\perp
}\partial \left\vert r_{\parallel }\right\vert ^{2}}\int_{0}^{2\pi }\!\!\!%
\frac{1}{\left\vert r_{\parallel }\right\vert -\mathrm{i}r_{\perp }\cos
\varphi }\mathrm{d}\varphi \right)  \label{D-7} \\
& =-\pi \left\vert r_{\parallel }\right\vert \left( -\pi \left\vert
r_{\parallel }\right\vert \frac{-2r_{\parallel }^{2}+3r_{\perp }^{2}}{\sqrt{%
r_{\boldsymbol{\parallel }}^{2}+r_{\perp }^{2}}^{7}}\right) -\mathrm{i}\frac{%
\pi }{2}r_{\perp }\left( \frac{\pi }{\mathrm{i}}r_{\perp }\frac{%
-4r_{\parallel }^{2}+r_{\perp }^{2}}{\sqrt{r_{\boldsymbol{\parallel }%
}^{2}+r_{\perp }^{2}}^{7}}\right)  \notag \\
& =-\frac{\pi ^{2}}{2r^{3}}\left( 1-\frac{12r_{\parallel }^{2}}{r^{2}}+\frac{%
15r_{\parallel }^{4}}{r^{4}}\right) ,  \label{D-12}
\end{align}%
where the integral (\ref{C-6}) is used in (\ref{D-7}), so that%
\begin{align}
\mathcal{F}^{-1}& \left[ \left( \boldsymbol{L}\cdot \boldsymbol{k}\right)
^{2}\frac{k_{y}^{2}}{k^{4}}\right] \left( \boldsymbol{r}\right) =\frac{L^{2}%
}{4\pi ^{3}}\int_{0}^{2\pi }I\left( \boldsymbol{r},\varphi \right) \mathrm{d}%
\varphi -\mathcal{F}^{-1}\left[ \left( \boldsymbol{L}\cdot \boldsymbol{k}%
\right) ^{2}\frac{k_{x}^{2}}{k^{4}}\right] \left( \boldsymbol{r}\right) 
\notag \\
& =-\frac{L^{2}}{4\pi ^{3}}\frac{\pi ^{2}}{2r^{2}}\left( 1-\frac{%
12r_{\parallel }^{2}}{r^{2}}+\frac{15r_{\parallel }^{4}}{r^{4}}\right) -%
\frac{L^{2}}{8\pi r^{3}}\left( -2+15\frac{r_{\parallel }^{2}}{r^{2}}-15\frac{%
r_{\parallel }^{4}}{r^{4}}\right)  \notag \\
& =\frac{L^{2}}{8\pi r^{3}}\left( 1-3\frac{r_{\parallel }^{2}}{r^{2}}\right)
.  \label{DD-2}
\end{align}

The inverse Fourier transform of the last matrix component on the main
diagonal is given by%
\begin{align}
\mathcal{F}^{-1}& \left[ \left( \boldsymbol{L}\cdot \boldsymbol{k}\right)
^{2}\frac{k_{z}^{2}}{k^{4}}\right] \left( \boldsymbol{r}\right) =\frac{1}{%
\left( 2\pi \right) ^{3}}\int_{%
\mathbb{R}
^{3}}\left( \boldsymbol{L}\cdot \boldsymbol{k}\right) ^{2}\frac{k_{z}^{2}}{%
k^{4}}\mathrm{e}^{\mathrm{i}\boldsymbol{k}\cdot \boldsymbol{r}}\mathrm{d}_{%
\boldsymbol{k}}V  \notag \\
& =\frac{1}{\left( 2\pi \right) ^{3}}\int_{%
\mathbb{R}
^{2}}\int_{%
\mathbb{R}
}\frac{k_{\parallel }^{4}}{\left( k_{\parallel }^{2}+\left\vert \boldsymbol{k%
}_{\perp }\right\vert ^{2}\right) ^{2}}\mathrm{e}^{\mathrm{i}k_{\boldsymbol{%
\parallel }}r_{\boldsymbol{\parallel }}}\mathrm{e}^{\mathrm{i}\boldsymbol{k}%
_{\perp }\cdot \boldsymbol{r}_{\perp }}\mathrm{d}k_{\parallel }\,\mathrm{d}_{%
\boldsymbol{k}_{\perp }}S  \notag \\
& =\frac{1}{\left( 2\pi \right) ^{3}}\int_{%
\mathbb{R}
^{2}}\int_{%
\mathbb{R}
}\frac{\left( k_{\parallel }^{2}+\left\vert \boldsymbol{k}_{\perp
}\right\vert ^{2}\right) ^{2}-2k_{\parallel }^{2}\left\vert \boldsymbol{k}%
_{\perp }\right\vert ^{2}-\left\vert \boldsymbol{k}_{\perp }\right\vert ^{2}%
}{\left( k_{\parallel }^{2}+\left\vert \boldsymbol{k}_{\perp }\right\vert
^{2}\right) ^{2}}\mathrm{e}^{\mathrm{i}k_{\boldsymbol{\parallel }}r_{%
\boldsymbol{\parallel }}}\mathrm{e}^{\mathrm{i}\boldsymbol{k}_{\perp }\cdot 
\boldsymbol{r}_{\perp }}\mathrm{d}k_{\parallel }\,\mathrm{d}_{\boldsymbol{k}%
_{\perp }}S  \notag \\
& =\frac{L^{2}}{\left( 2\pi \right) ^{3}}\int_{%
\mathbb{R}
^{3}}\mathrm{e}^{\mathrm{i}\boldsymbol{k}\cdot \boldsymbol{r}}\mathrm{d}_{%
\boldsymbol{k}}V-\frac{1}{\left( 2\pi \right) ^{3}}\int_{%
\mathbb{R}
^{2}}\int_{%
\mathbb{R}
}\frac{2k_{\parallel }^{2}\left\vert \boldsymbol{k}_{\perp }\right\vert
^{2}+\left\vert \boldsymbol{k}_{\perp }\right\vert ^{2}}{\left( k_{\parallel
}^{2}+\left\vert \boldsymbol{k}_{\perp }\right\vert ^{2}\right) ^{2}}\mathrm{%
e}^{\mathrm{i}k_{\boldsymbol{\parallel }}r_{\boldsymbol{\parallel }}}\mathrm{%
e}^{\mathrm{i}\boldsymbol{k}_{\perp }\cdot \boldsymbol{r}_{\perp }}\mathrm{d}%
k_{\parallel }\,\mathrm{d}_{\boldsymbol{k}_{\perp }}S  \notag \\
& =L^{2}\delta \left( \boldsymbol{r}\right) -\frac{1}{\left( 2\pi \right)
^{3}}\int_{0}^{2\pi }\!\!\!\int_{0}^{\infty }k_{\perp }\mathrm{e}^{\mathrm{i}%
k_{\perp }r_{\perp }\cos \varphi }\int_{-\infty }^{\infty }\frac{%
2k_{\parallel }^{2}k_{\perp }^{2}+k_{\perp }^{4}}{\left( k_{\parallel
}^{2}+k_{\perp }^{2}\right) ^{2}}\mathrm{e}^{\mathrm{i}k_{\parallel
}r_{\parallel }}\mathrm{d}k_{\parallel }\mathrm{d}k_{\perp }\mathrm{d}\varphi
\notag \\
& =L^{2}\delta \left( \boldsymbol{r}\right) -\frac{1}{4\pi ^{3}}%
\int_{0}^{2\pi }\!\!\!\int_{0}^{\infty }k_{\perp }\mathrm{e}^{\mathrm{i}%
k_{\perp }r_{\perp }\cos \varphi }\left( 2k_{\perp }^{2}\int_{0}^{\infty }%
\frac{k_{\parallel }^{2}\mathrm{\cos }\left( k_{\parallel }r_{\parallel
}\right) }{\left( k_{\parallel }^{2}+k_{\perp }^{2}\right) ^{2}}\mathrm{d}%
k_{\parallel }+k_{\perp }^{4}\int_{0}^{\infty }\frac{\mathrm{\cos }\left(
k_{\parallel }r_{\parallel }\right) }{\left( k_{\parallel }^{2}+k_{\perp
}^{2}\right) ^{2}}\mathrm{d}k_{\parallel }\right) \mathrm{d}k_{\perp }%
\mathrm{d}\varphi  \notag \\
& =L^{2}\delta \left( \boldsymbol{r}\right) -\frac{1}{4\pi ^{3}}%
\int_{0}^{2\pi }\left( 2I\left( \boldsymbol{r},\varphi \right) +J\left( 
\boldsymbol{r},\varphi \right) \right) \mathrm{d}\varphi \,\,\text{%
with\thinspace \thinspace }J\left( \boldsymbol{r},\varphi \right)
=\int_{0}^{\infty }k_{\perp }^{5}\mathrm{e}^{\mathrm{i}k_{\perp }r_{\perp
}\cos \varphi }\int_{0}^{\infty }\frac{\mathrm{\cos }\left( k_{\parallel
}r_{\parallel }\right) }{\left( k_{\parallel }^{2}+k_{\perp }^{2}\right) ^{2}%
}\mathrm{d}k_{\parallel }\mathrm{d}k_{\perp }  \label{D-10} \\
& =L^{2}\delta \left( \boldsymbol{r}\right) -\frac{1}{4\pi ^{3}}\left( -2%
\frac{\pi ^{2}}{2r^{3}}\left( 1-\frac{12r_{\parallel }^{2}}{r^{2}}+\frac{%
15r_{\parallel }^{4}}{r^{4}}\right) +\frac{\pi ^{2}}{2r^{3}}\left( -1-\frac{%
6r_{\parallel }^{2}}{r^{2}}+\frac{15r_{\parallel }^{4}}{r^{4}}\right) \right)
\notag \\
& =L^{2}\delta \left( \boldsymbol{r}\right) +\frac{3}{8\pi r^{3}}\left( 1-%
\frac{6r_{\parallel }^{2}}{r^{2}}+\frac{5r_{\parallel }^{4}}{r^{4}}\right) ,
\label{DD-3}
\end{align}%
where the integral $I$, defined by (\ref{D-1})$_{2}$, appears in (\ref{D-10}%
), while (\ref{D-12}) and (\ref{D-11}) are used in (\ref{D-10}) for
integrals $\int_{0}^{2\pi }I\left( \boldsymbol{r},\varphi \right) \mathrm{d}%
\varphi $ and $\int_{0}^{2\pi }J\left( \boldsymbol{r},\varphi \right) 
\mathrm{d}\varphi $, respectively, since the integral $J$, defined by (\ref%
{D-10})$_{2}$ is calculated as%
\begin{align}
J\left( \boldsymbol{r},\varphi \right) & =\int_{0}^{\infty }k_{\perp }^{5}%
\mathrm{e}^{\mathrm{i}k_{\perp }r_{\perp }\cos \varphi }\left(
\int_{0}^{\infty }\frac{\mathrm{\cos }\left( k_{\parallel }r_{\parallel
}\right) }{\left( k_{\parallel }^{2}+k_{\perp }^{2}\right) ^{2}}\mathrm{d}%
k_{\parallel }\right) \mathrm{d}k_{\perp }  \label{D-13} \\
& =\int_{0}^{\infty }k_{\perp }^{5}\mathrm{e}^{\mathrm{i}k_{\perp }r_{\perp
}\cos \varphi }\left( \frac{\pi }{4k_{\perp }^{3}}\left( 1+\left\vert
r_{\parallel }\right\vert k_{\perp }\right) \mathrm{e}^{-\left\vert
r_{\parallel }\right\vert k_{\perp }}\right) \mathrm{d}k_{\perp }  \notag \\
& =\frac{\pi }{4}\int_{0}^{\infty }k_{\perp }^{2}\left( 1+\left\vert
r_{\parallel }\right\vert k_{\perp }\right) \mathrm{e}^{-k_{\perp }\left(
\left\vert r_{\parallel }\right\vert -\mathrm{i}r_{\perp }\cos \varphi
\right) }\mathrm{d}k_{\perp }\mathrm{d}\varphi  \label{D-14} \\
& =\frac{\pi }{4}\left( \frac{2}{\left( \left\vert r_{\parallel }\right\vert
-\mathrm{i}r_{\perp }\cos \varphi \right) ^{3}}+\left\vert r_{\parallel
}\right\vert \frac{6}{\left( \left\vert r_{\parallel }\right\vert -\mathrm{i}%
r_{\perp }\cos \varphi \right) ^{4}}\right)  \notag \\
& =\frac{\pi }{2}\frac{4\left\vert r_{\parallel }\right\vert -\mathrm{i}%
r_{\perp }\cos \varphi }{\left( \left\vert r_{\parallel }\right\vert -%
\mathrm{i}r_{\perp }\cos \varphi \right) ^{4}},  \label{D-11}
\end{align}%
see (\ref{D-8}) for the expression resulting from the calculation of
integral in (\ref{D-13}), as well as for (\ref{C-5}) and (\ref{D-15}) being
results of integrals in (\ref{D-14}), so that%
\begin{align}
\int_{0}^{2\pi }J\left( \boldsymbol{r},\varphi \right) \mathrm{d}\varphi & =%
\frac{\pi }{2}\int_{0}^{2\pi }\frac{4\left\vert r_{\parallel }\right\vert -%
\mathrm{i}r_{\perp }\cos \varphi }{\left( \left\vert r_{\parallel
}\right\vert -\mathrm{i}r_{\perp }\cos \varphi \right) ^{4}}\mathrm{d}\varphi
\notag \\
& =2\pi \left\vert r_{\parallel }\right\vert \int_{0}^{2\pi }\frac{1}{\left(
\left\vert r_{\parallel }\right\vert -\mathrm{i}r_{\perp }\cos \varphi
\right) ^{4}}\mathrm{d}\varphi -\mathrm{i}\frac{\pi }{2}r_{\perp
}\int_{0}^{2\pi }\frac{\cos \varphi }{\left( \left\vert r_{\parallel
}\right\vert -\mathrm{i}r_{\perp }\cos \varphi \right) ^{4}}\mathrm{d}\varphi
\notag \\
& =2\pi \left\vert r_{\parallel }\right\vert \left( -\frac{1}{6}\frac{%
\partial ^{3}}{\partial \left\vert r_{\parallel }\right\vert ^{3}}%
\int_{0}^{2\pi }\frac{1}{\left\vert r_{\parallel }\right\vert -\mathrm{i}%
r_{\perp }\cos \varphi }\mathrm{d}\varphi \right) \!\!-\mathrm{i}\frac{\pi }{%
2}r_{\perp }\left( \frac{1}{6\mathrm{i}}\frac{\partial ^{3}}{\partial
r_{\perp }\partial \left\vert r_{\parallel }\right\vert ^{2}}\int_{0}^{2\pi }%
\frac{1}{\left\vert r_{\parallel }\right\vert -\mathrm{i}r_{\perp }\cos
\varphi }\mathrm{d}\varphi \right)  \label{D-16} \\
& =2\pi \left\vert r_{\parallel }\right\vert \left( -\pi \left\vert
r_{\parallel }\right\vert \frac{-2r_{\parallel }^{2}+3r_{\perp }^{2}}{\sqrt{%
r_{\boldsymbol{\parallel }}^{2}+r_{\boldsymbol{\perp }}^{2}}^{7}}\right) -%
\mathrm{i}\frac{\pi }{2}r_{\perp }\left( \frac{\pi }{\mathrm{i}}r_{\perp }%
\frac{-4r_{\parallel }^{2}+r_{\perp }^{2}}{\sqrt{r_{\boldsymbol{\parallel }%
}^{2}+r_{\boldsymbol{\perp }}^{2}}^{7}}\right)  \notag \\
& =\frac{\pi ^{2}}{2r^{3}}\left( -1-\frac{6r_{\parallel }^{2}}{r^{2}}+\frac{%
15r_{\parallel }^{4}}{r^{4}}\right) ,  \notag
\end{align}%
where integral (\ref{C-6}) is used in (\ref{D-16}).

Due to the fact that matrix (\ref{operator-D-fd}), representing the tensor $%
\left( \boldsymbol{L}\cdot \boldsymbol{k}\right) ^{2}\frac{\boldsymbol{k}%
\otimes \boldsymbol{k}}{k^{4}},$ is symmetric, the inverse Fourier transform
of the only off-diagonal matrix component that yields a non-zero solution is
given as follows%
\begin{align}
\mathcal{F}^{-1}& \left[ \left( \boldsymbol{L}\cdot \boldsymbol{k}\right)
^{2}\frac{k_{x}k_{z}}{k^{4}}\right] \left( \boldsymbol{r}\right) =\frac{1}{%
\left( 2\pi \right) ^{3}}\int_{%
\mathbb{R}
^{3}}\left( \boldsymbol{L}\cdot \boldsymbol{k}\right) ^{2}\frac{k_{x}k_{z}}{%
k^{4}}\mathrm{e}^{\mathrm{i}\boldsymbol{k}\cdot \boldsymbol{r}}\mathrm{d}_{%
\boldsymbol{k}}V  \notag \\
& =\frac{L^{2}}{\left( 2\pi \right) ^{3}}\int_{0}^{2\pi
}\!\!\!\int_{0}^{\infty }\!\!\!\int_{-\infty }^{\infty }\frac{k_{\parallel
}^{3}k_{\perp }}{\left( k_{\parallel }^{2}+k_{\perp }^{2}\right) ^{2}}\cos
\varphi \,\mathrm{e}^{\mathrm{i}k_{\boldsymbol{\parallel }}r_{\boldsymbol{%
\parallel }}}\mathrm{e}^{\mathrm{i}k_{\perp }r_{\boldsymbol{\perp }}\cos
\varphi }k_{\perp }\mathrm{d}k_{\parallel }\mathrm{d}k_{\perp }\mathrm{d}%
\varphi  \notag \\
& =\mathrm{i}\frac{L^{2}}{4\pi ^{3}}\int_{0}^{2\pi }\cos \varphi
\int_{0}^{\infty }k_{\perp }^{2}\mathrm{e}^{\mathrm{i}k_{\perp }r_{\perp
}\cos \varphi }\int_{0}^{\infty }\frac{k_{\parallel }^{3}\sin \left(
k_{\parallel }r_{\parallel }\right) }{\left( k_{\parallel }^{2}+k_{\perp
}^{2}\right) ^{2}}\mathrm{d}k_{\parallel }\mathrm{d}k_{\perp }\mathrm{d}%
\varphi  \notag \\
& =\mathrm{i}\frac{L^{2}}{4\pi ^{3}}\int_{0}^{2\pi }\cos \varphi
\int_{0}^{\infty }k_{\perp }^{2}\mathrm{e}^{\mathrm{i}k_{\perp }r_{\perp
}\cos \varphi }\left( \frac{\partial ^{3}}{\partial r_{\parallel }^{3}}%
\int_{0}^{\infty }\frac{\cos \left( k_{\parallel }r_{\parallel }\right) }{%
\left( k_{\parallel }^{2}+k_{\perp }^{2}\right) ^{2}}\mathrm{d}k_{\parallel
}\right) \mathrm{d}k_{\perp }\mathrm{d}\varphi  \label{D-18} \\
& =\mathrm{i}\frac{L^{2}}{4\pi ^{3}}\int_{0}^{2\pi }\cos \varphi
\int_{0}^{\infty }k_{\perp }^{2}\mathrm{e}^{\mathrm{i}k_{\perp }r_{\perp
}\cos \varphi }\left( \frac{\pi }{4}\left( 2\,\mathrm{sgn\,}r_{\parallel
}-r_{\parallel }k_{\perp }\right) \mathrm{e}^{-\left\vert r_{\parallel
}\right\vert k_{\perp }}\right) \mathrm{d}k_{\perp }\mathrm{d}\varphi  \notag
\\
& =\mathrm{i}\frac{L^{2}}{16\pi ^{2}}\int_{0}^{2\pi }\cos \varphi
\int_{0}^{\infty }k_{\perp }^{2}\left( 2\,\mathrm{sgn\,}r_{\parallel
}-r_{\parallel }k_{\perp }\right) \mathrm{e}^{-k_{\perp }\left( \left\vert
r_{\parallel }\right\vert -\mathrm{i}r_{\perp }\cos \varphi \right) }\mathrm{%
d}k_{\perp }\mathrm{d}\varphi  \label{D-19} \\
& =\mathrm{i}\frac{L^{2}}{16\pi ^{2}}\int_{0}^{2\pi }\left( 2\,\mathrm{sgn\,}%
r_{\parallel }\frac{2}{\left( \left\vert r_{\parallel }\right\vert -\mathrm{i%
}r_{\perp }\cos \varphi \right) ^{3}}-r_{\parallel }\frac{6}{\left(
\left\vert r_{\parallel }\right\vert -\mathrm{i}r_{\perp }\cos \varphi
\right) ^{4}}\right) \cos \varphi \,\mathrm{d}\varphi  \notag \\
& =\mathrm{i}\frac{L^{2}}{8\pi ^{2}}\left( 2\,\mathrm{sgn\,}r_{\parallel
}\left( -\frac{1}{2\mathrm{i}}\frac{\partial ^{2}}{\partial \left\vert
r_{\parallel }\right\vert \partial r_{\perp }}\int_{0}^{2\pi } \! \! \frac{1%
}{\left\vert r_{\parallel }\right\vert -\mathrm{i}r_{\perp }\cos \varphi }%
\mathrm{d}\varphi \right) -3r_{\parallel }\left( \frac{1}{6\mathrm{i}}\frac{%
\partial ^{3}}{\partial \left\vert r_{\parallel }\right\vert ^{2}\partial
r_{\perp }}\int_{0}^{2\pi } \! \! \frac{1}{\left\vert r_{\parallel
}\right\vert -\mathrm{i}r_{\perp }\cos \varphi }\mathrm{d}\varphi \right)
\right)  \label{D-20} \\
& =-\frac{L^{2}}{8\pi ^{2}}\left( \mathrm{sgn\,}r_{\parallel }\left( 6\pi 
\frac{\left\vert r_{\parallel }\right\vert r_{\perp }}{\sqrt{\left\vert
r_{\parallel }\right\vert ^{2}+r_{\perp }^{2}}^{5}}\right) +\frac{1}{2}%
r_{\parallel }\left( 6\pi r_{\perp }\frac{-4\left\vert r_{\parallel
}\right\vert ^{2}+r_{\perp }^{2}}{\sqrt{\left\vert r_{\parallel }\right\vert
^{2}+r_{\perp }^{2}}^{7}}\right) \right)  \notag \\
& =-\frac{3L^{2}}{8\pi r^{3}}\left( 3-5\frac{r_{\parallel }^{2}}{r^{2}}%
\right) \frac{r_{\parallel }r_{\perp }}{r^{2}},  \label{DD-4}
\end{align}%
where the integral (\ref{D-8}) is used in (\ref{D-18}), the integrals (\ref%
{C-5}) and (\ref{D-15}) are used in (\ref{D-19}), while the integral (\ref%
{C-6}) is used in (\ref{D-20}).

One can show that the inverse Fourier transform of the remaining
off-diagonal matrix components yield zero, i.e.,%
\begin{align}
\mathcal{F}^{-1}&\left[ \left( \boldsymbol{L}\cdot \boldsymbol{k}\right) ^{2}%
\frac{k_{y}k_{z}}{k^{4}}\right] \left( \boldsymbol{r}\right) =\frac{1}{%
\left( 2\pi \right) ^{3}}\int_{%
\mathbb{R}
^{3}}\left( \boldsymbol{L}\cdot \boldsymbol{k}\right) ^{2}\frac{k_{y}k_{z}}{%
k^{4}}\mathrm{e}^{\mathrm{i}\boldsymbol{k}\cdot \boldsymbol{r}}\mathrm{d}_{%
\boldsymbol{k}}V  \notag \\
& =\frac{L^{2}}{\left( 2\pi \right) ^{3}}\int_{0}^{2\pi
}\!\!\!\int_{0}^{\infty }\!\!\!\int_{-\infty }^{\infty }\frac{k_{\parallel
}^{3}k_{\perp }}{\left( k_{\parallel }^{2}+k_{\perp }^{2}\right) ^{2}}\sin
\varphi \,\mathrm{e}^{\mathrm{i}k_{\boldsymbol{\parallel }}r_{\boldsymbol{%
\parallel }}}\mathrm{e}^{\mathrm{i}k_{\perp }r_{\boldsymbol{\perp }}\cos
\varphi }k_{\perp }\mathrm{d}k_{\parallel }\mathrm{d}k_{\perp }\mathrm{d}%
\varphi  \notag \\
& =\mathrm{i}\frac{L^{2}}{4\pi ^{3}}\int_{0}^{2\pi }\sin \varphi
\int_{0}^{\infty }k_{\perp }^{2}\mathrm{e}^{\mathrm{i}k_{\perp }r_{\perp
}\cos \varphi }\int_{0}^{\infty }\frac{k_{\parallel }^{3}\sin \left(
k_{\parallel }r_{\parallel }\right) }{\left( k_{\parallel }^{2}+k_{\perp
}^{2}\right) ^{2}}\mathrm{d}k_{\parallel }\mathrm{d}k_{\perp }\mathrm{d}%
\varphi  \notag \\
& =\mathrm{i}\frac{L^{2}}{4\pi ^{3}}\int_{0}^{2\pi }\sin \varphi
\int_{0}^{\infty }k_{\perp }^{2}\mathrm{e}^{\mathrm{i}k_{\perp }r_{\perp
}\cos \varphi }\left( \frac{\partial ^{3}}{\partial r_{\parallel }^{3}}%
\int_{0}^{\infty }\frac{\cos \left( k_{\parallel }r_{\parallel }\right) }{%
\left( k_{\parallel }^{2}+k_{\perp }^{2}\right) ^{2}}\mathrm{d}k_{\parallel
}\right) \mathrm{d}k_{\perp }\mathrm{d}\varphi  \label{D-17} \\
& =\mathrm{i}\frac{L^{2}}{4\pi ^{3}}\int_{0}^{2\pi }\sin \varphi
\int_{0}^{\infty }k_{\perp }^{2}\mathrm{e}^{\mathrm{i}k_{\perp }r_{\perp
}\cos \varphi }\left( \frac{\pi }{4}\left( 2\,\mathrm{sgn\,}r_{\parallel
}-r_{\parallel }k_{\perp }\right) \mathrm{e}^{-\left\vert r_{\parallel
}\right\vert k_{\perp }}\right) \mathrm{d}k_{\perp }\mathrm{d}\varphi  \notag
\\
& =\mathrm{i}\frac{L^{2}}{16\pi ^{2}}\int_{0}^{2\pi }\sin \varphi
\int_{0}^{\infty }k_{\perp }^{2}\left( 2\,\mathrm{sgn\,}r_{\parallel
}-r_{\parallel }k_{\perp }\right) \mathrm{e}^{-k_{\perp }\left( \left\vert
r_{\parallel }\right\vert -\mathrm{i}r_{\perp }\cos \varphi \right) }\mathrm{%
d}k_{\perp }\mathrm{d}\varphi  \label{D-21} \\
& =\mathrm{i}\frac{L^{2}}{16\pi ^{2}}\int_{0}^{2\pi }\sin \varphi \left( 2\,%
\mathrm{sgn\,}r_{\parallel }\frac{2}{\left( \left\vert r_{\parallel
}\right\vert -\mathrm{i}r_{\perp }\cos \varphi \right) ^{3}}-r_{\parallel }%
\frac{6}{\left( \left\vert r_{\parallel }\right\vert -\mathrm{i}r_{\perp
}\cos \varphi \right) ^{4}}\right) \mathrm{d}\varphi =0,  \label{DD-5}
\end{align}%
since cosine is an even function on the interval $\varphi \in \left( 0,2\pi
\right) $ with respect to $\varphi =\pi $, while sine is an odd function,
where the integral (\ref{C-6}) is used in (\ref{D-17}), while the integrals (%
\ref{C-5}) and (\ref{D-15}) are used in (\ref{D-21}).

Similarly, for the other off-diagonal matrix component one has%
\begin{align}
\mathcal{F}^{-1} &\left[ \left( \boldsymbol{L}\cdot \boldsymbol{k}\right)
^{2}\frac{k_{x}k_{y}}{k^{4}}\right] \left( \boldsymbol{r}\right) =\frac{1}{%
\left( 2\pi \right) ^{3}}\int_{%
\mathbb{R}
^{3}}\left( \boldsymbol{L}\cdot \boldsymbol{k}\right) ^{2}\frac{k_{x}k_{y}}{%
k^{4}}\mathrm{e}^{\mathrm{i}\boldsymbol{k}\cdot \boldsymbol{r}}\mathrm{d}_{%
\boldsymbol{k}}V  \notag \\
& =\frac{L^{2}}{\left( 2\pi \right) ^{3}}\int_{0}^{2\pi
}\!\!\!\int_{0}^{\infty }\!\!\!\int_{-\infty }^{\infty }\frac{k_{\parallel
}^{2}k_{\perp }^{2}}{\left( k_{\parallel }^{2}+k_{\perp }^{2}\right) ^{2}}%
\sin \varphi \cos \varphi \,\mathrm{e}^{\mathrm{i}k_{\boldsymbol{\parallel }%
}r_{\boldsymbol{\parallel }}}\mathrm{e}^{\mathrm{i}k_{\perp }r_{\perp }\cos
\varphi }k_{\perp }\mathrm{d}k_{\parallel }\mathrm{d}k_{\perp }\mathrm{d}%
\varphi  \notag \\
& =\frac{L^{2}}{4\pi ^{3}}\int_{0}^{2\pi }\sin \varphi \cos \varphi
\int_{0}^{\infty }k_{\perp }^{3}\mathrm{e}^{\mathrm{i}k_{\perp }r_{\perp
}\cos \varphi }\int_{0}^{\infty }\frac{k_{\parallel }^{2}\cos \left(
k_{\parallel }r_{\parallel }\right) }{\left( k_{\parallel }^{2}+k_{\perp
}^{2}\right) ^{2}}\mathrm{d}k_{\parallel }\mathrm{d}k_{\perp }\mathrm{d}%
\varphi  \notag \\
& =\frac{L^{2}}{4\pi ^{3}}\int_{0}^{2\pi }I\left( \boldsymbol{r},\varphi
\right) \sin \varphi \cos \varphi \,\mathrm{d}\varphi  \label{D-22} \\
& =\frac{L^{2}}{16\pi ^{2}}\int_{0}^{2\pi }\left( \frac{1}{\left( \left\vert
r_{\parallel }\right\vert -\mathrm{i}r_{\perp }\cos \varphi \right) ^{3}}%
-\left\vert r_{\parallel }\right\vert \frac{3}{\left( \left\vert
r_{\parallel }\right\vert -\mathrm{i}r_{\perp }\cos \varphi \right) ^{4}}%
\right) \sin \left( 2\varphi \right) \,\mathrm{d}\varphi =0,  \label{DD-6}
\end{align}%
since cosine is an even function on the interval $\varphi \in \left( 0,2\pi
\right) $ with respect to $\varphi =\pi $, while sine is an odd function,
where the integral $I$ defined by (\ref{D-1})$_{2}$, calculated as (\ref%
{D-23}), is used in (\ref{D-22}).

Collecting all calculated inversions of the matrix components, see (\ref%
{DD-1}), (\ref{DD-2}), (\ref{DD-3}), (\ref{DD-4}), (\ref{DD-5}), and (\ref%
{DD-6}), and by plugging them into the matrix (\ref{operator-D-fd}),
corresponding to the inverse Fourier transform of the tensor $\left( 
\boldsymbol{L}\cdot \boldsymbol{k}\right) ^{2}\frac{\boldsymbol{k}\otimes 
\boldsymbol{k}}{k^{4}},$ one obtains%
\begin{align}
\mathcal{F}^{-1}& \left[ \left( \boldsymbol{L}\cdot \boldsymbol{k}\right)
^{2}\frac{\boldsymbol{k}\otimes \boldsymbol{k}}{k^{4}}\right] \left( 
\boldsymbol{r}\right)  \notag \\
& =\delta \left( \boldsymbol{r}\right) \left( \boldsymbol{L}\otimes 
\boldsymbol{L}\right) +\frac{L^{2}}{8\pi r^{3}}\left[ 
\begin{array}{ccc}
-2+15\frac{r_{\parallel }^{2}}{r^{2}}-15\frac{r_{\parallel }^{4}}{r^{4}} & 0
& -3\left( 3-5\frac{r_{\parallel }^{2}}{r^{2}}\right) \frac{r_{\parallel
}r_{\perp }}{r^{2}} \\ 
0 & 1-3\frac{r_{\parallel }^{2}}{r^{2}} & 0 \\ 
-3\left( 3-5\frac{r_{\parallel }^{2}}{r^{2}}\right) \frac{r_{\parallel
}r_{\perp }}{r^{2}} & 0 & 3\left( 1-6\frac{r_{\parallel }^{2}}{r^{2}}+5\frac{%
r_{\parallel }^{4}}{r^{4}}\right)%
\end{array}%
\right]  \notag \\
& =\delta \left( \boldsymbol{r}\right) \left( \boldsymbol{L}\otimes 
\boldsymbol{L}\right) +\frac{L^{2}}{8\pi r^{3}}\left( 1-3\frac{r_{\parallel
}^{2}}{r^{2}}\right) \boldsymbol{\hat{I}}+\frac{L^{2}}{8\pi r^{3}}\left[ 
\begin{array}{ccc}
-3+18\frac{r_{\parallel }^{2}}{r^{2}}-15\frac{r_{\parallel }^{4}}{r^{4}} & 0
& -3\left( 3-5\frac{r_{\parallel }^{2}}{r^{2}}\right) \frac{r_{\parallel
}r_{\perp }}{r^{2}} \\ 
0 & 0 & 0 \\ 
-3\left( 3-5\frac{r_{\parallel }^{2}}{r^{2}}\right) \frac{r_{\parallel
}r_{\perp }}{r^{2}} & 0 & 2-15\frac{r_{\parallel }^{2}}{r^{2}}+15\frac{%
r_{\parallel }^{4}}{r^{4}}%
\end{array}%
\right]  \label{matrix} \\
& =\delta \left( \boldsymbol{r}\right) \left( \boldsymbol{L}\otimes 
\boldsymbol{L}\right) +\frac{L^{2}}{8\pi r^{3}}\left( \left( 1-3\frac{%
r_{\parallel }^{2}}{r^{2}}\right) \boldsymbol{\hat{I}}+2\frac{\boldsymbol{L}%
\otimes \boldsymbol{L}}{L^{2}}-6\frac{r_{\parallel }}{r}\left( \frac{%
\boldsymbol{r}}{r}\otimes \frac{\boldsymbol{L}}{L}+\frac{\boldsymbol{L}}{L}%
\otimes \frac{\boldsymbol{r}}{r}\right) -3\left( 1-5\frac{r_{\parallel }^{2}%
}{r^{2}}\right) \frac{\boldsymbol{r}\otimes \boldsymbol{r}}{r^{2}}\right) , 
\notag
\end{align}%
since the matrix in the expression (\ref{matrix}) can be represented by%
\begin{align*}
& \alpha \frac{\boldsymbol{L}\otimes \boldsymbol{L}}{L^{2}}+\beta \left( 
\frac{\boldsymbol{L}}{L}\otimes \frac{\boldsymbol{r}}{r}+\frac{\boldsymbol{r}%
}{r}\otimes \frac{\boldsymbol{L}}{L}\right) +\gamma \frac{\boldsymbol{r}%
\otimes \boldsymbol{r}}{r^{2}} \\
& \quad =\alpha \left( \boldsymbol{e}_{z}\otimes \boldsymbol{e}_{z}\right)
+\beta \left( \boldsymbol{e}_{z}\otimes \left( \frac{r_{\perp }}{r}%
\boldsymbol{e}_{x}+\frac{r_{\parallel }}{r}\boldsymbol{e}_{z}\right) +\left( 
\frac{r_{\perp }}{r}\boldsymbol{e}_{x}+\frac{r_{\parallel }}{r}\boldsymbol{e}%
_{z}\right) \otimes \boldsymbol{e}_{z}\right) +\gamma \left( \left( \frac{%
r_{\perp }}{r}\boldsymbol{e}_{x}+\frac{r_{\parallel }}{r}\boldsymbol{e}%
_{z}\right) \otimes \left( \frac{r_{\perp }}{r}\boldsymbol{e}_{x}+\frac{%
r_{\parallel }}{r}\boldsymbol{e}_{z}\right) \right) \\
& \quad =\gamma \frac{r_{\perp }^{2}}{r^{2}}\left( \boldsymbol{e}_{x}\otimes 
\boldsymbol{e}_{x}\right) +\left( \beta \frac{r_{\perp }}{r}+\gamma \frac{%
r_{\parallel }r_{\perp }}{r^{2}}\right) \left( \boldsymbol{e}_{x}\otimes 
\boldsymbol{e}_{z}+\boldsymbol{e}_{z}\otimes \boldsymbol{e}_{x}\right)
+\left( \alpha +2\beta \frac{r_{\parallel }}{r}+\gamma \frac{r_{\parallel
}^{2}}{r^{2}}\right) \left( \boldsymbol{e}_{z}\otimes \boldsymbol{e}%
_{z}\right) ,
\end{align*}%
implying%
\begin{equation*}
\left[ 
\begin{array}{ccc}
\gamma \frac{r_{\perp }^{2}}{r^{2}} & 0 & \beta \frac{r_{\perp }}{r}+\gamma 
\frac{r_{\parallel }r_{\perp }}{r^{2}} \\ 
0 & 0 & 0 \\ 
\beta \frac{r_{\perp }}{r}+\gamma \frac{r_{\parallel }r_{\perp }}{r^{2}} & 0
& \alpha +2\beta \frac{r_{\parallel }}{r}+\gamma \frac{r_{\parallel }^{2}}{%
r^{2}}%
\end{array}%
\right] =\left[ 
\begin{array}{ccc}
-3+18\frac{r_{\parallel }^{2}}{r^{2}}-15\frac{r_{\parallel }^{4}}{r^{4}} & 0
& -3\left( 3-5\frac{r_{\parallel }^{2}}{r^{2}}\right) \frac{r_{\parallel
}r_{\perp }}{r^{2}} \\ 
0 & 0 & 0 \\ 
-3\left( 3-5\frac{r_{\parallel }^{2}}{r^{2}}\right) \frac{r_{\parallel
}r_{\perp }}{r^{2}} & 0 & 2-15\frac{r_{\parallel }^{2}}{r^{2}}+15\frac{%
r_{\parallel }^{4}}{r^{4}}%
\end{array}%
\right] ,
\end{equation*}%
yielding%
\begin{equation*}
\alpha =2,\quad \beta =-6\frac{r_{\parallel }}{r},\quad \gamma =-3\left( 1-5%
\frac{r_{\parallel }^{2}}{r^{2}}\right) .
\end{equation*}

\noindent \textbf{CRediT authorship contribution statement}\smallskip

\noindent \textbf{Sla\dj an Jeli\'{c}}: Conceptualization, Formal analysis,
Investigation, Methodology, Visualization, Writing original draft, Writing -
review \& editing.

\noindent \textbf{Du\v{s}an Zorica}: Conceptualization, Formal analysis,
Investigation, Methodology, Visualization, Writing original draft, Writing -
review \& editing. \smallskip

\noindent \textbf{Declaration of Competing Interest} \smallskip

\noindent The authors declare that they have no conflict of
interest.\smallskip

\noindent \textbf{Data availability }\smallskip

\noindent No data was used for the research described in the article.
\smallskip

\noindent \textbf{Acknowledgements} \smallskip

\noindent The work is supported by the Ministry of Science, Technological
Development and Innovation of the Republic of Serbia under grants
451-03-33/2026-03/200125 \& 451-03-34/2026-03/200125 (SJ and DZ).

\smallskip

\noindent The authors are grateful to Dr. Hassan Abdalla for the valuable
discussions and insightful comments that helped in improving the quality of
paper.

\bibliographystyle{plain}
\bibliography{dz}

\end{document}